\documentclass[12pt,preprint,a4paper]{aastex}
\usepackage{times}
\usepackage{graphics,epsf}
\usepackage{amsmath}                
\usepackage{amsfonts}               
\usepackage{amssymb}                
\usepackage{epsfig}                 
\usepackage{graphicx}
\usepackage{url}
\usepackage{tabularx}
\usepackage{booktabs}
\usepackage{epstopdf}
\usepackage{subfigure}

\def \cm{~\rm{cm}}
\def \s{~\rm{s}}
\def \km{~\rm{km}}

\def \K{~\rm{K}}
\def \g{~\rm{g}}

\def \AU{~\rm{AU}}

\def \yrs{~\rm{yrs}}
\def \yr{~\rm{yr}}

\begin{document}

\title{Forming H-shaped and barrel-shaped nebulae with interacting jets}

\author{Muhammad Akashi\altaffilmark{1}, Ealeal Bear\altaffilmark{1}, and Noam Soker\altaffilmark{1,2}}

\altaffiltext{1}{Department of Physics, Technion -- Israel
Institute of Technology, Haifa 32000, Israel; akashi@physics.technion.ac.il; ealealbh@gmail.com; 
soker@physics.technion.ac.il}

\altaffiltext{2}{Guangdong Technion Israel Institute of Technology, Shantou, Guangdong Province, China}

\begin{abstract}
We conduct three-dimensional hydrodynamical simulations of two opposite jets launched from a binary stellar system into a previously ejected shell and show that the interaction can form barrel-like and H-like shapes in the descendant nebula. Such features are observed in planetary nebulae and supernova remnants. Under our assumption the dense shell is formed by a short instability phase of the giant star as it interacts with a stellar companion, and the jets are then launched by the companion as it accretes mass through an accretion disk from the giant star. 
We find that the H-shaped and barrel-shaped morphological features that the jets form evolve with time, and that there are complicated flow patterns, such as vortices, instabilities, and caps moving ahead along the symmetry axis. 
We compare our numerical results with images of 12 planetary nebulae, and show that jet-shell interaction that we simulate can account for the barrel-like or H-like morphologies that are observed in these PNe. 
\end{abstract}


\section{INTRODUCTION}
\label{sec:intro}

An interacting binary system can supply gravitational energy from mass transfer and from the spiraling-in process that releases gravitational orbital energy. The mass that streams from one star to the other has a relatively large specific angular momentum and in many cases it is accreted through an accretion disk. One of the expectations is that the accretion disk launches two opposite jets that shape the nebula that is formed by the mass-loss process from the binary system. 
It is not surprising therefore that the finding that most, and probably all, planetary nebulae (PNe) are shaped by binary interaction (e.g, \citealt{Bondetal1978, BondLivio1990, SokerHarpaz1992, NordhausBlackman2006, GarciaSeguraetal2014, DeMarco2015, Zijlstra2015}) and the understanding that in many cases jets shape the PNe have evolved together (e.g., \citealt{Morris1987, Soker1990AJ, SahaiTrauger1998, Boffinetal2012, HuarteEspinosaetal2012, Balicketal2013, Miszalskietal2013, Tocknelletal2014, Huangetal2016, Sahaietal2016, RechyGarciaetal2016, GarciaSeguraetal2016}). 
 
Recent years have seen further development. Most important is the hard work of many observing groups and theoretical studies that put the binary-shaping model of PNe on an extremely solid ground  
(e.g., some papers from 2015 on, \citealt{Akrasetal2015, Alleretal2015a, Boffin2015, Corradietal2015, Decinetal2015, DeMarcoetal2015, Douchinetal2015, Fangetal2015, Gorlovaetal2015, Jonesetal2015, Manicketal2015, Martinezetal2015, Miszalskietal2015, Mocniketal2015, Montezetal2015, Akrasetal2016, Alietal2016, Bondetal2016, Chenetal2016, Chiotellisetal2016, GarciaRojasetal2016, Hillwigetal2016a, Jones2016, Jonesetal2016, Madappattetal2016, Chenetal2017, Hillwigetal2017,JonesBoffin2017}).
The establishment of the binary shaping mechanism of PNe has promoted further research in different directions, such as comparison to other types of binary systems, e.g., to symbiotic stars (e.g., \citealt{Clyneetal2015, RamosLariosetal2017}), and the shaping of PNe by triple stellar systems (e.g., \citealt{BearSoker2017, Hilleletal2017}), including jets in triple stellar systems \citep{AkashiSoker2017}. 
Very impressive is the similarity of some bipolar PNe to binary systems composed of a post-asymptotic giant branch (post-AGB) primary star and a main sequence binary companion with a bipolar nebula around them, e.g., the Red Rectangle (e.g., \citealt{VanWinckel2014}). 

The orbital separation of these post-AGB intermediate binaries (post-AGBIBs) is roughly $1 \AU$, intermediate between post-common envelope binary systems and binary systems that their orbital separation has increased because of mass-loss during the AGB evolution. There are strong observational indications (e.g., \citealt{Gorlovaetal2015, VanWinckel2017}) that in probably most post-AGBIBs the secondary main sequence star  launches two opposite jets. These jets have wide opening angles \citep{Thomasetal2013, Bollenetal2017}, as we have simulated in the past 
(e.g., \citealt{AkashiSoker2008b}). Similar arguments seem to hold for post-red giant branch (RGB) intermediate binaries (e.g., \citealt{Kamathetal2016}). 

There are two approaches to simulate the shaping of nebulae by jets. One approach is to consider specific systems and try to simulate their specific structure. Examples include the simulations of the PNe (or and proto-PNe) CRL~618 by \cite{Leeetal2009} and \cite{Balicketal2013}, of  OH~231.8+04.2 by \cite{Balicketal2017}, and of the knotty jet in Hen 2-90 by \cite{LeeSahai2004}. The second approach is to conduct general studies of features common to many PNe. Examples, to list a small number, include the role of magnetic fields \citep{Dennisetal2009}, formation of butterfly and elliptical PNe \citep{HuarteEspinosaetal2012}, simulations of the formation of a clumpy equatorial ring \citep{Akashietal2015}, and simulations to explain a bulge on the front of a main lobe  \citep{AkashiSoker2008a}.  

In the present study we examine the formation H-like and/or barrel-like morphological features that are observed in some PNe and other types of nebulae, such as supernova remnants of core collapse supernovae (CCSNe). The supernova remnant W49B (e.g., \citealt{Lopezetal2013}) has an H-like shape in its inner region as revealed by X-ray lines of, e.g., Ar and Ca, while its outer region seen in X-ray continuum has a barrel-like shape. \cite{BearSoker2017} compare these morphological features to those of some PNe, and suggest that two opposite jets launched at explosion are responsible for these morphological features.

We describe simulations of jets that interact with a spherical circumstellar matter (CSM). We assume that the primary giant star experienced a high mass-loss rate episode that lasted for a limited period of time and formed a dense spherical shell. We further assume that jets are launched by a secondary star that accretes mass from the primary giant star. The scenario we assume is within the binary shaping regime where jets play a major role in the shaping process. Due to numerical limitations we do not include the orbital motion of the secondary star around the centre of mass. We do conduct full 3D hydrodynamical simulations to follow instabilities when they develop.  

We describe the initial setting and the numerical code in section \ref{sec:numerical}. In section \ref{sec:flow} we describe the basic properties of the flow, and in section \ref{sec:morphologies} we present the different morphologies we have obtained. In section \ref{sec:PNe} we compare our results to observations of 12 PNe that contain H-like or barrel-like morphological features.  
We summarize our results in section \ref{sec:summary}. 
  
\section{NUMERICAL SET-UP}
 \label{sec:numerical}
\subsection{The approach to nebular shaping}
 \label{subsec:nebular}
 
 There is a very large parameter space to simulate the shaping of PNe. We limit ourselves to a specific setting that follows our decade-old approach of launching jets into a dense nebula. In most cases the pre-jet nebulae have a spherical morphology (e.g., \citealt{AkashiSoker2008a, AkashiSoker2008b, AkashiSoker2013}). We follow the earlier simulations in setting the principles of the simulations. (1) A large fraction of the mass, in most cases most of it, was ejected at early times in a spherical geometry. It could have been also an elliptical morphology, but we have limited most of our simulations to a spherical slow ejecta. (2) Due to a binary interaction, in some cases there is a phase of very high mass-loss rate from the AGB progenitor, but still at a low velocity. This forms a shell. In some cases we launch the jets after the ejection of a dense shell and after the AGB wind has weakened, and in some cases there is no shell and the jets are launched simultaneously with the slow AGB wind (e.g.,  \citealt{GarciaArredondoFrank2004}). (3) We inject two opposite jets into the dense slow wind/shell. The jets can be launched in one (relatively short) episode, can be continuous for the entire duration of the simulation, or can be intermittent. The jets can precess or can maintain a constant axis.

Although we simulate jets having a large opening angle, a half opening angle  $\alpha = 30^\circ$ in the present study and up to $70 ^\circ$ in past studies, we keep referring to them as jets. Firstly, there is no quantitative definition on a limiting opening angle for a collimate outflow to be called a jet. Secondly, these wide bipolar outflows obey our criteria for jets, as they (a) have an axisymmetrical outflow, (b) have a mirror symmetry about the equatorial plane, namely two opposite jets, and (c) there is no outflow in a substantial solid angle about the equatorial plane.  If we follow these criteria, we would also term jets the ejection of two opposite bullets. What \cite{Balicketal2013} term a set of two opposite bullets we would call intermittent jets.

The setting of the flow in the present study follows our long-term objective to show that jets can lead to a rich varieties of morphologies. In this study, we focus on H-shaped and barrel-shaped morphologies. This does not say that other types of flow setting, some that involve no jets, cannot lead to some of the morphologies we simulate with jets. It also does not mean that other flow settings with jets do not lead to H-shaped and barrel-shaped morphologies. We here choose the simplest setting we could think of (not considering the different values of the shell mass and size and jets' opening angles, etc.), namely, one shell and on pair of jets. We did not check intermittent jets or two shells or an elliptical rather than a spherical shell.

Indeed, there are other approaches to the shaping of nebulae. Contrary to our approach of spherical slow wind and aspherical fast wind (jets), there is the setting of an aspherical slow wind and a spherical fast wind (e.g., \citealt{MellemaFrank1995, Steffenetal2013}). This is generally termed the  generalized interacting stellar winds (GISW) scenario.  Such a setting can also lead to barrel-shaped and H-shaped nebulae (e.g., \citealt{Franketal1993, FrankMellema1994}). Interacting winds models with magnetic fields as simulated by \cite{Steffenetal2009} can also form these shapes. 
 
\subsection{Code and initial conditions}
 \label{subsec:code}

We use version 4.2.2 of the hydrodynamical FLASH code \citep{Fryxell2000} to perform our 3D simulations. The FLASH code is an adaptive-mesh refinement (AMR) modular code used for solving hydrodynamics or magnetohydrodynamics problems. Here we use the unsplit PPM
(piecewise-parabolic method) solver of FLASH. We do not include gravity as velocities are much above the escape speed in the regions we simulate. We include radiative cooling of optically thin gas in some of the runs, and ignore radiative cooling in others. The reason not to include radiative cooling in some runs is to allow future scaling of our results to cases where radiative cooling is not important, e.g., much lower densities.

We employ a full 3D AMR (9 levels; $2^{12}$ cells in each direction) using a Cartesian grid $(x,y,z)$ with outflow boundary conditions at all boundary surfaces. We take the $x-y$ plane with
$z=0$ to be in the equatorial plane of the binary system, that is also the equatorial plane of the nebula, and we simulate the whole space (the two sides of the equatorial plane).

\begin{figure}[h!]
\begin{center}
\includegraphics[width=120mm]{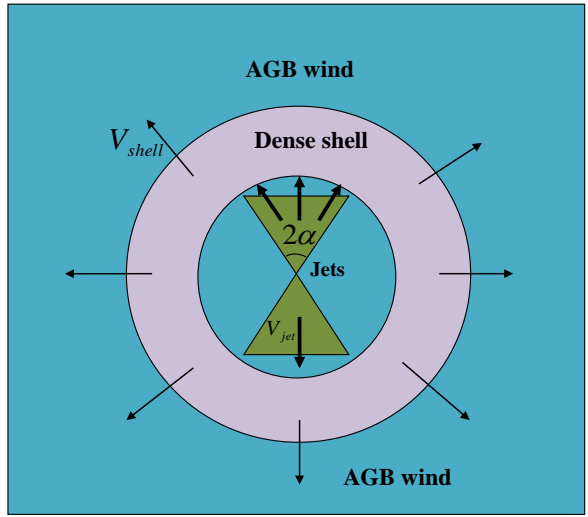}
\vskip -6.0 cm
\caption{A schematic drawing of the outflow from the binary star in the meridional plane. 
The binary system is much smaller than the grid, and it is at the center of the grid $(x,y,z)=(0,0,0)$ during the entire simulation.}
 \label{fig:schematic}
\end{center}
%
%
\end{figure}

At time $t=0$ we place a spherical dense shell around the center, as we schematically present in Fig. \ref{fig:schematic}. We simulate cases for two shell sizes. In each case we use only one of the two shells. 
We take the large shell to be in zone $10^{17} \cm < r <  1.1 \times 10^{17} \cm$ and with a density profile of $\rho_s = 1.6 \times 10^{-19} (r/10^{17} \cm)^{-2} \g \cm^{-3}$, 
such that the total mass in the shell is $0.1M_\odot$. 
For the small shell we take the spherical zone and the density profile to be $ 5 \times 10^{16} \cm < r <  5.3 \times 10^{16} \cm$ and $\rho_s = 5.3 \times 10^{-20} (r/10^{17}\cm)^{-2} \g \cm^{-3}$, respectively, such that the total mass in the small shell is $0.01M_\odot$. 
The gas in the shells has an initial radial velocity of $v_s = 10 \km \s^{-1}$. 
The large shell was formed during a time of about 300 years with an average mass-loss rate of about $3 \times 10^{-4} M_\odot \yr^{-1}$, while for the small shell these numbers are about 100 years and 
$10^{-4} M_\odot \yr^{-1}$. 

 The regions outside and inside the dense shell are filled with a much lower density gas, the spherically slow wind, with a density profile of $\rho_{\rm wind}(r) = 7.5 \times 10^{-22} (r/10^{17} \cm)^{-2}$ and a velocity of $v_{\rm wind}=v_s= 10 \km \s^{-1}$. This wind was formed by a constant mass-loss rate of $\dot M_{\rm wind}=1.5 \times 10^{-6}{\rm M_\odot \yr^{-1}}$.
 
We launch the two opposite jets from the inner $10^{16} \cm $ zone along the $z$-axis and  within a half opening angle of $\alpha = 30^\circ$. By the term `jets' we refer also to wide
outflows, as we simulate here. More generally, we simulate slow-massive-wide (SMW) outflows. 
We simulate cases in which we either stop the launching of the jets at $t=48\yrs$ or we continue launching to the end of the run. 
The jets' initial velocity is $v_{\rm jet}=800 \km \s^{-1}$, just a little above the escape speed form a main sequence star. The mass-loss rate into the two jets together is 
$\dot M_{\rm 2jets} = 2.2 \times 10^{-5} M_\odot \yr^{-1}$. 
 
The slow wind, dense shell, and the ejected jets start with a temperature of $10000 \K$. The initial jets' temperature has no influence on the results (as long as it is highly supersonic) because the jets rapidly cool due to adiabatic expansion. For numerical reasons a very weak slow wind is injected in the sector $\alpha<\theta<90^\circ$ (more numerical details are in \citealt{AkashiSoker2013}).

We simulate different cases, by taking either a small or a large shell, by either including or not radiative cooling, and by either launching jets for the entire duration of the simulation of only during the first 48 years. The present study is a general study aiming to extend the general morphological features that interacting jets can form, and it is not aiming to fit a specific nebula. For that we use representative initial parameters, as we summarize in Table \ref{table:Runs}. 
\begin{table}[ht]
\centering
\begin{tabular}{l*{6}{c}r}
Run              & Shell & Jets' launching & $\dot E_{\rm rad}$ \\
\hline
R1    & Small  &  48 yr       & No  \\
R2    & Small  &  48 yr       & Yes \\
R3    & Large  &  Continuous  & No  \\
R4    & Large  &  Continuous  & Yes  \\
R5    & Large  &  48 yr       & No \\
\hline
\end{tabular}
\caption{Initial parameters of the simulations. $\dot E_{\rm rad}$ stands for radiative cooling in optically thin gas. }
\label{table:Runs}
\end{table}

\section{THE FLOW STRUCTURE}
 \label{sec:flow}

\begin{figure}
\begin{center}
\subfigure[$t=50$~\yr]{\includegraphics[height=3.6in,width=3.6in,angle=0]{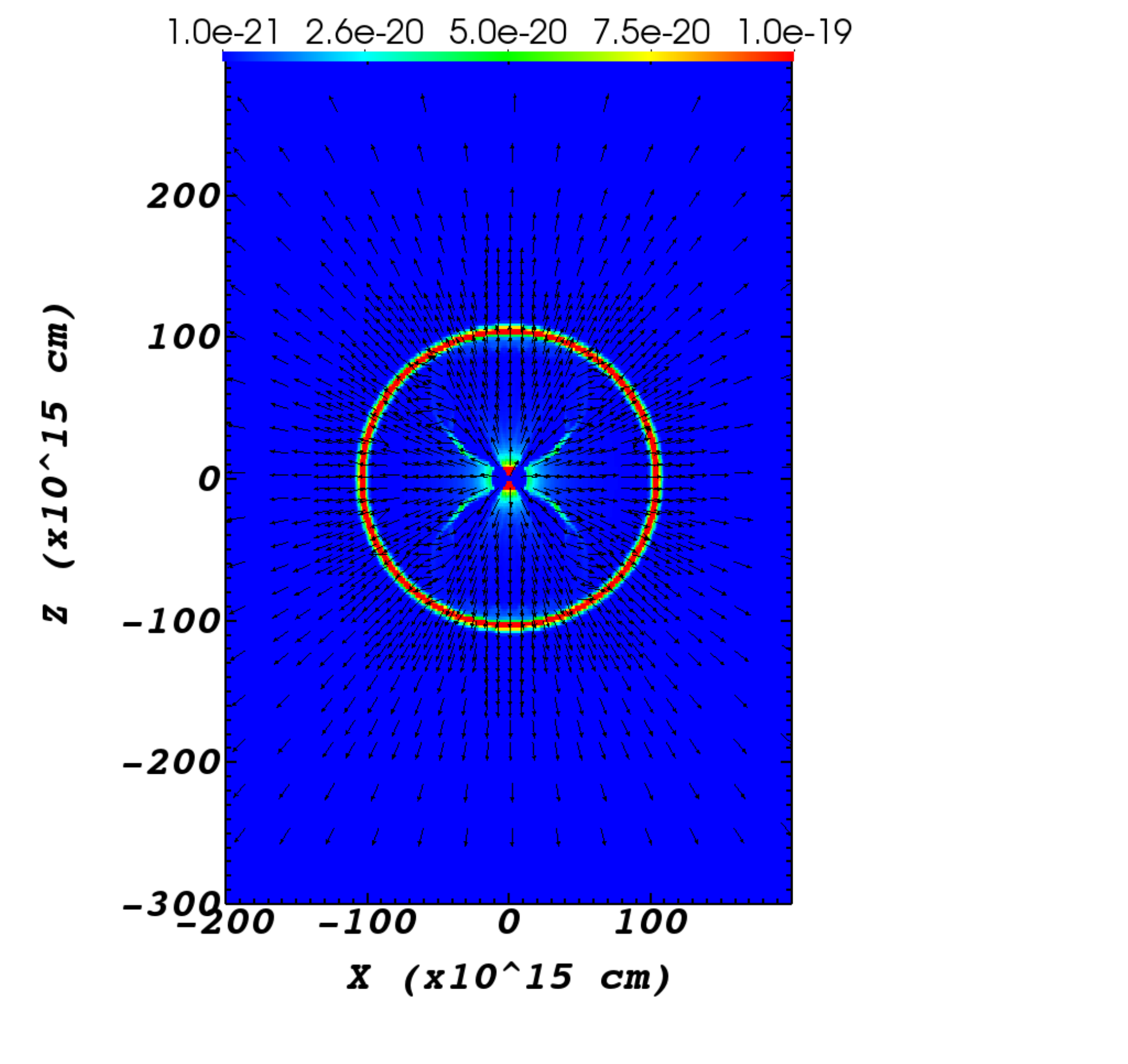}}
\hskip -2.0 cm
\subfigure[$t=108$~\yr]{\includegraphics[height=3.6in,width=3.6in,angle=0]{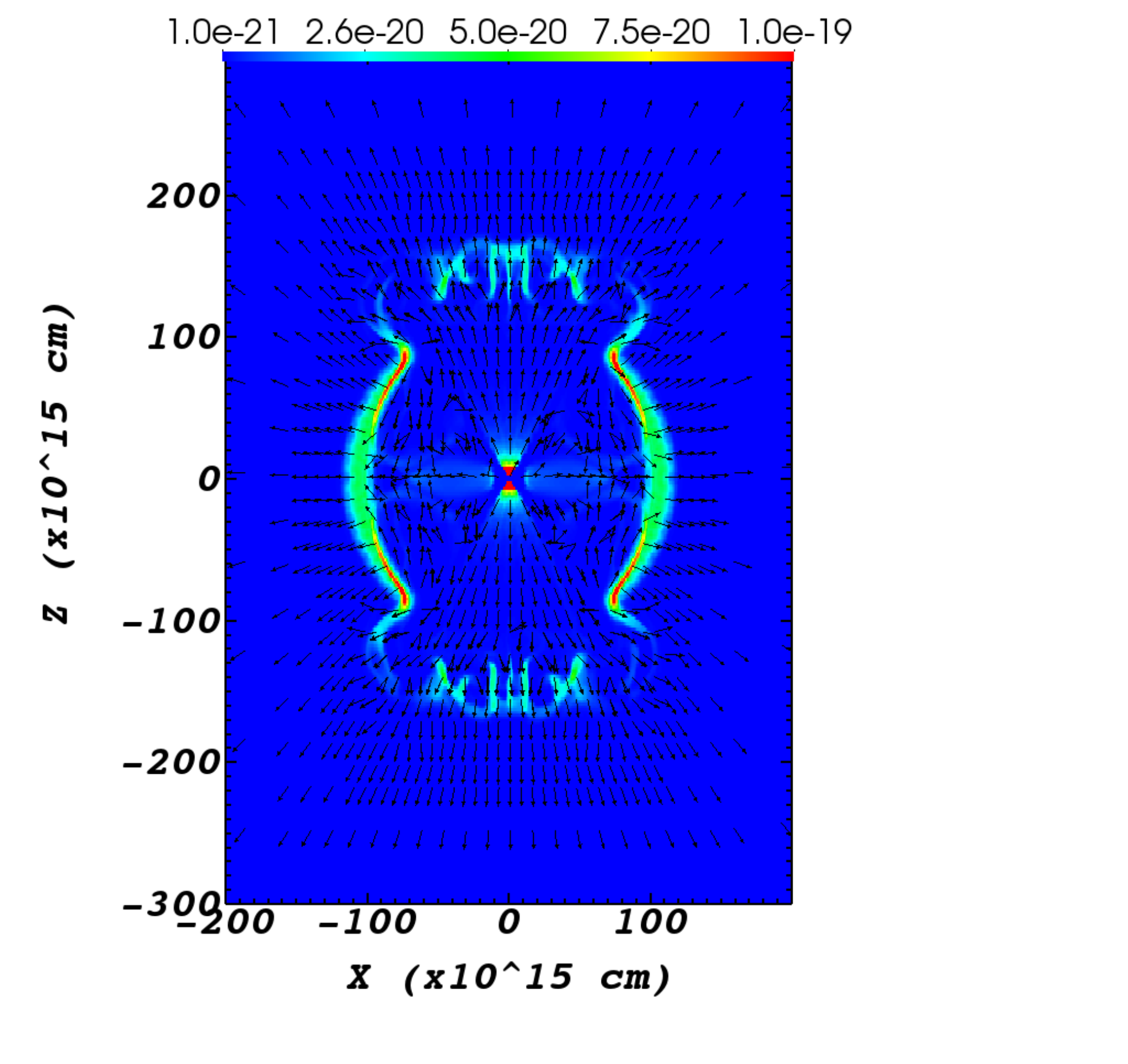}}
\subfigure[$t=184$~\yr]{\includegraphics[height=3.6in,width=3.6in,angle=0]{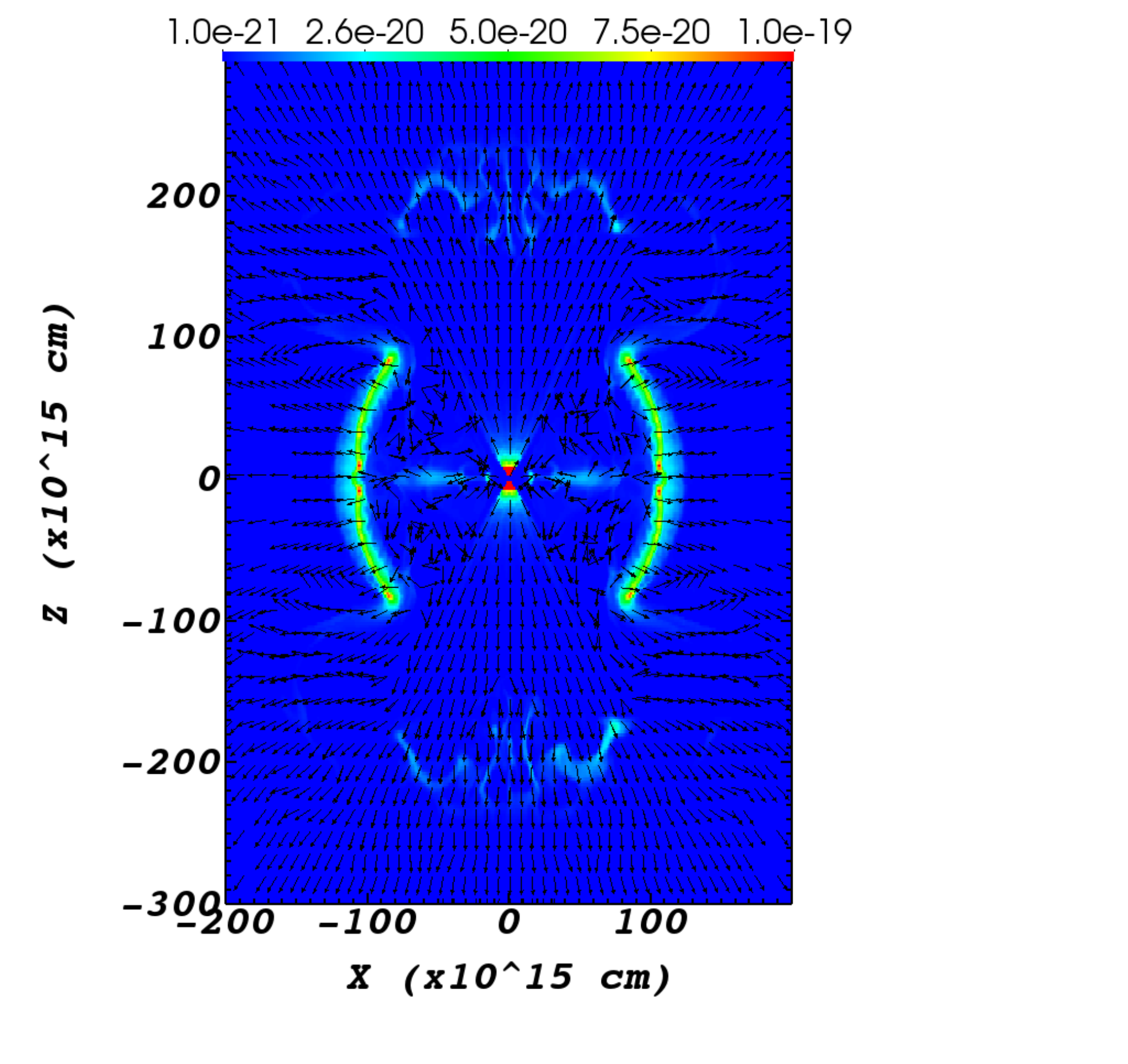}}
\hskip -2.0 cm
\subfigure[$t=245$~\yr]{\includegraphics[height=3.6in,width=3.6in,angle=0]{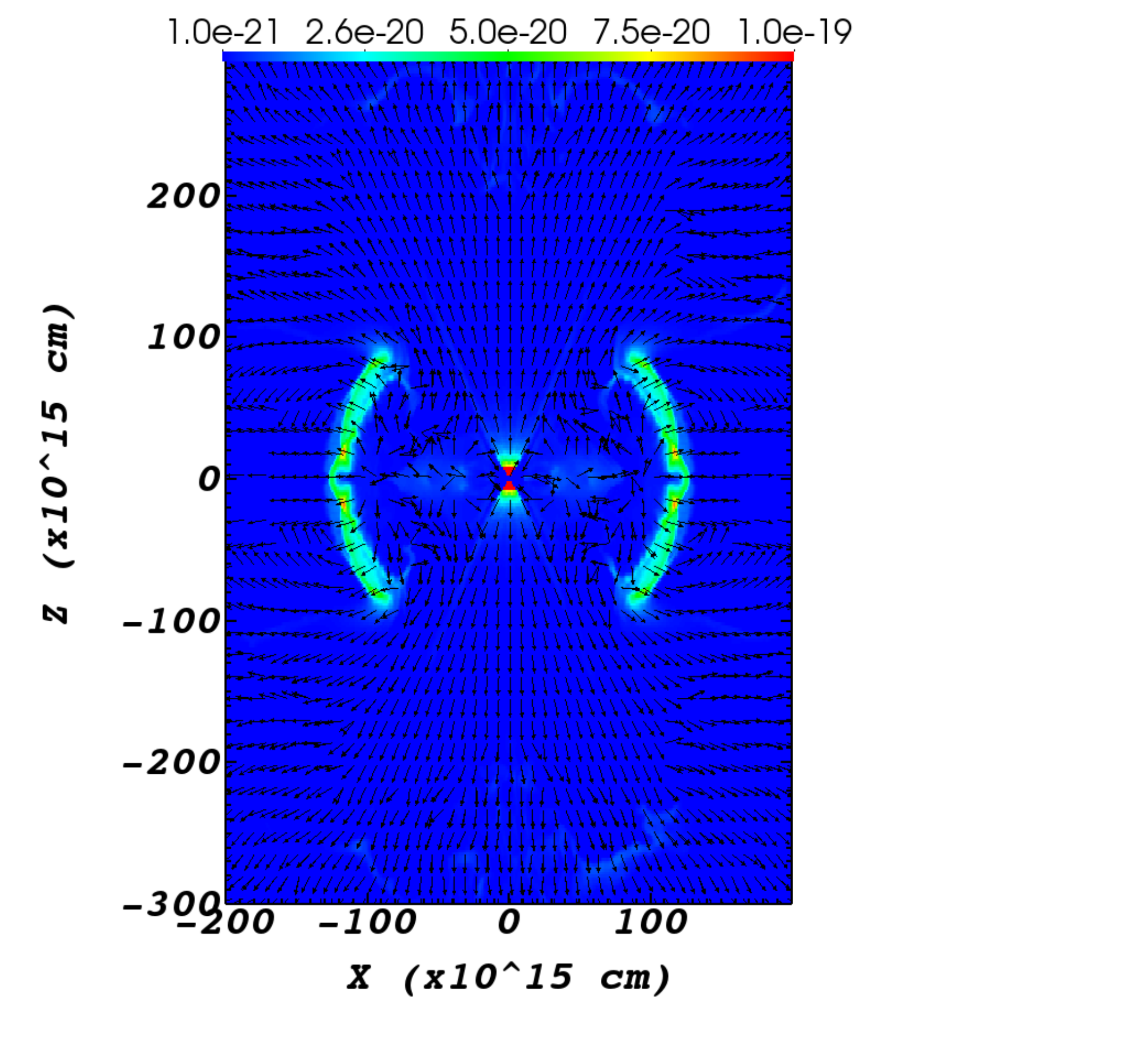}}
\caption{The density and velocity directions (black arrows) maps of run R3 in the meridional plane $y=0$ at four times (for the initial parameters see table \ref{table:Runs}).  
The density scale is given by the colour-bar in units of $\g \cm^{-3}$. }
\label{fig:Flow1xz}
\end{center}
\end{figure}

\begin{figure}
\begin{center}
\subfigure[$t=108$~\yr]{\includegraphics[height=3.6in,width=3.6in,angle=0]{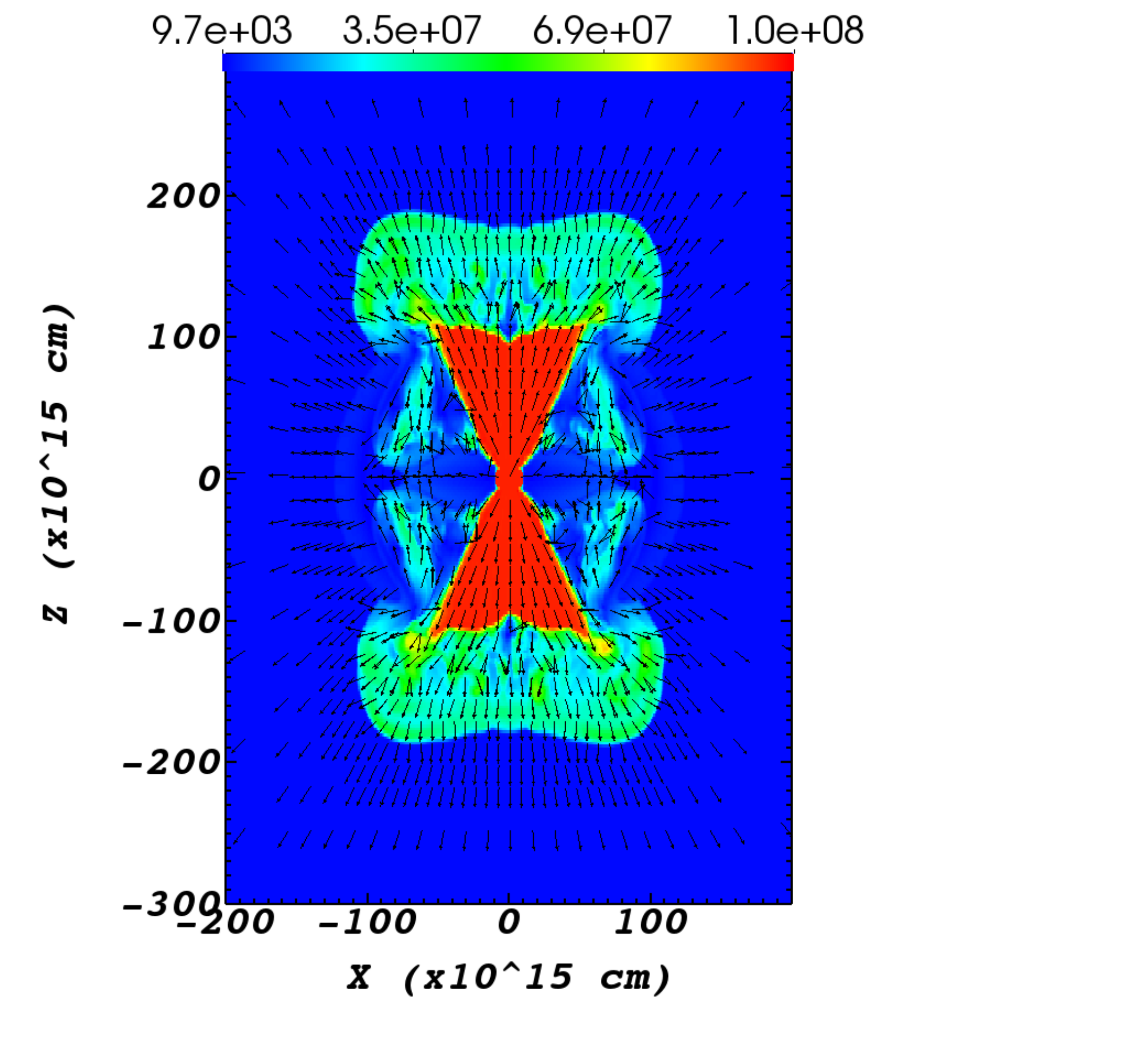}}
\hskip -2.0 cm
\subfigure[$t=245$~\yr]{\includegraphics[height=3.6in,width=3.6in,angle=0]{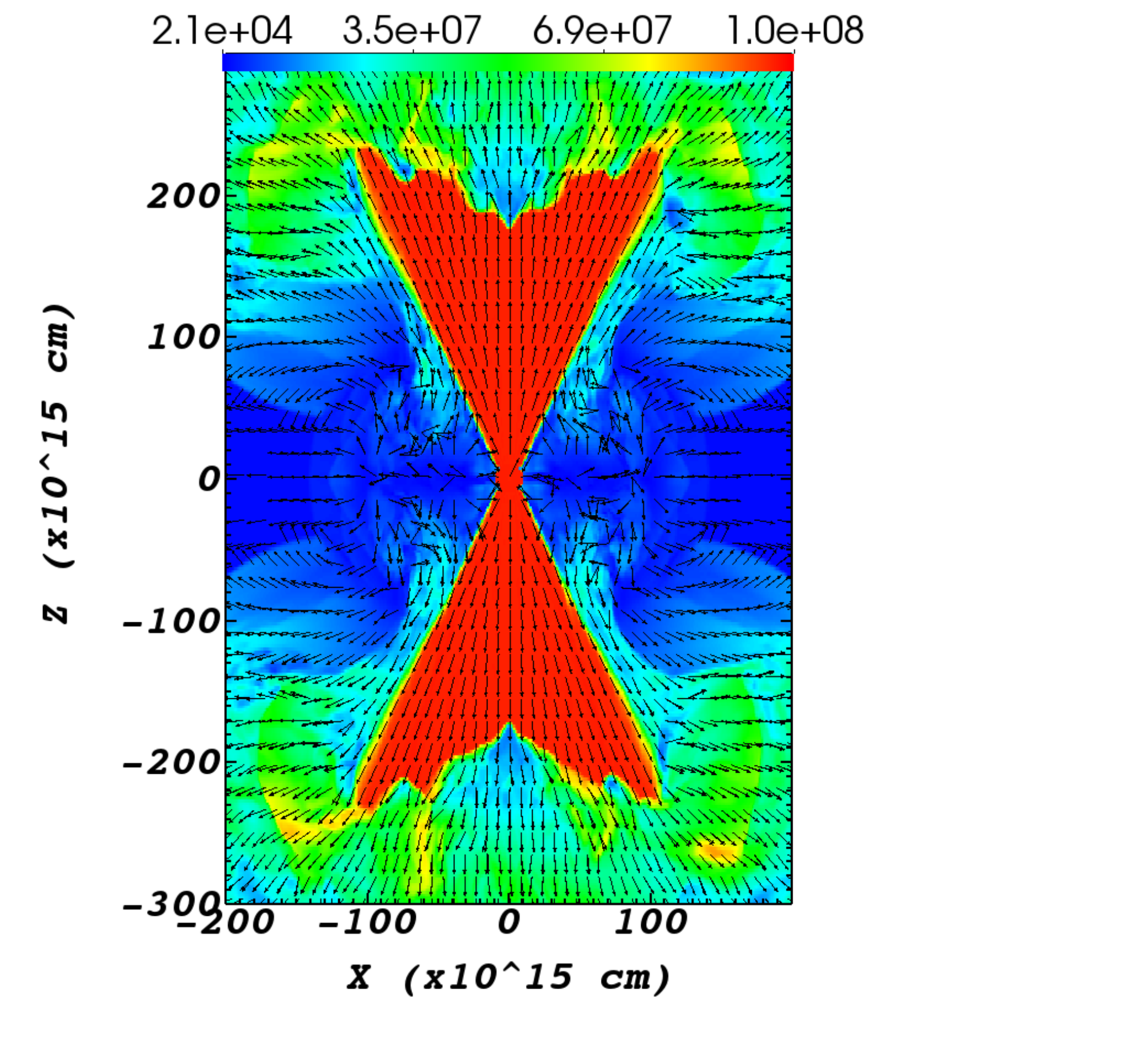}}
\caption{The velocity directions (black arrows) and magnitude of run R3 in the meridional plane at two times. The magnitude of the velocity is given by the colour-bar in units of $\cm \s^{-1}$.  }
\label{fig:Flow1xzVel}
\end{center}
\end{figure}

Before turning to describe the morphological features that we have obtained in the different simulations, for one case of run R3 we present the evolution of the flow with time. In Fig. \ref{fig:Flow1xz} we present the density and velocity maps at four times in the meridional plane, and in Fig. \ref{fig:Flow1xzVel} we show the magnitude of the velocity at two times. 
   
Before the jets hit the dense shell they interact with the slow wind inner to the dense shell. This interaction leads to the opening of two opposite cones with dense walls. In the meridional plane this has a curved `X' shape (convex in upper part and concave in lower part; green region in panel a of Fig. \ref{fig:Flow1xz}). Such a structure is observed in many bipolar PNe, but it is not the focus of the present study. The two green cones in the density maps that touch the center are the pre-shock regions of the jets close to the center. The full regions of the pre-shocked jets are seen as red zones in the velocity maps of Fig. \ref{fig:Flow1xzVel}. Then the gas is shocked and the velocity drops.

Panel b of Fig. \ref{fig:Flow1xz} at $t=108 \yr$ presents the flow after the jets have removed the two opposite portions of the shell in the polar directions. 
The two regions are seen as two opposite `caps' above and below the barrel-shaped main nebula (coloured green and red in panel b). Due to strong Rayleigh-Taylor instabilities the caps are broken to several `fingers'. The interaction with the dense shells leads to high pressure regions that accelerate some gas backwards. At later times (panel c and d of Fig. \ref{fig:Flow1xz}) the jets continue to disperse the caps, and the dense parts form a barrel-like structure.

\begin{figure}
\begin{center}
\subfigure[$t=50$~\yr]{\includegraphics[height=2.6in,width=2.6in,angle=0]{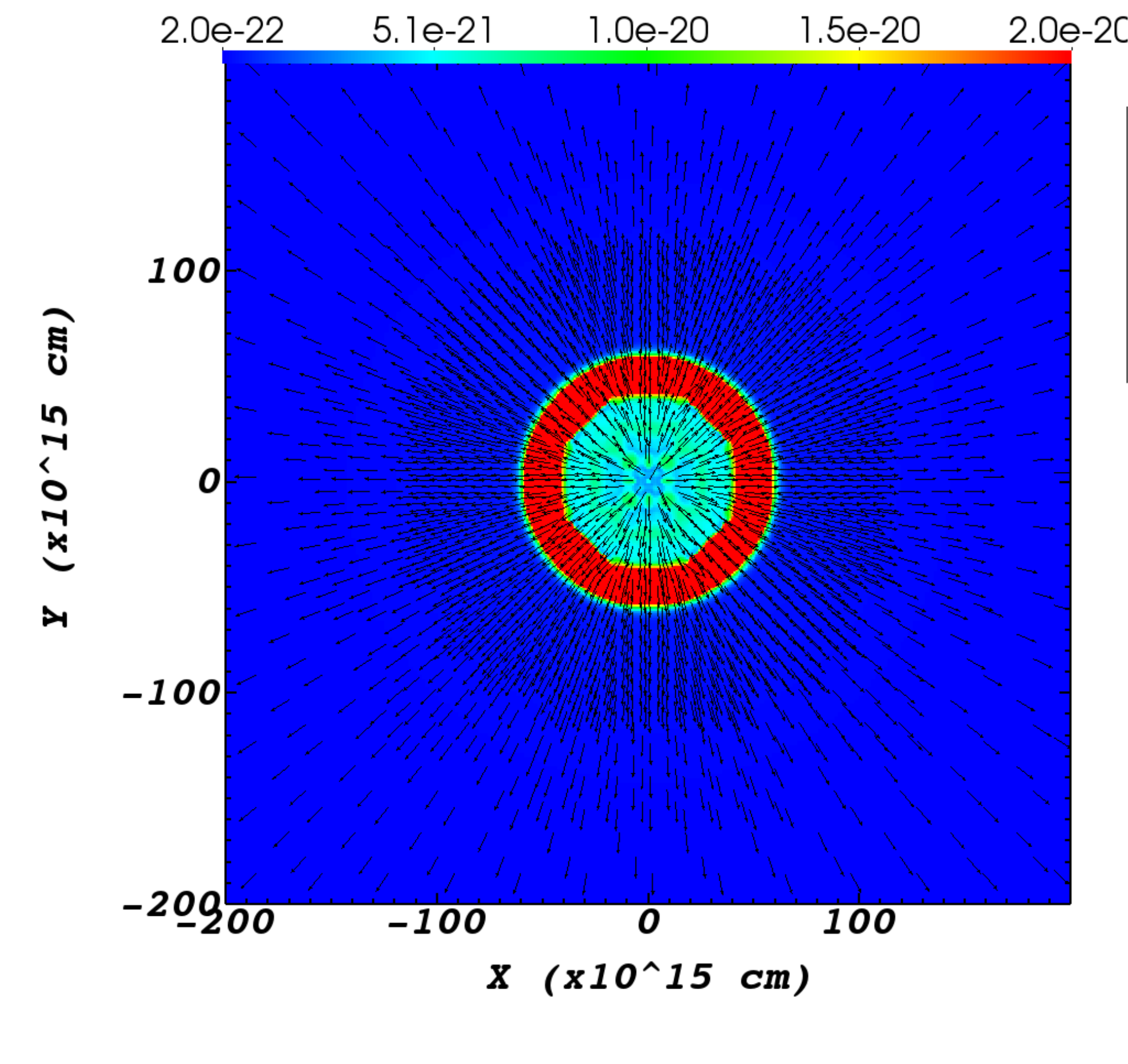}}
\subfigure[$t=108$~\yr]{\includegraphics[height=2.6in,width=2.6in,angle=0]{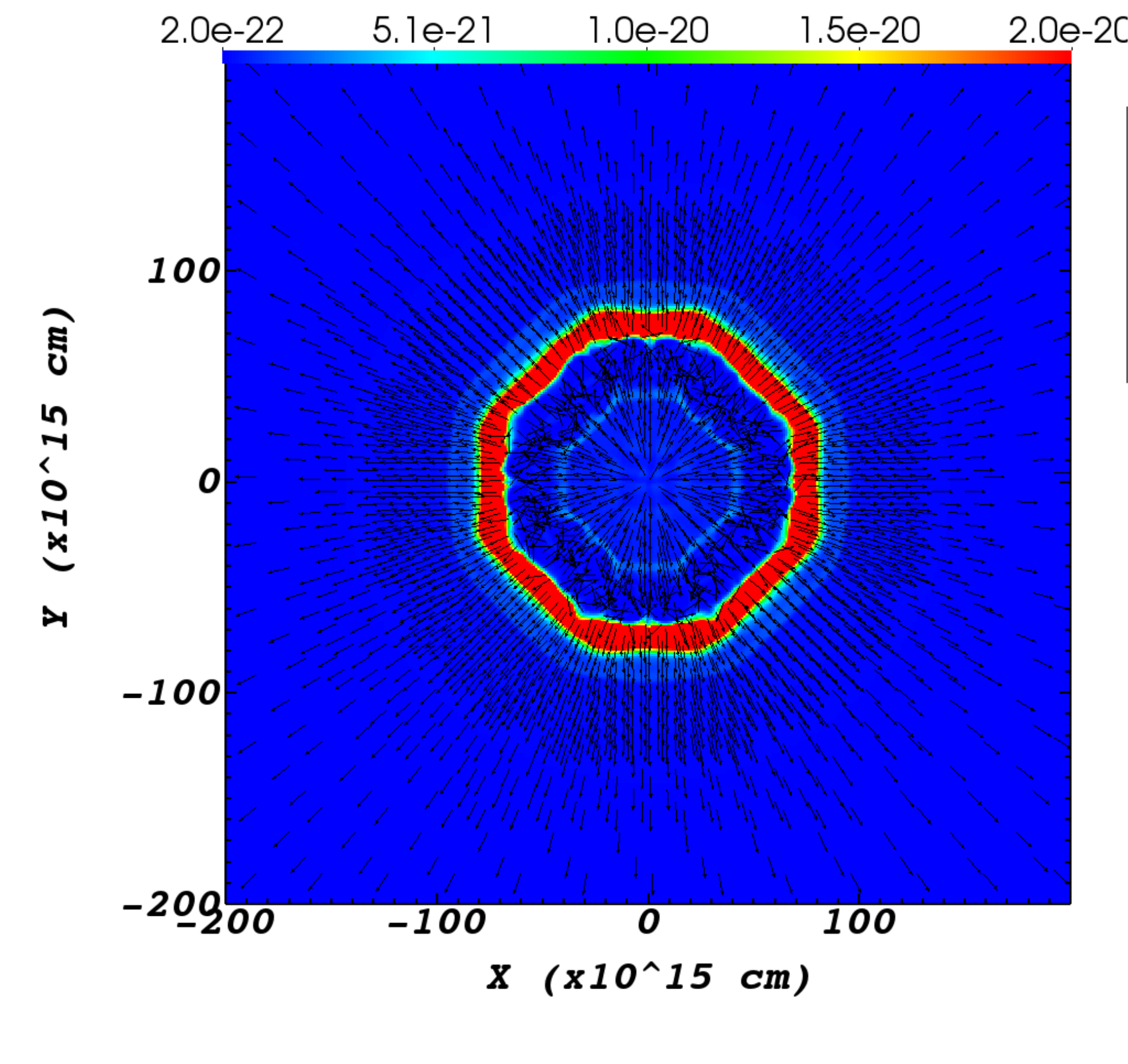}}
\subfigure[$t=184$~\yr]{\includegraphics[height=2.6in,width=2.6in,angle=0]{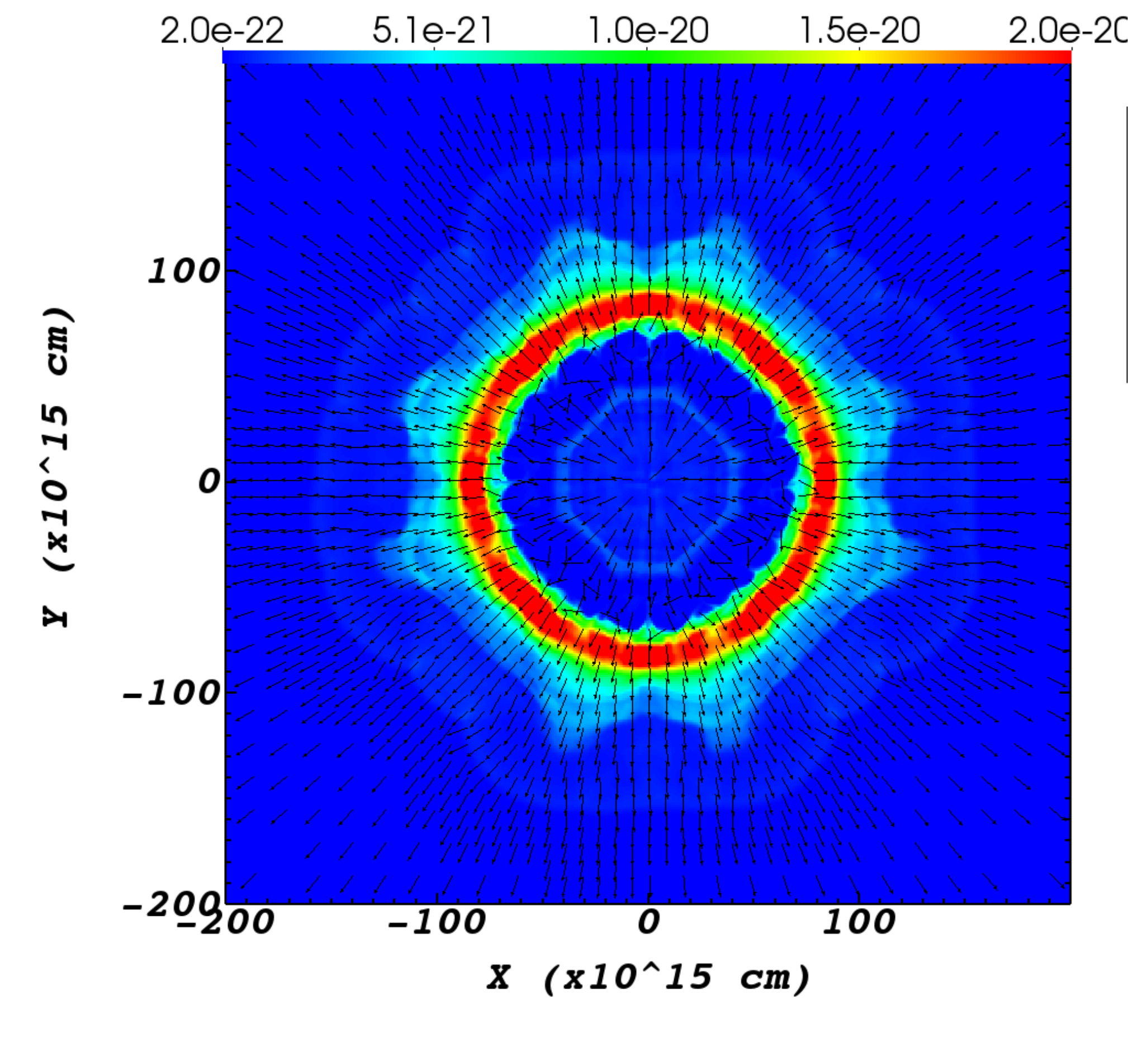}}
\subfigure[$t=245$~\yr]{\includegraphics[height=2.6in,width=2.6in,angle=0]{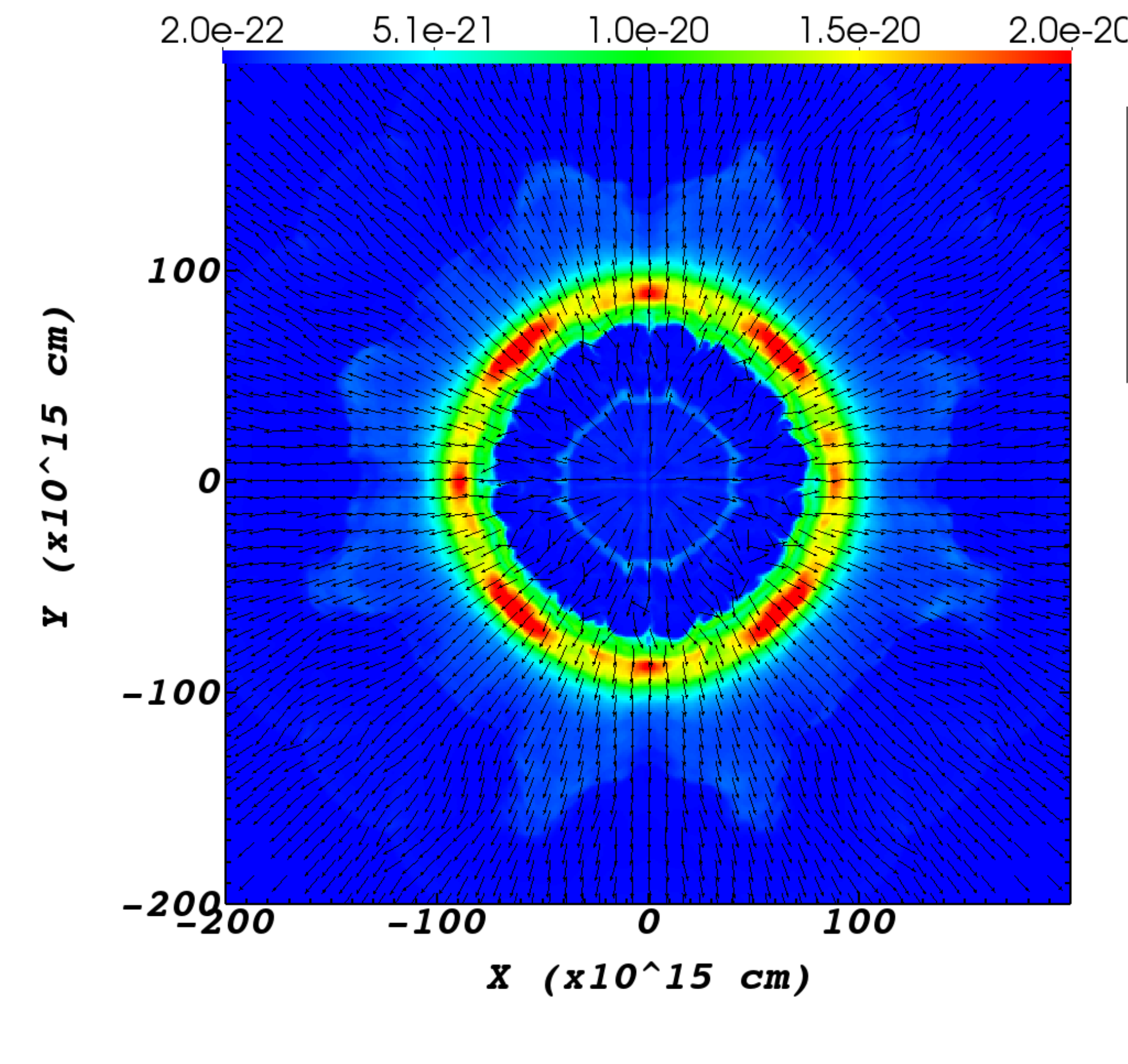}}
\caption{The density and velocity directions (black arrows) maps of run R3 in the  plane $z=9\times 10^{16} \cm$ at four times as in Fig. \ref{fig:Flow1xz}.
The density scale is given by the colour-bar in units of $\g \cm^{-3}$.  }
\label{fig:Flow1xy}
\end{center}
\end{figure}

In Fig. \ref{fig:Flow1xy} we present density and velocity maps in the $z=9 \times 10^{16} \cm$ plane, where the equatorial plane is the $z=0$ plane, for the same run as in the previous two figures. We clearly see the disruption of this region of the shell by the jets. The instabilities are physical and real, but their location and initial perturbations are determine by the shape of the grid and the finite resolution. 

\section{MORPHOLOGIES}
 \label{sec:morphologies}
 
\begin{figure}
\begin{center}
\vskip -0.8 cm
\subfigure{\includegraphics[height=2.5in,width=2.5in,angle=0]{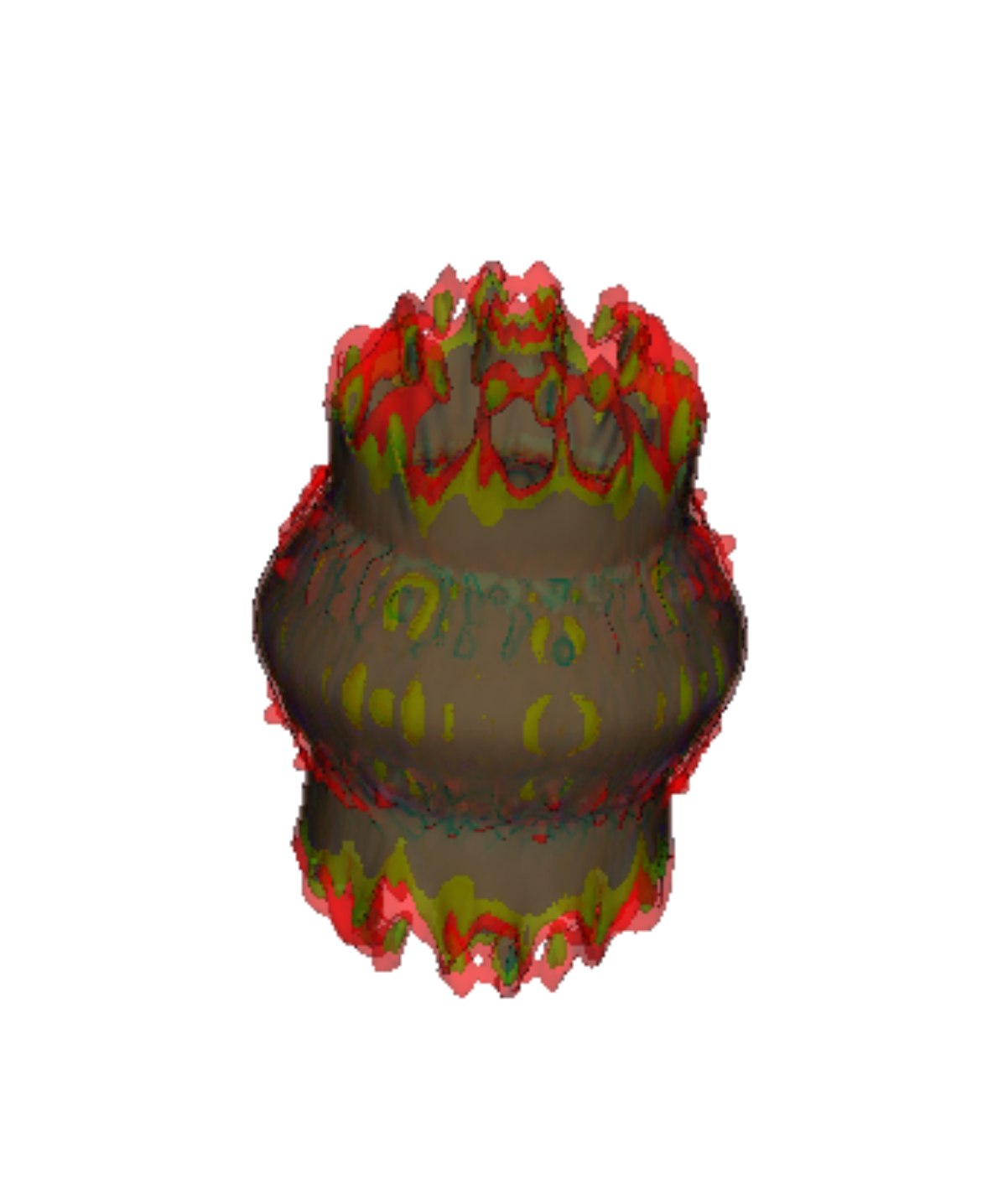}}
\subfigure{\includegraphics[height=2.5in,width=2.5in,angle=0]{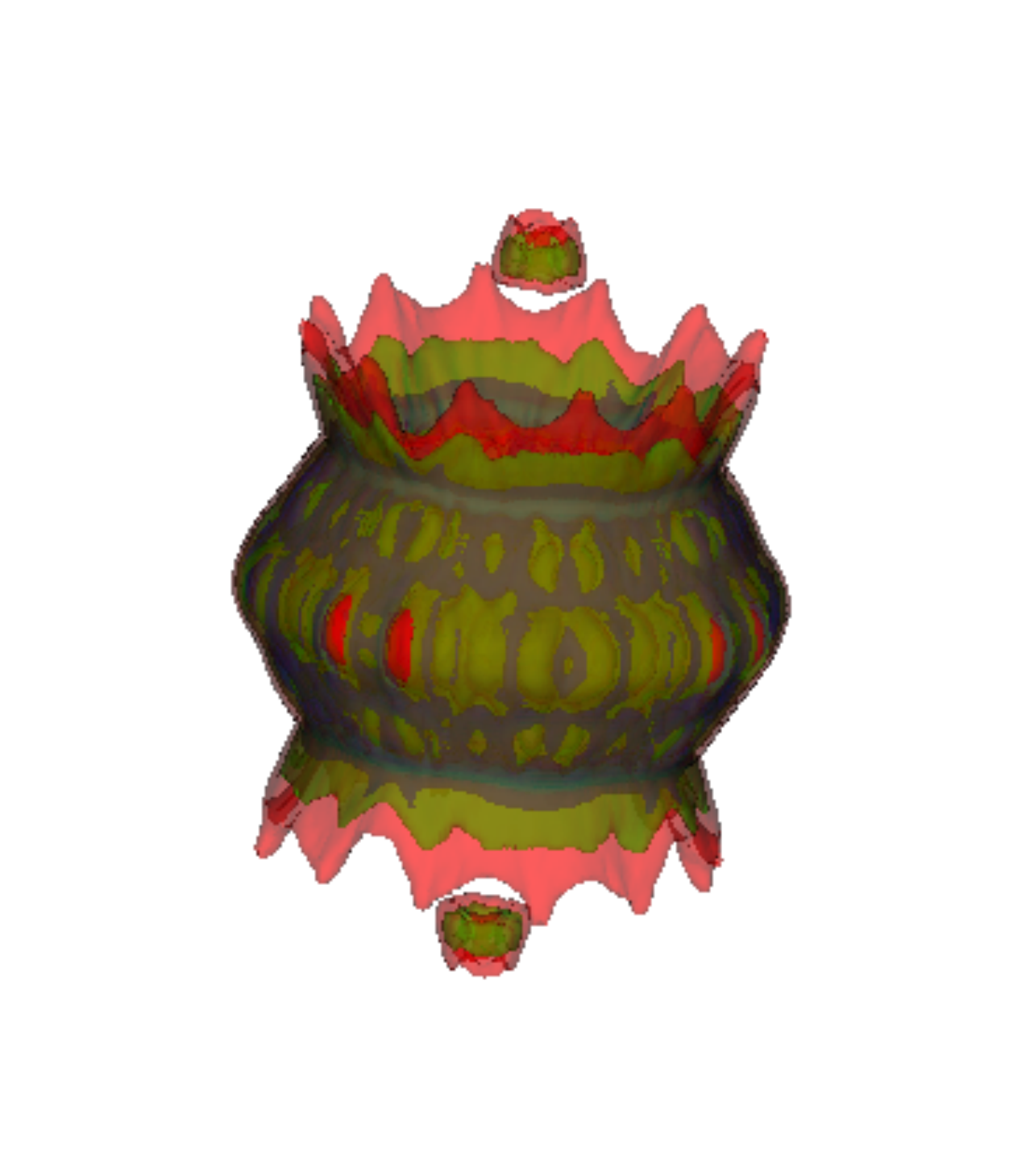}}
\vskip -0.3 cm
\hskip 1.1 cm
\subfigure{\includegraphics[height=2.2in,width=2.5in,angle=0]{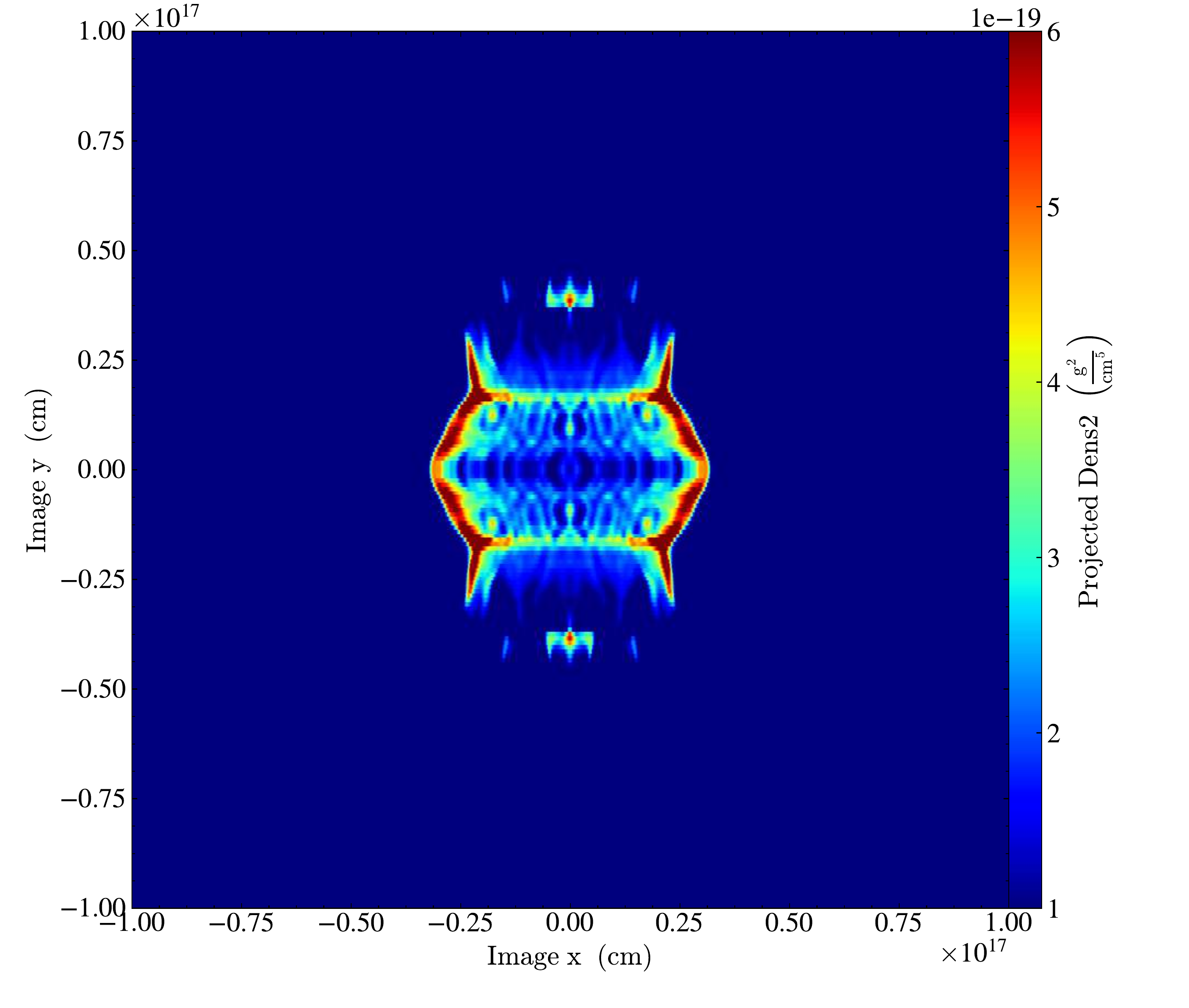}}
\subfigure{\includegraphics[height=2.2in,width=2.5in,angle=0]{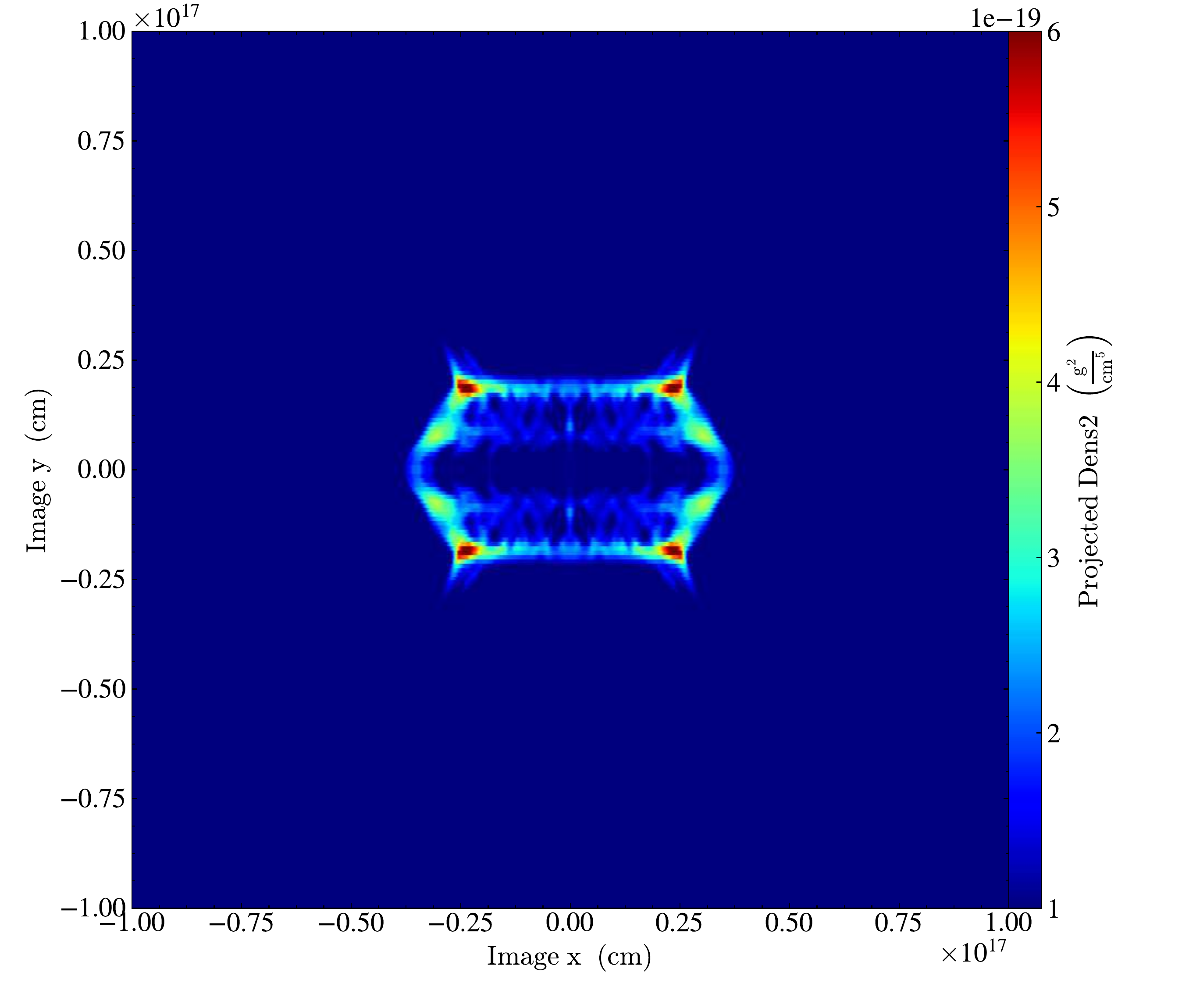}}
\vskip -0.3 cm
\subfigure{\includegraphics[height=2.5in,width=2.5in,angle=0]{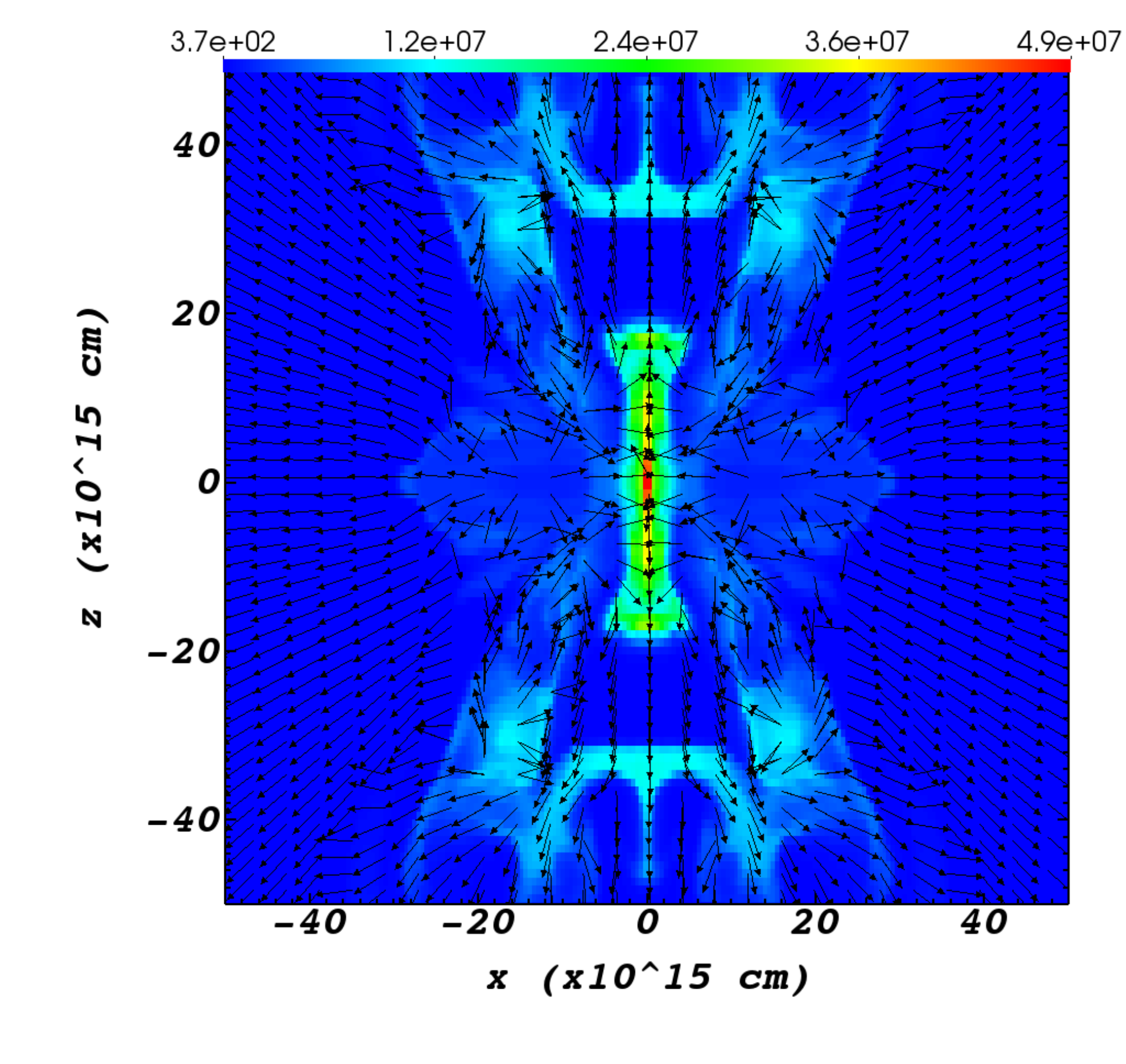}}
\subfigure{\includegraphics[height=2.5in,width=2.5in,angle=0]{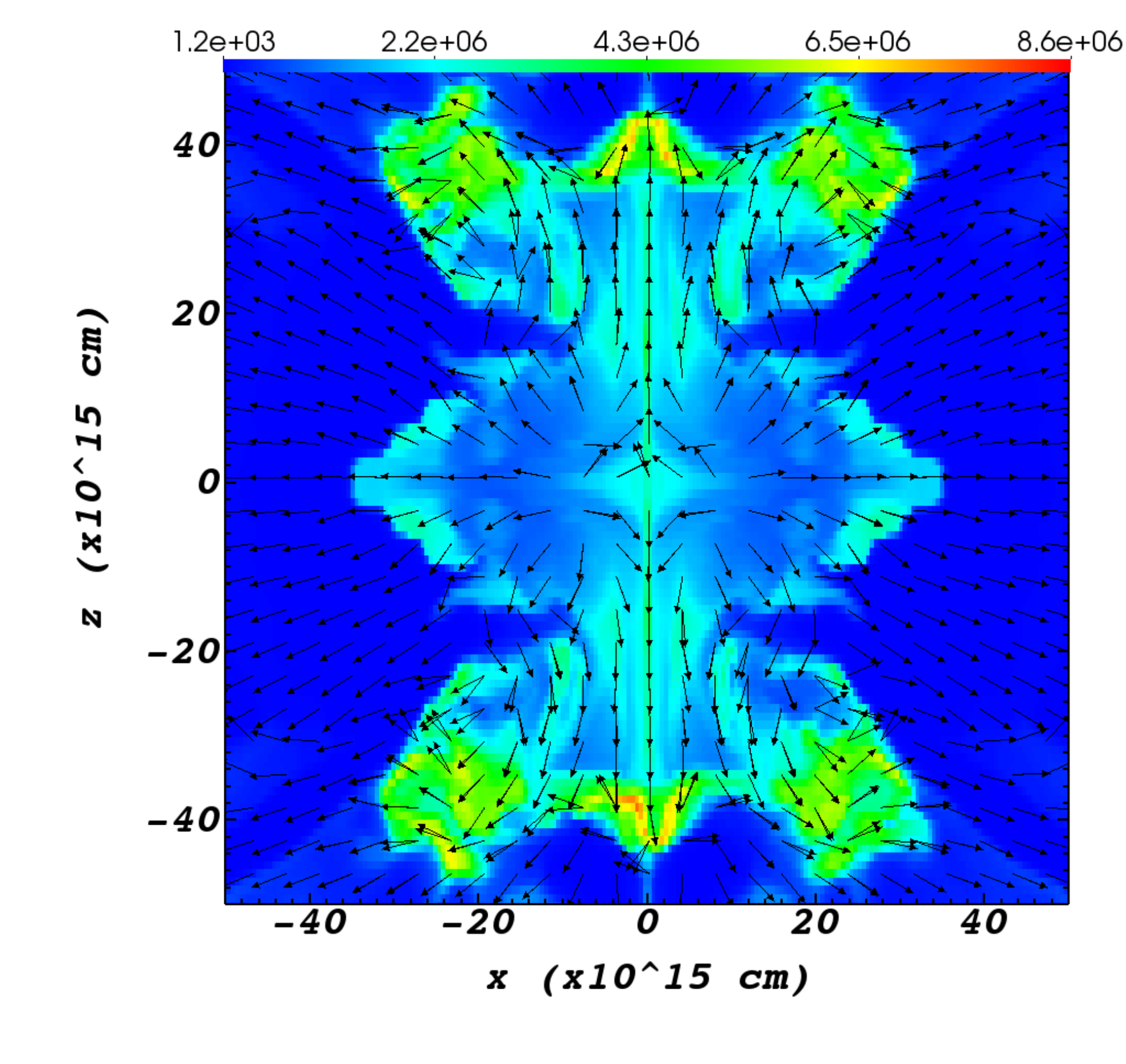}}
\vskip -0.8 cm
\caption{Results of run R1 (see table \ref{table:Runs}) at two times, $107 \yr$ and $150 \yr$ in the left and right column, respectively. 
Upper panels are three dimensional maps of the nebula, where colours from red (lowest density) to green indicate increasing density.  
Middle-row panels present the integration of the density square along the line of sight $\int \rho^2 dy$. The scale (bar on top of each panel) in the left and right panels run from zero (blue) to upper values (red) of $4.1 \times 10^{-31}\g^2 \cm^{-5}$ and $2.2 \times 10^{-31} \g^2 \cm^{-5}$, respectively (the numbers written on the bars are numerical units). 
The lower panels present the temperature and velocity maps in the meridional plane $y=0$. Arrows present only the direction of the flow. The colours correspond to the temperature as indicated by the bar.
}
\label{fig:R1}
\end{center}
\end{figure}

In Fig. \ref{fig:R1} we present the results of run R1 at two times as indicated. This run has no radiative cooling and the jets' launching phase lasts for only 48 years. In the 3D plots in the upper row we can identify a barrel shape nebula that was formed by the interaction of the jets with a dense shell. 
By barrel-like shape we refer to an axisymmetrical nebula with an almost cylindrically shaped dense envelope, but the envelope on the sides has a small curvature out, and the entire structure is trimmed on both ends. 
We also notice condensations along the symmetry axis on both sides and outside the main nebula (sometimes termed ansae). 
The two panels in the second row of Fig. \ref{fig:R1} show the integrated square density along the y axis, i.e., $\int \rho^2 dy$. This quantity mimics the brightness of the nebula at each point on the sky when the symmetry axis of the nebula is on the plane of the sky.  
 In the lower-row panels of Fig. \ref{fig:R1} we present the temperature maps and velocity-direction maps in the meridional plane $y=0$. The arrows have one length, and they depict only the direction of the velocity. We notice hot gas that was shock-heated to millions of degrees.
The most important result of this run is the formation of a general barrel-like morphology. 

\begin{figure}
\begin{center}
\vskip -0.2 cm
\subfigure{\includegraphics[height=2.7in,width=2.7in,angle=0]{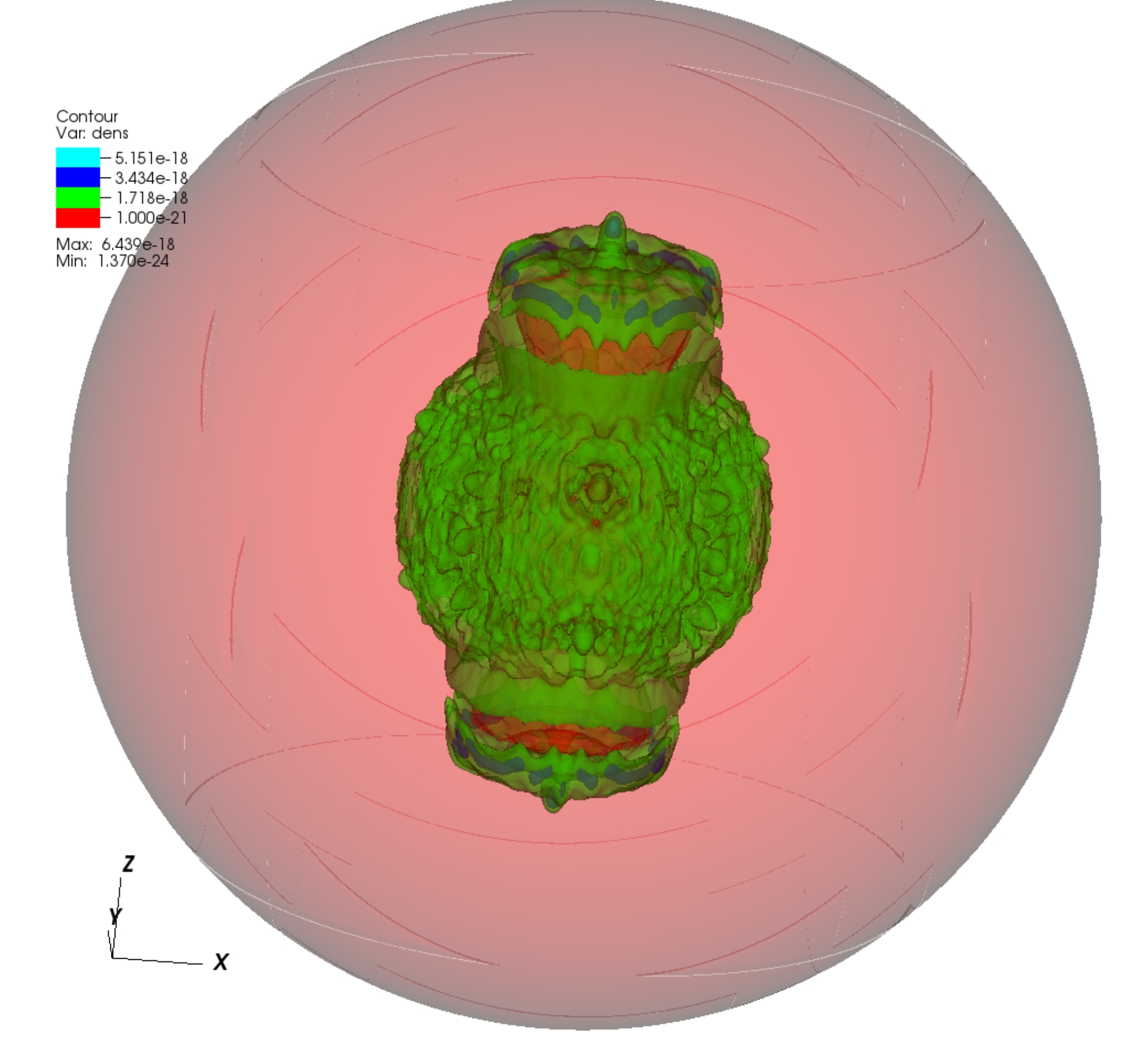}}
\subfigure{\includegraphics[height=2.7in,width=2.7in,angle=0]{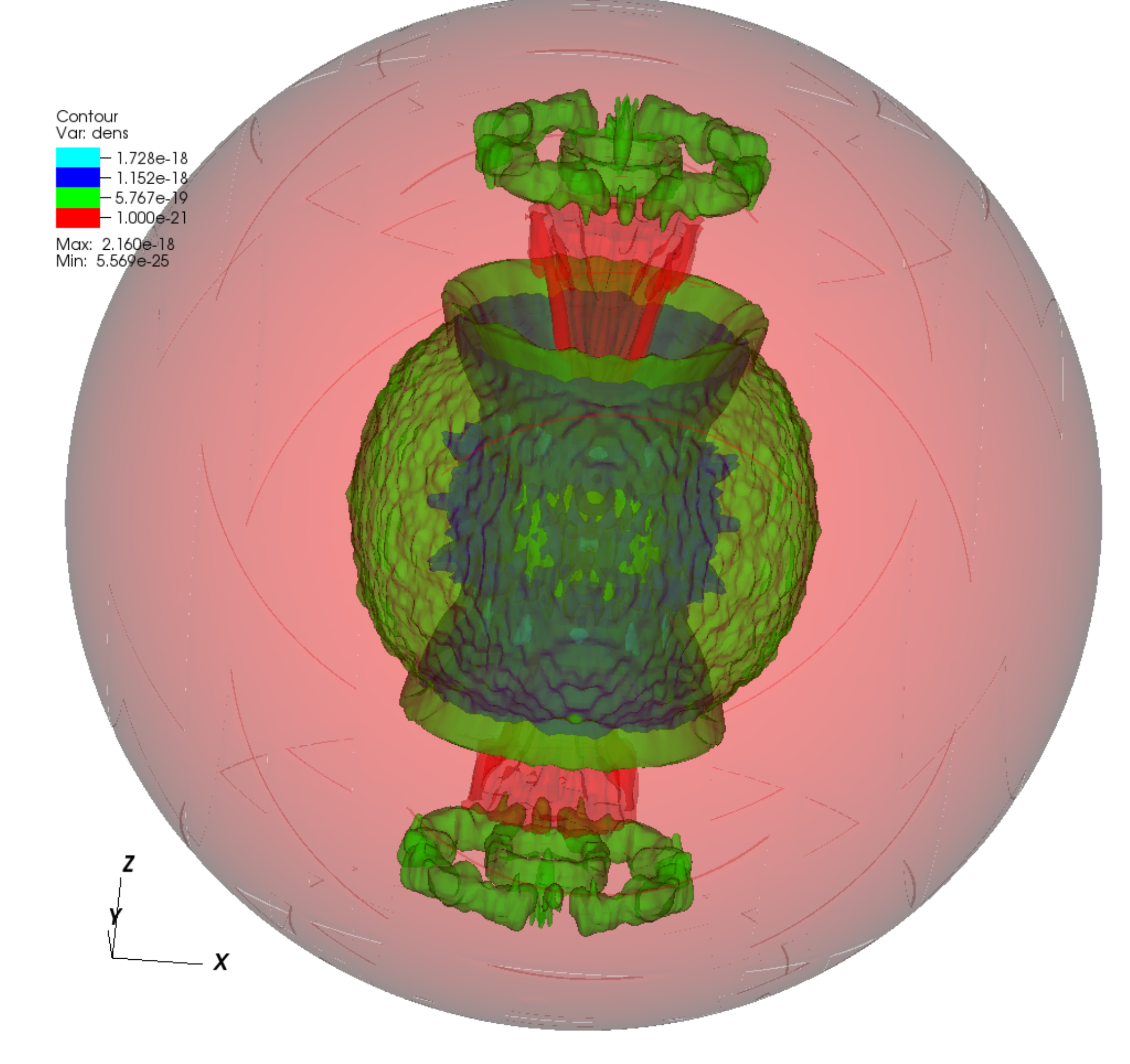}}
\vskip -0.1 cm
\hskip 1.1 cm
\subfigure{\includegraphics[height=2.3in,width=2.7in,angle=0]{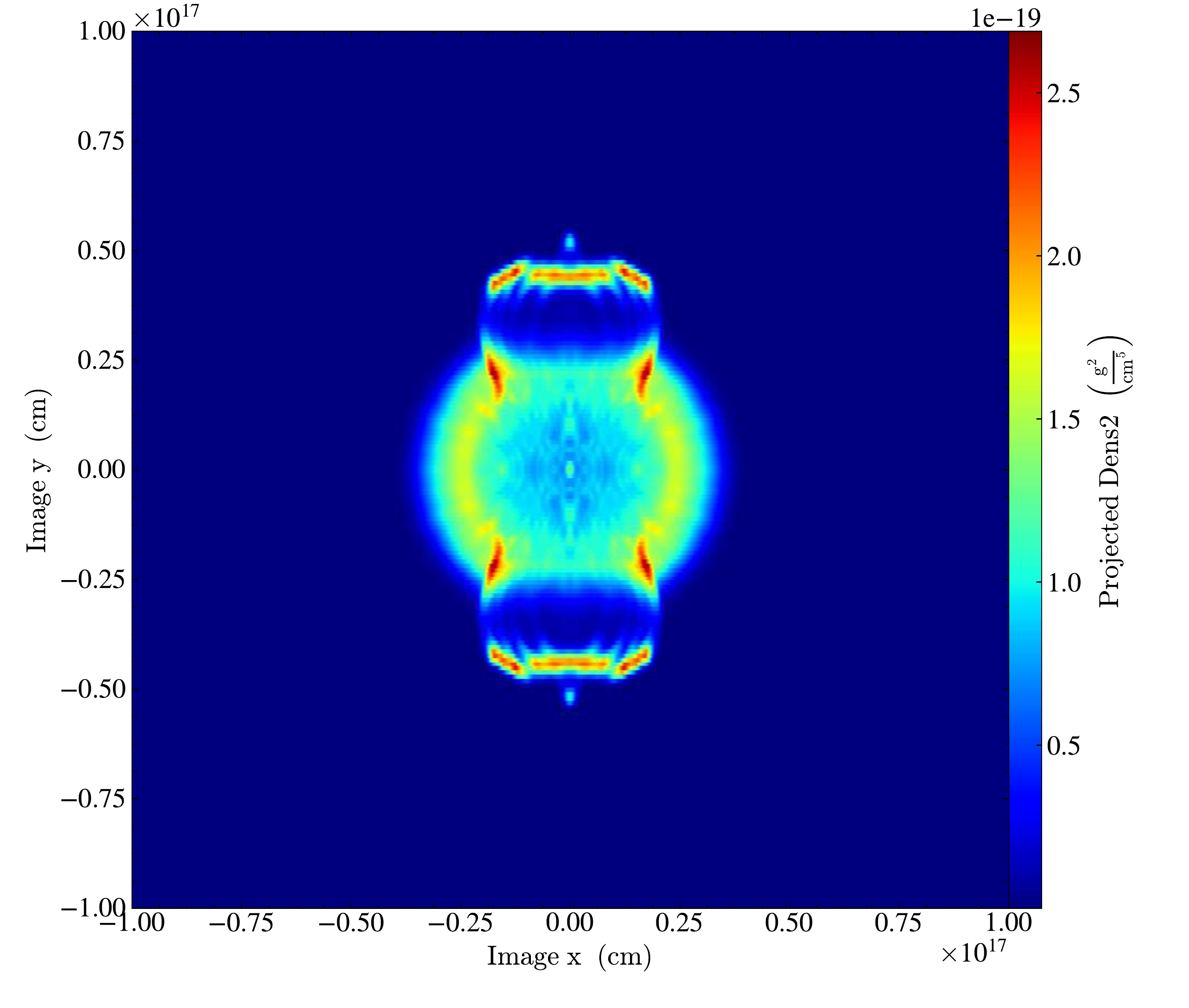}}
\subfigure{\includegraphics[height=2.3in,width=2.7in,angle=0]{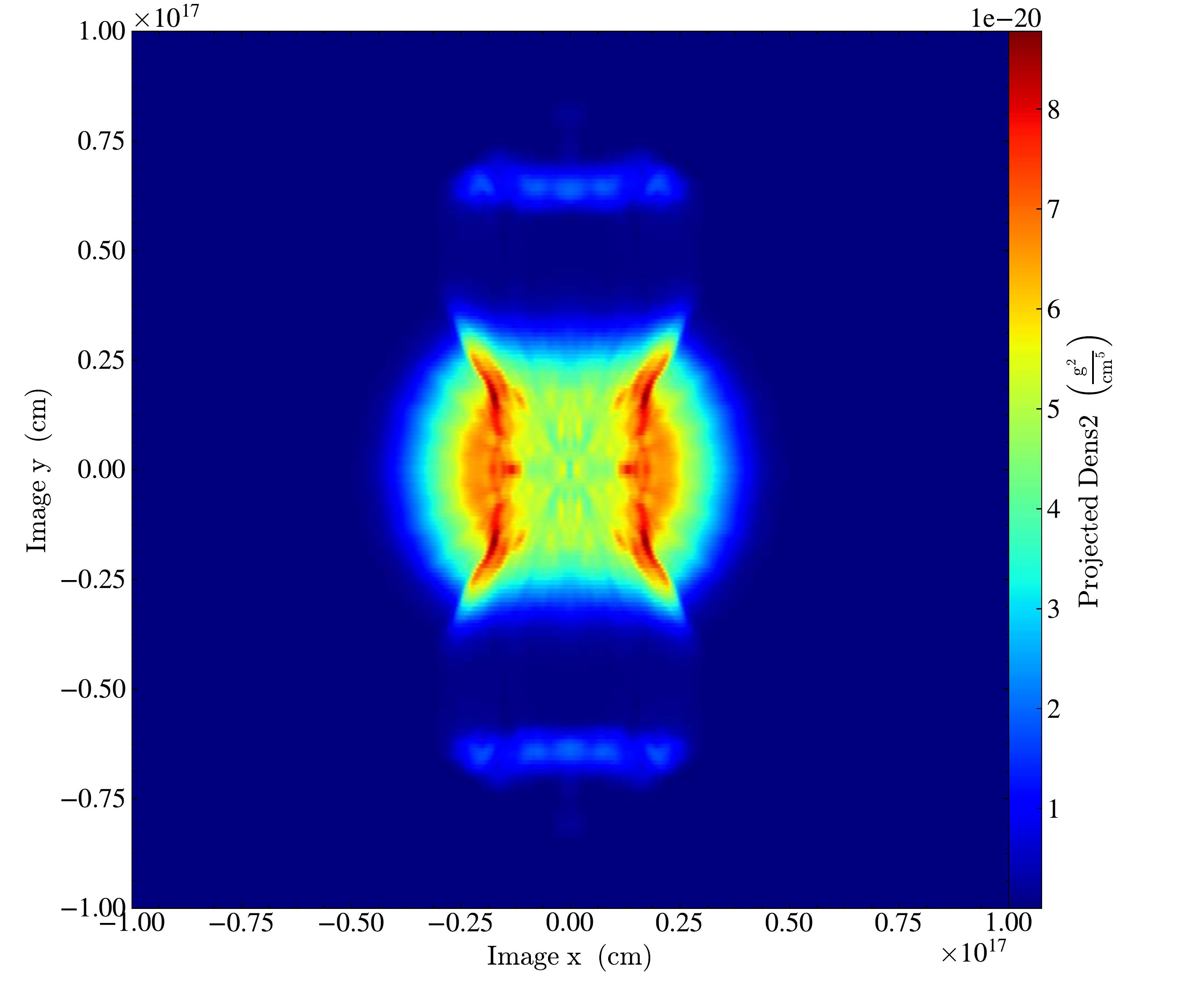}}
\vskip -0.3 cm
\subfigure{\includegraphics[height=2.7in,width=2.7in,angle=0]{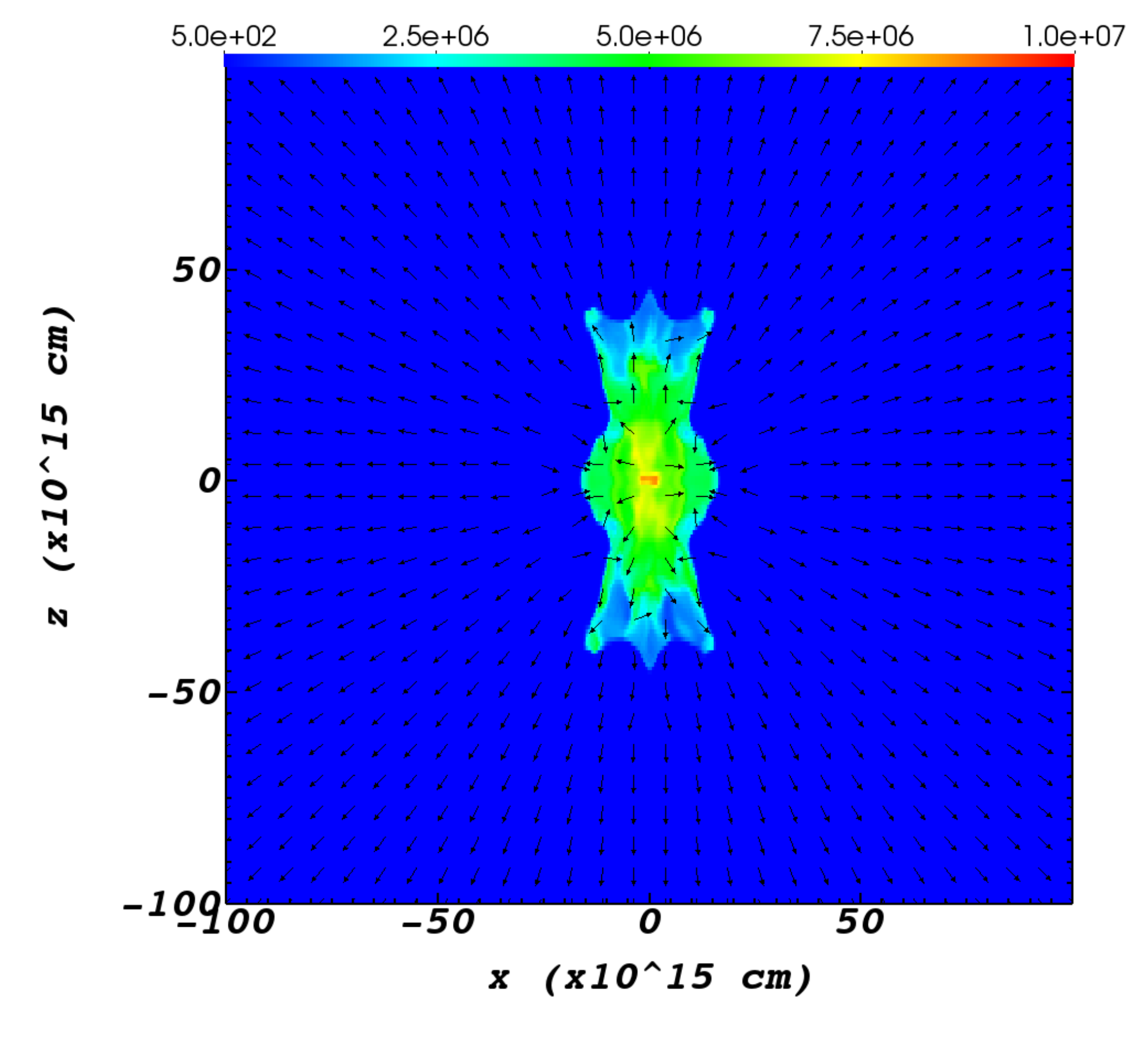}}
\subfigure{\includegraphics[height=2.7in,width=2.7in,angle=0]{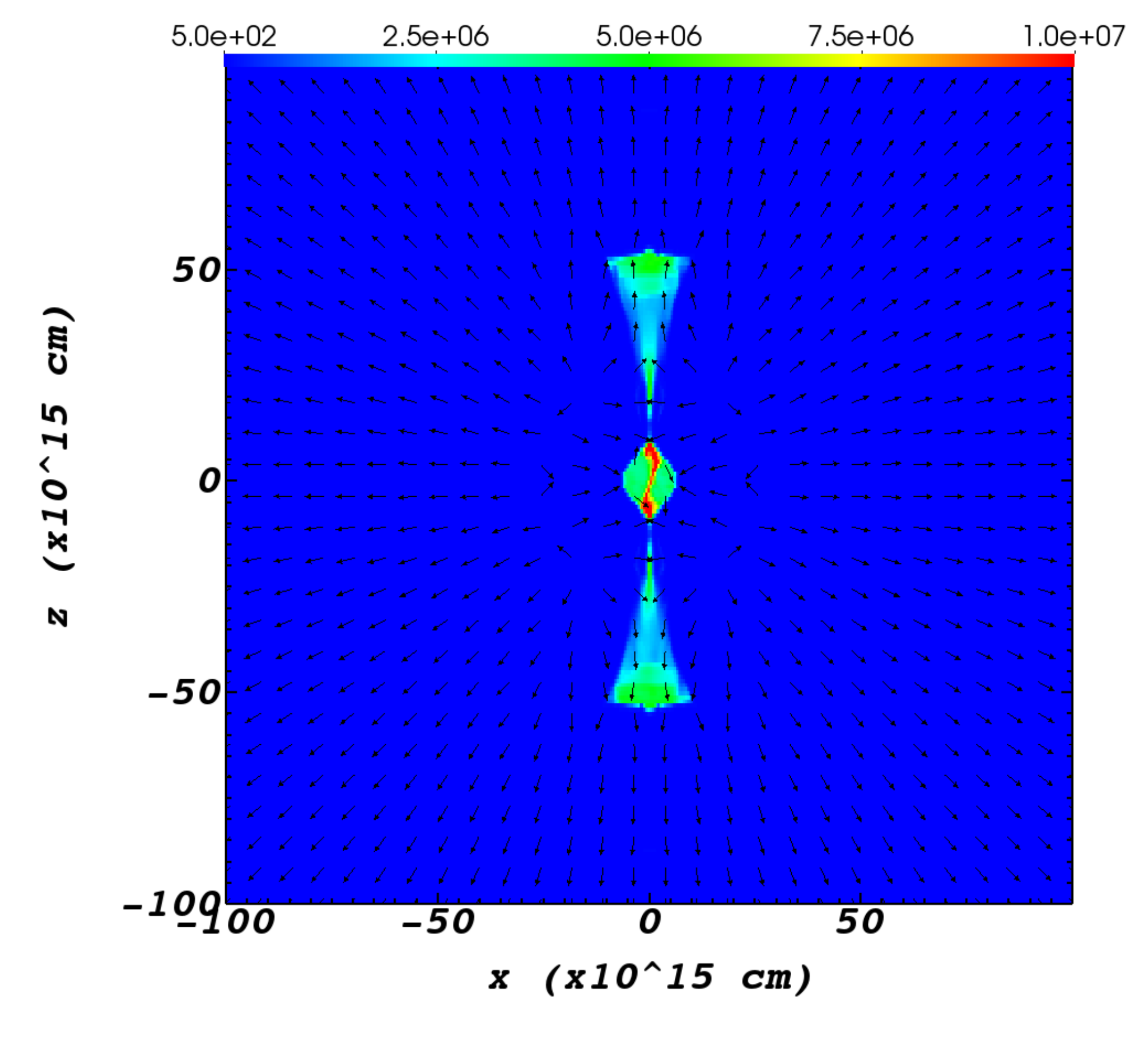}}
\vskip -0.8 cm
\caption{Like Fig. \ref{fig:R1} but for the run R2 and at two later times of $152 \yr$ and $247 \yr$ in the left and right column, respectively. The density in units of $\g \cm^{-3}$ of the 4-colour coding of upper-left panel is $10^{-21}$ (red) $1.7 \times 10^{-18}$ (green), $3.4 \times 10^{-18}$ (blue), and $5.2 \times 10^{-18}$ (pale blue). For the upper-right panel the respective coding is $10^{-21}$ (red) $5.8 \times 10^{-19}$ (green), $1.2 \times 10^{-18}$ (blue), and $1.7 \times 10^{-18}$ (pale blue).
 }
 \label{fig:R2}
\end{center}
\end{figure}

In Fig. \ref{fig:R2} we present the results of run R2 that includes radiative cooling, but otherwise is identical to run R1. The figure is similar to Fig. \ref{fig:R1}, but the two columns are at later times.   
We can identify a barrel-like shape at t=247 years in the 3D image (upper right panel) and by the green intensity colour of the right panel in the middle row. In addition to the barrel-like shape, we can see the two `cups' that are moving at faster speeds along the symmetry axis.  
If we were to examine the high intensity regions as depicted by the red colour in middle-right panel, we could see the development of an H-like shape. An H-like shape is a projected shape on the plane of the sky with two parallel bars on opposite sides of the center, and with a bright bar connecting them.  
The lower-row panels show that when radiative cooling is efficient, the hot gas is confined in tight cones where there are shocks. 
    
\begin{figure}
\begin{center}
\vskip -0.3 cm
\subfigure{\includegraphics[height=2.9in,width=2.9in,angle=0]{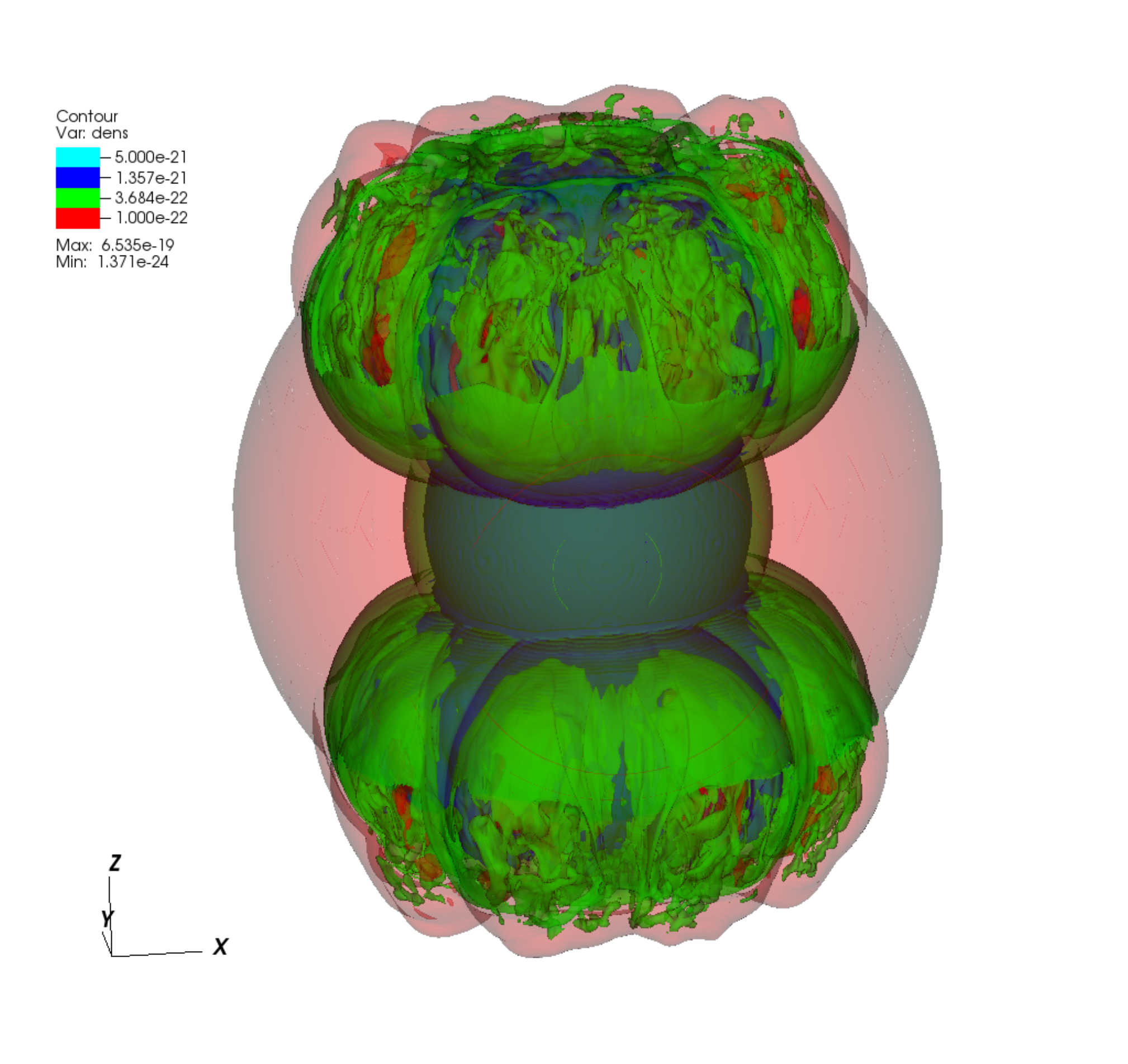}}
\subfigure{\includegraphics[height=2.9in,width=2.9in,angle=0]{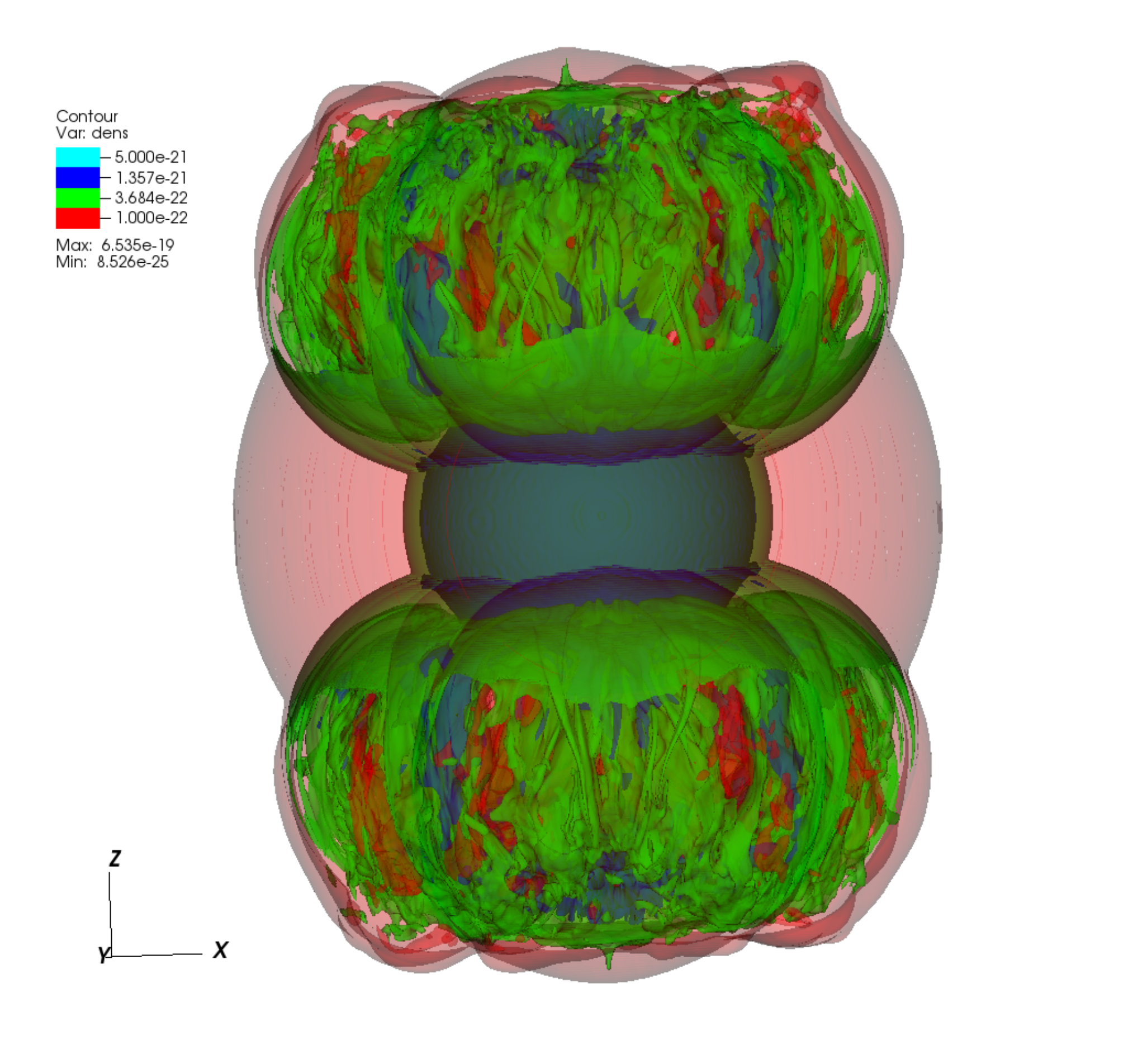}}
\vskip -0.3 cm
\hskip 1.1 cm
\subfigure{\includegraphics[height=2.4in,width=2.9in,angle=0]{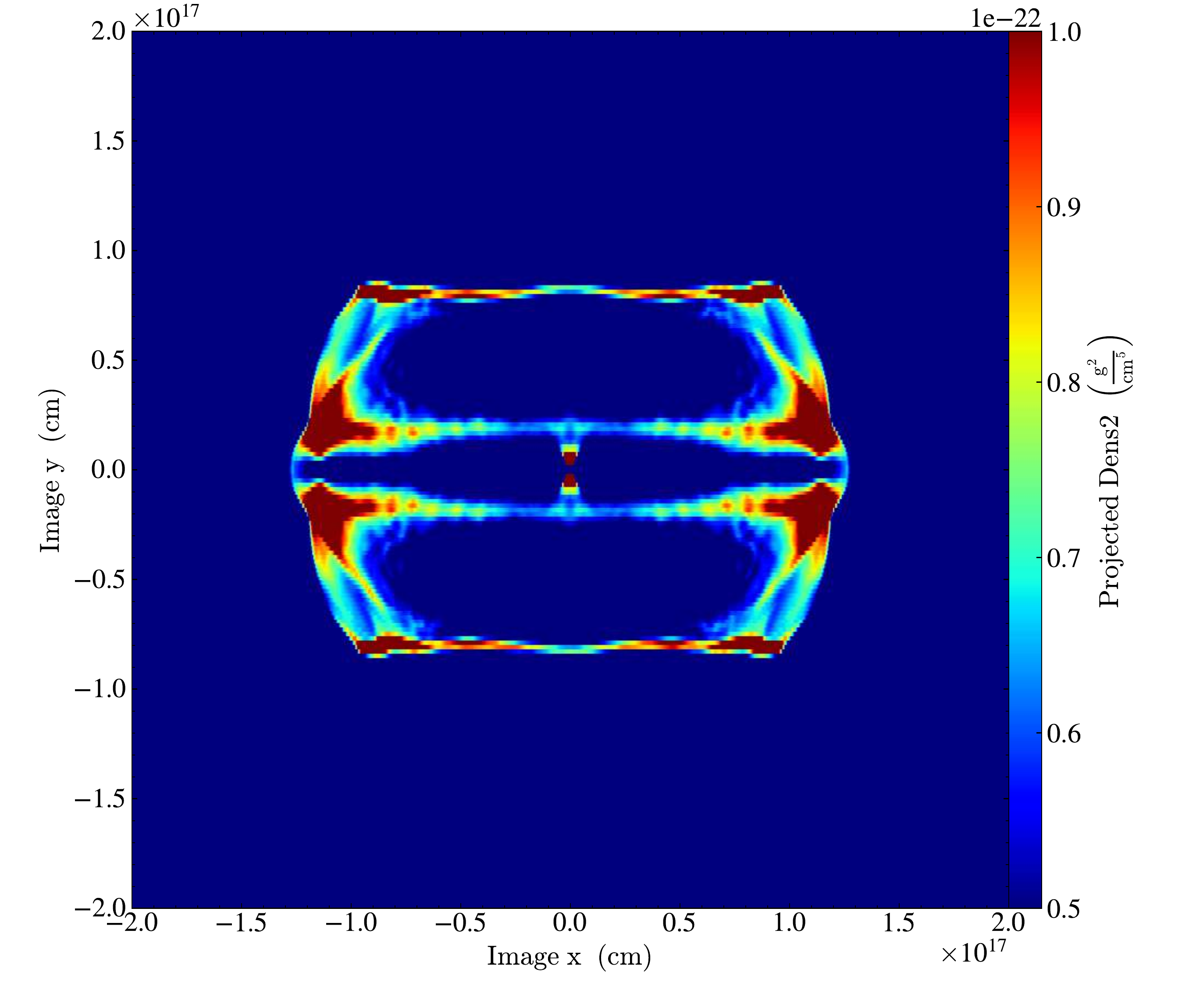}}
\subfigure{\includegraphics[height=2.4in,width=2.9in,angle=0]{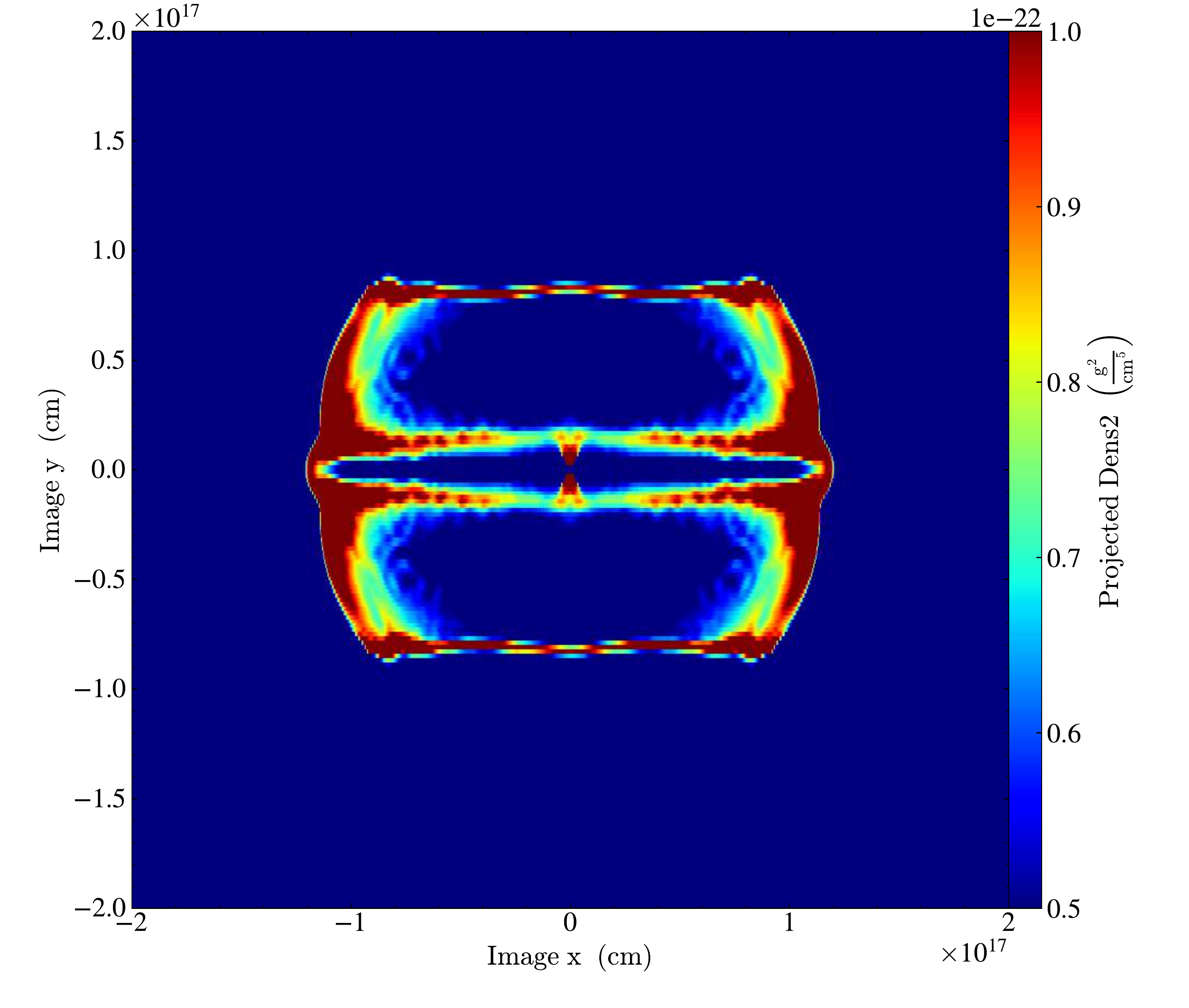}}
\vskip -0.3 cm
\subfigure{\includegraphics[height=2.9in,width=2.9in,angle=0]{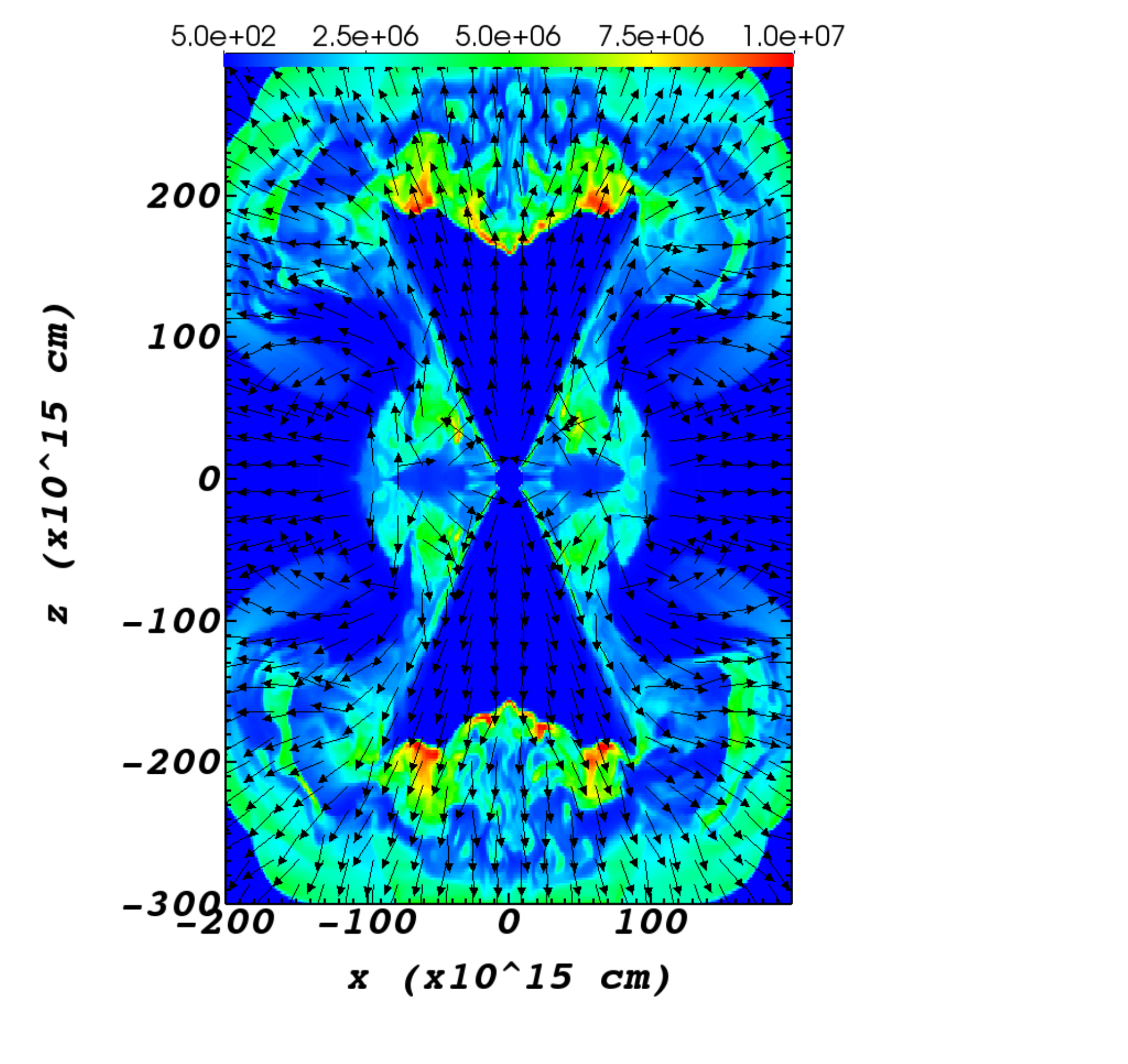}}
\subfigure{\includegraphics[height=2.9in,width=2.9in,angle=0]{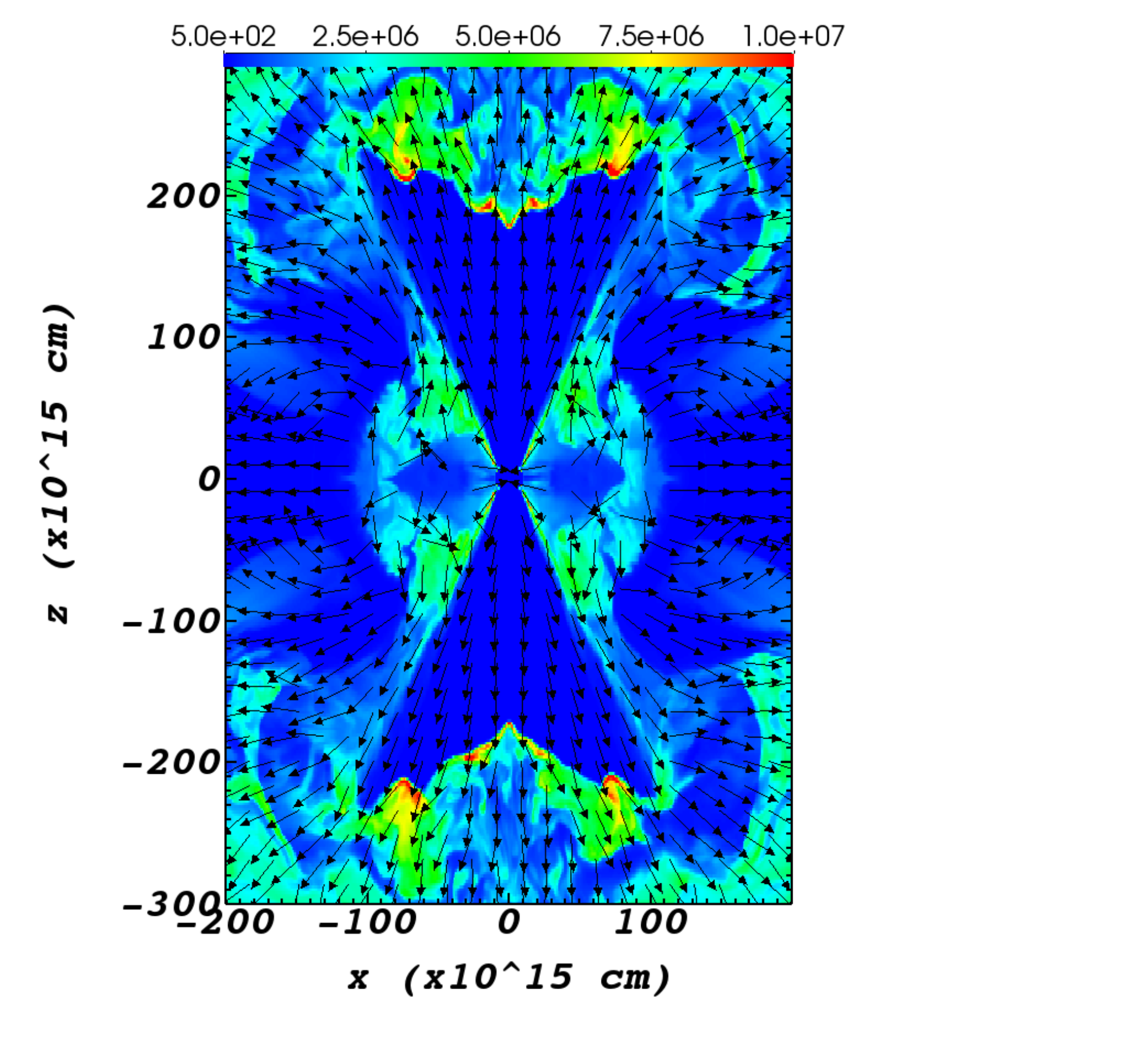}}
\vskip -0.6 cm
\caption{Like Fig. \ref{fig:R1} but for the run R3 and at two later times of $215 \yr$ and $245 \yr$ in the left and right column, respectively. The 4-colour coding for both upper panels is 
$10^{-22}$ (red) $3.7 \times 10^{-22}$ (green), $1.4 \times 10^{-21}$ (blue), and $5.0 \times 10^{-21}$ (pale blue). 
}
 \label{fig:R3}
\end{center}
\end{figure}

In Fig. \ref{fig:R3} we show the results of run R3, that is a case of a large shell, no radiative cooling, and the jets are active for the entire duration of the run. We analyzed the properties of this run in section \ref{sec:flow}. 
In the two panels in the middle row that mimic the observations, we can see a barrel-like shape of the nebula. The dense two opposite arcs seen in these panels are the projection of the dense envelope of the barrel. The dense envelope is the waist of the 3D image seen in the upper row in dark-green.  
We also see the bright regions that for a H-like shape, although the central bar is composed of two bars. As with all other cases studied here, the appearance of the nebula evolves with time during the hundreds years of our simulations.   
The two lower panels present a complicated flow as seen in the meridional plane. The temperature structure show instabilities that have been developing during the interaction of the jets with the dense shell. The velocity structure contains vortices near the equatorial plane. 
We note that although our flow setting has a cylindrical symmetry, the development of instabilities in regions that are highly unstable depend on numerical noise, and hence these instabilities might break the cylindrical symmetry on small scales. This is seen for example in the red regions on the lobes in the upper row of Fig. \ref{fig:R3}. 
           
\begin{figure}
\begin{center}
\subfigure{\includegraphics[height=2.9in,width=2.9in,angle=0]{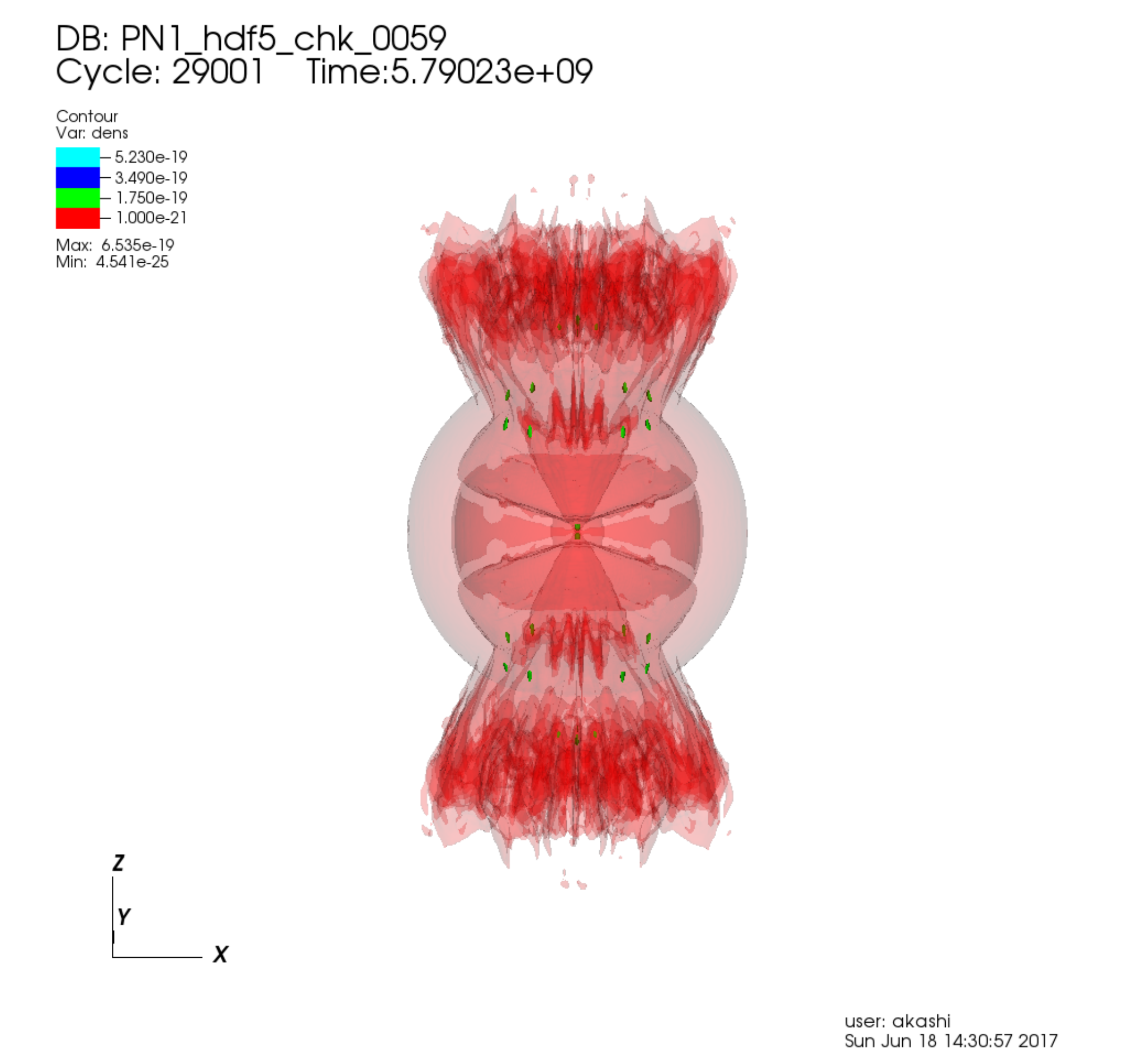}}
\subfigure{\includegraphics[height=2.9in,width=2.9in,angle=0]{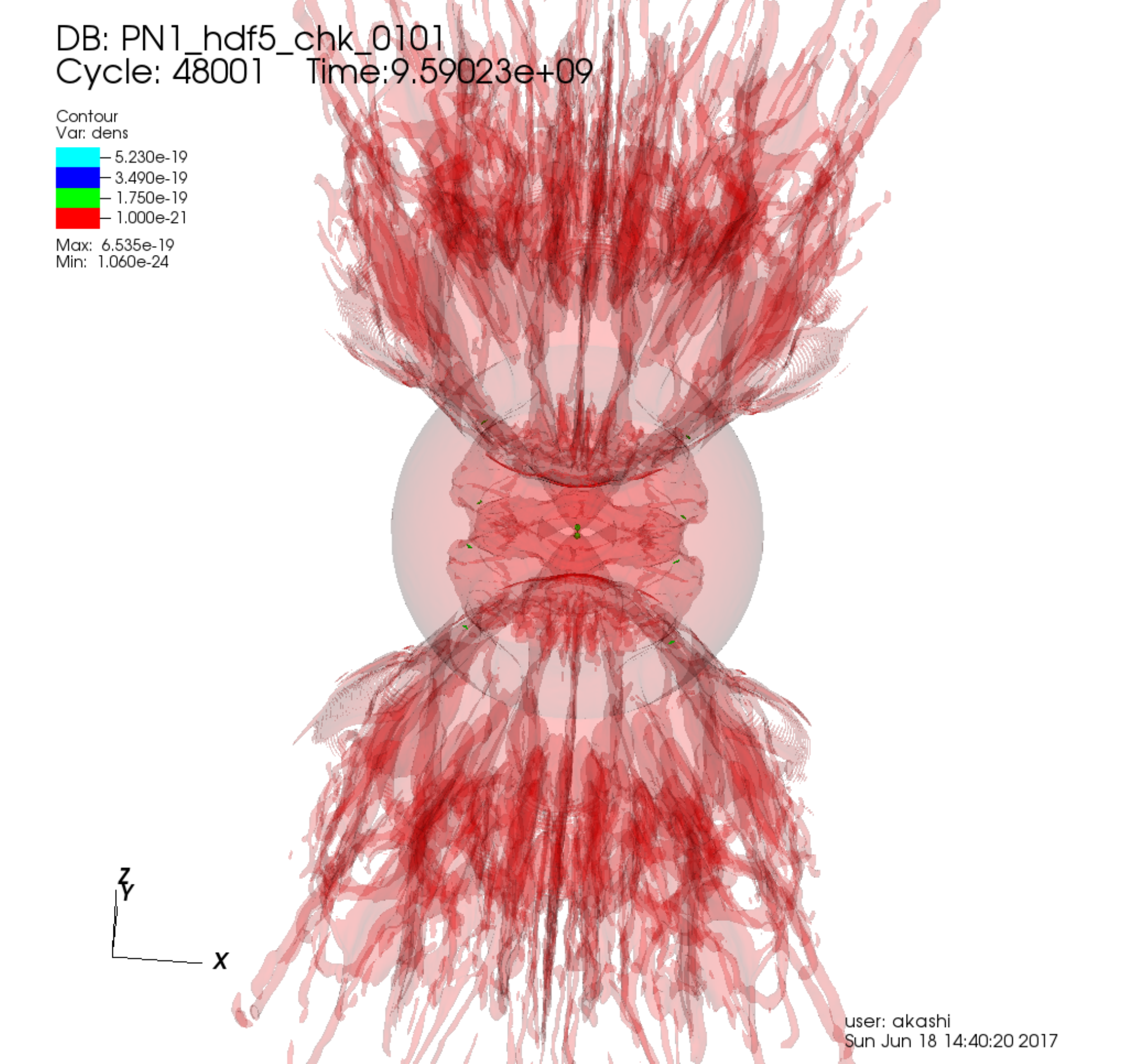}}
\subfigure{\includegraphics[height=2.45in,width=2.9in,angle=0]{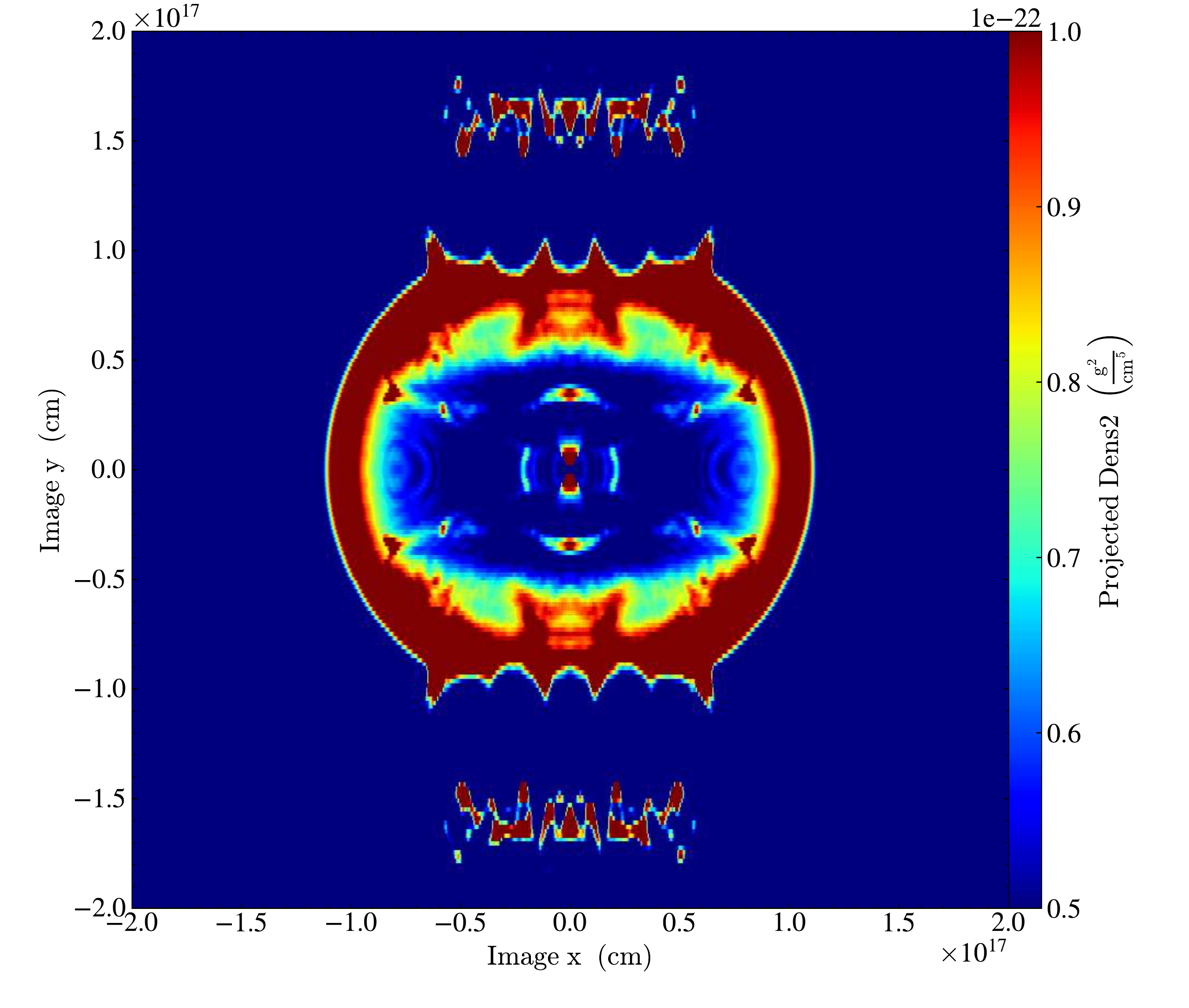}}
\subfigure{\includegraphics[height=2.45in,width=2.9in,angle=0]{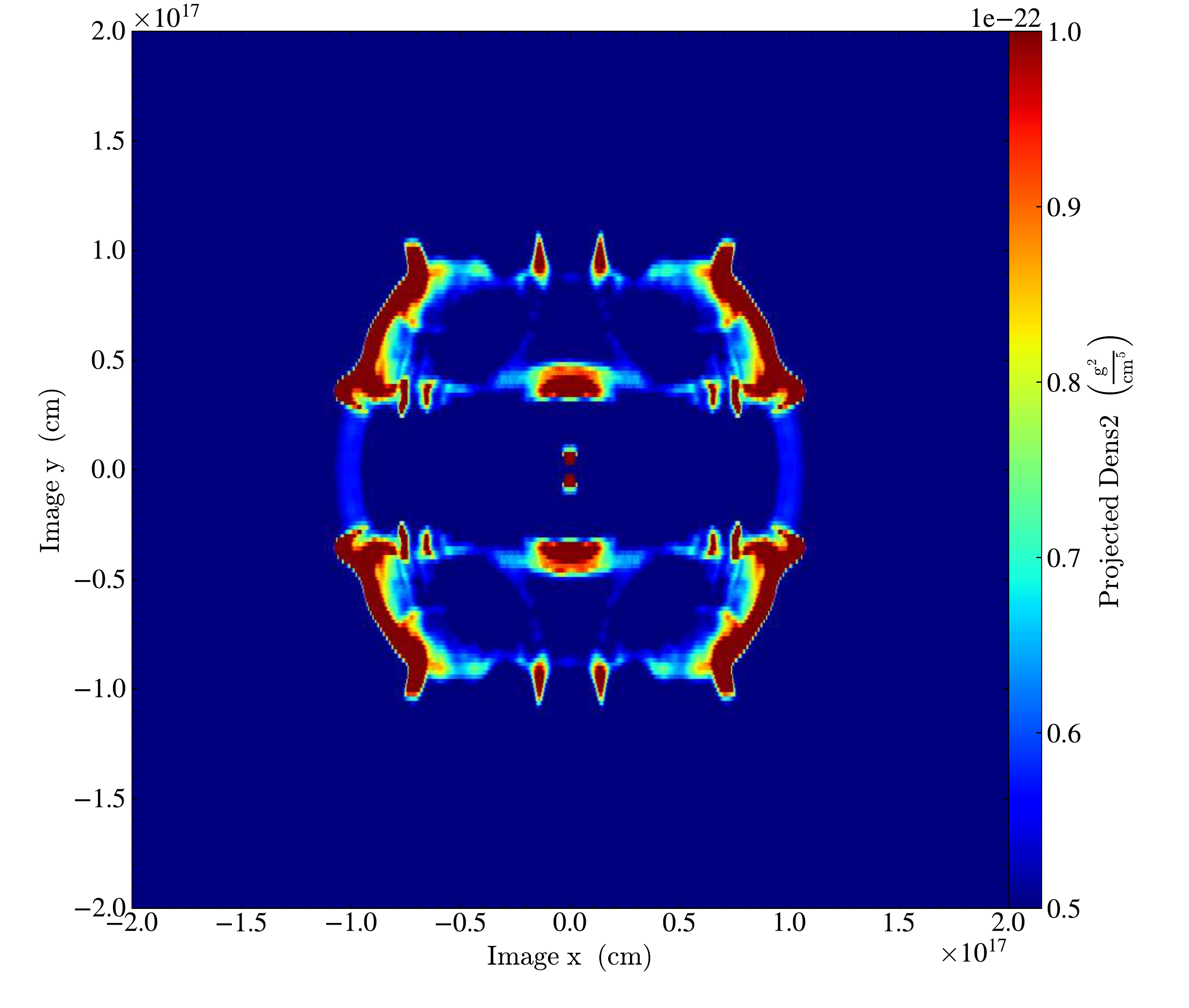}}
\vskip -0.3 cm
\subfigure{\includegraphics[height=2.9in,width=2.9in,angle=0]{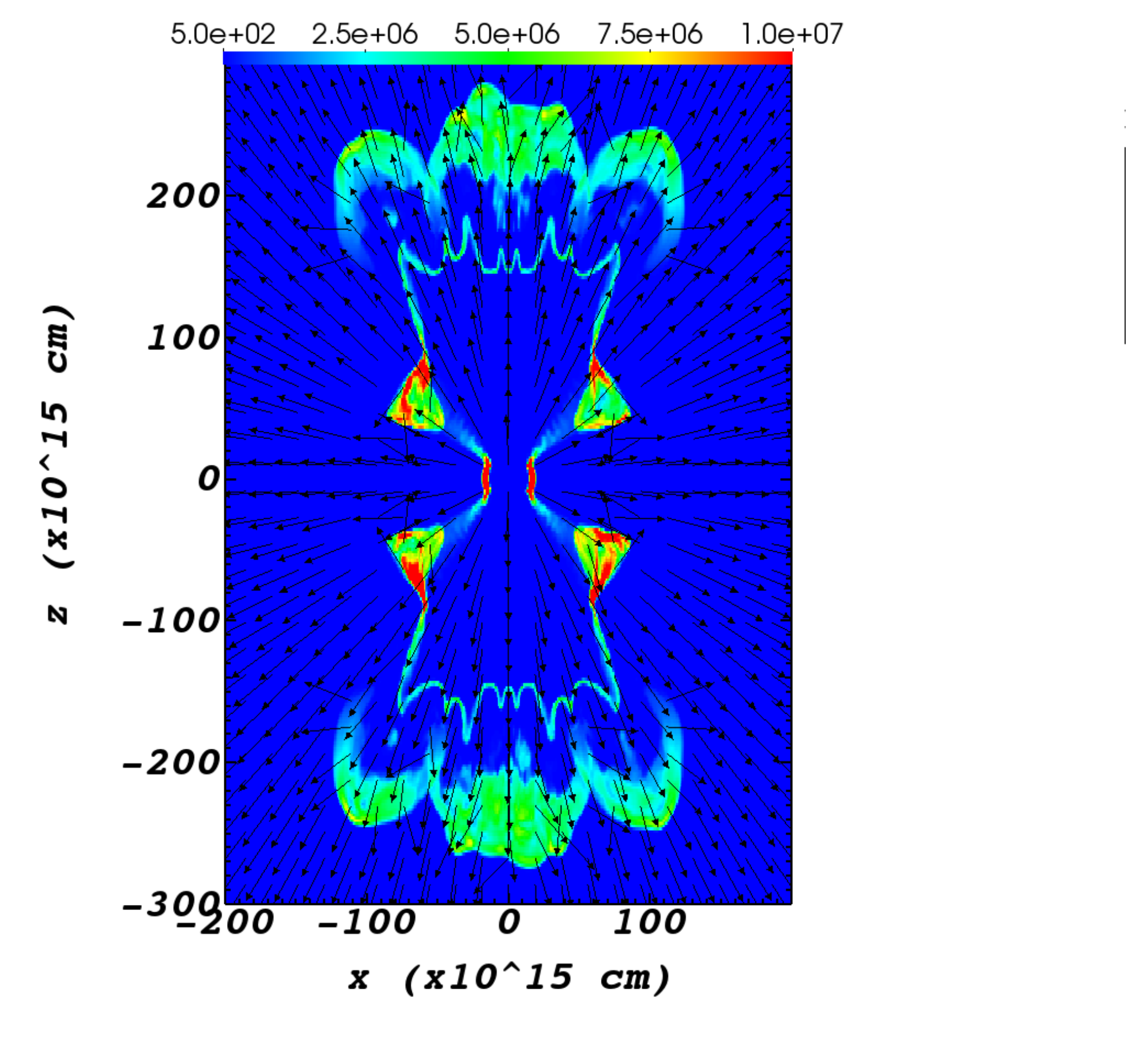}}
\subfigure{\includegraphics[height=2.9in,width=2.9in,angle=0]{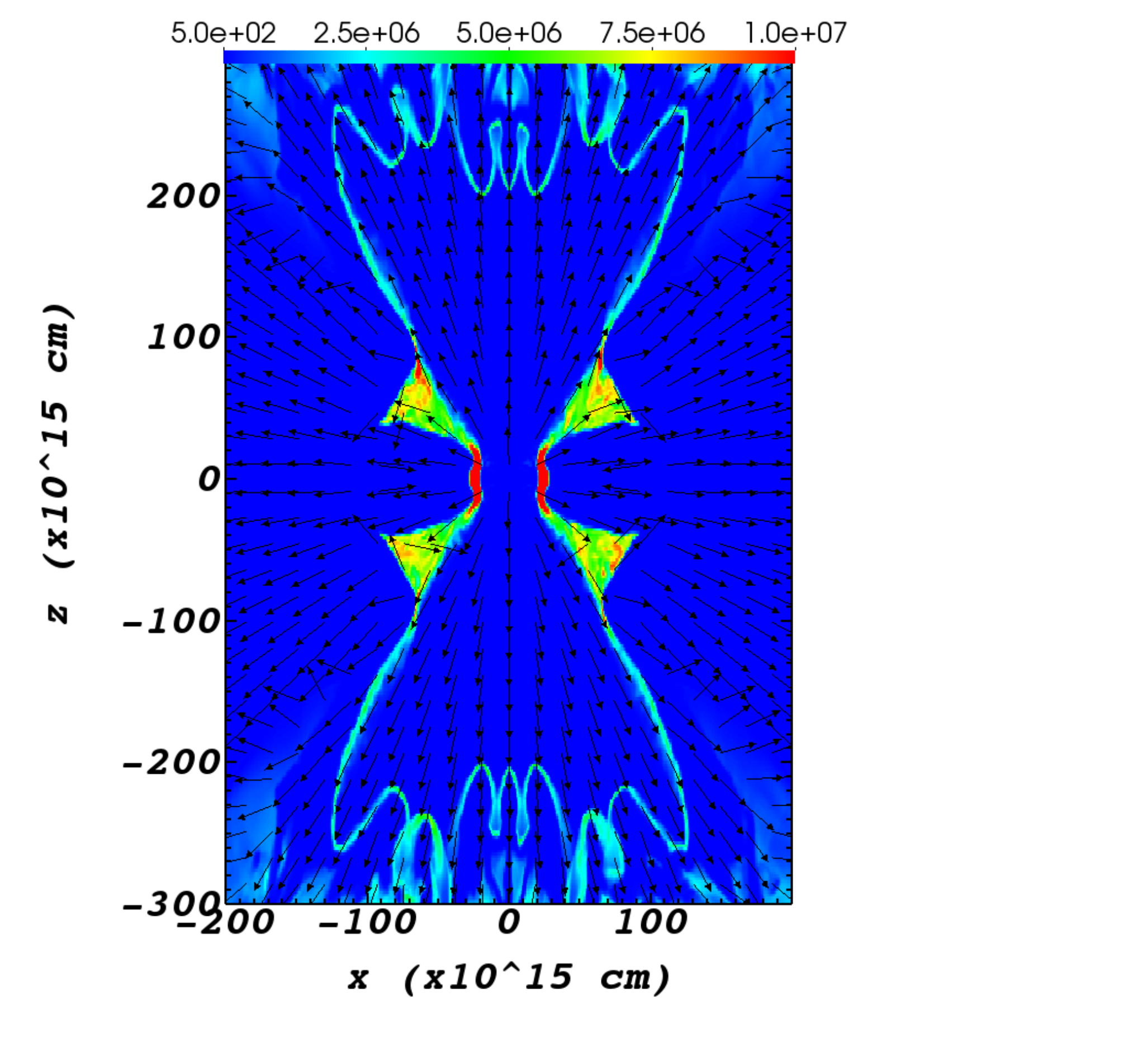}}
\vskip -0.7 cm
\caption{Like Fig. \ref{fig:R1} but for the run R4 at at later times of $185 \yr$ and $305 \yr$ in the left and right column, respectively. The 4-colour coding for both upper panels is $10^{-21}$ (red) $1.7 \times 10^{-19}$ (green), $3.5 \times 10^{-19}$ (blue), and $5.2 \times 10^{-19}$ (pale blue). }
 \label{fig:R4}
\end{center}
\end{figure}

We show in Fig. \ref{fig:R4} the results of run R4, which is like run R3, but radiative cooling is included. The intensity maps in the middle-row panels reveal a clear development of a barrel-like nebula, with two opposite `caps' that are detached from the main part of the nebula along the polar directions. 
Like in run R3, the two lower panels present a complicated flow in the meridional plane. Again, the temperature structure shows instabilities that have been developing during the interaction of the jets with the dense shell. The velocity structure contains vortices, but at mid-latitudes rather than near the equatorial plane.
         
\begin{figure}
\begin{center}
\vskip -0.7 cm
{\includegraphics[height=2.9in,width=2.9in,angle=0]{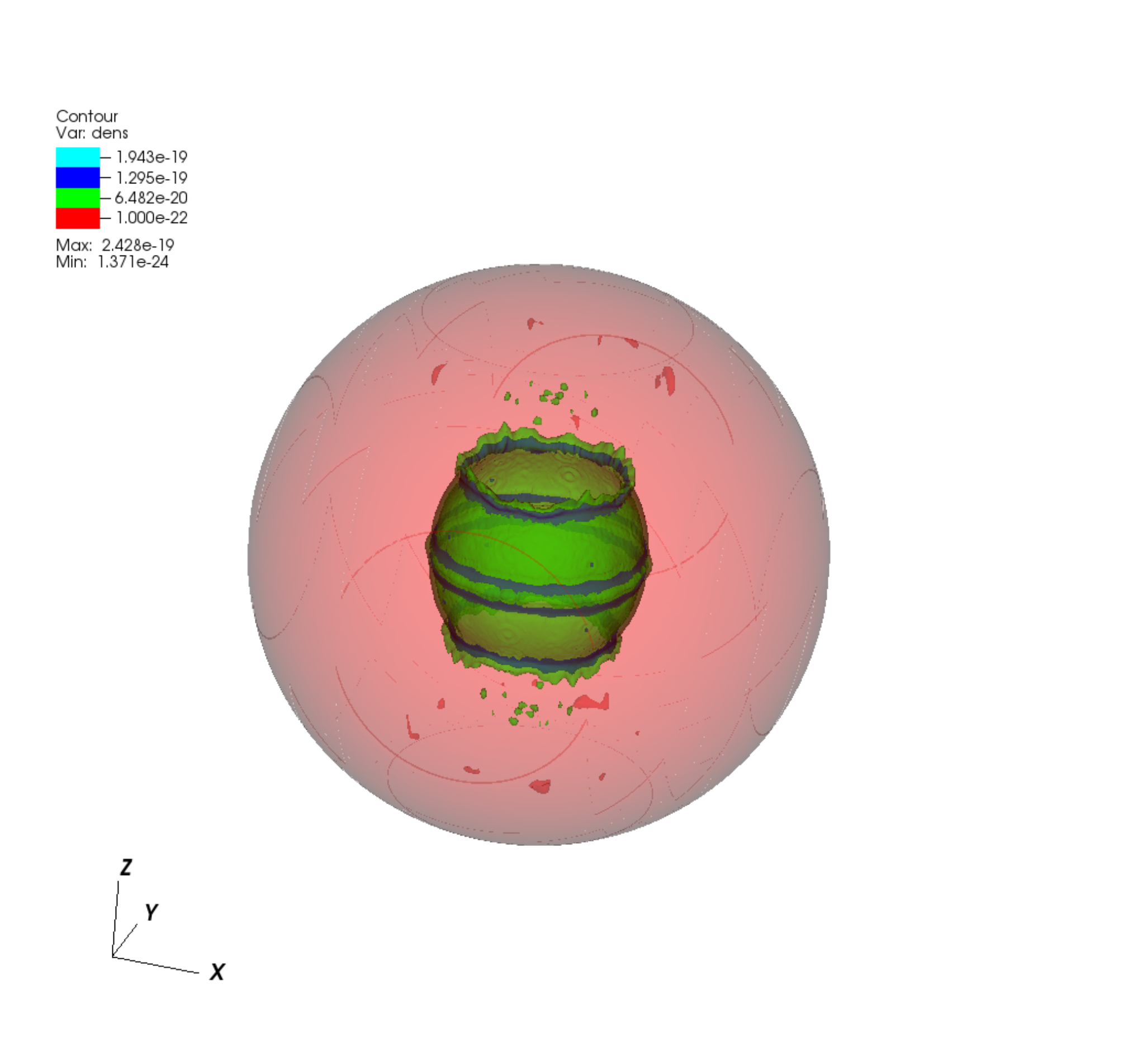}}
{\includegraphics[height=2.9in,width=2.9in,angle=0]{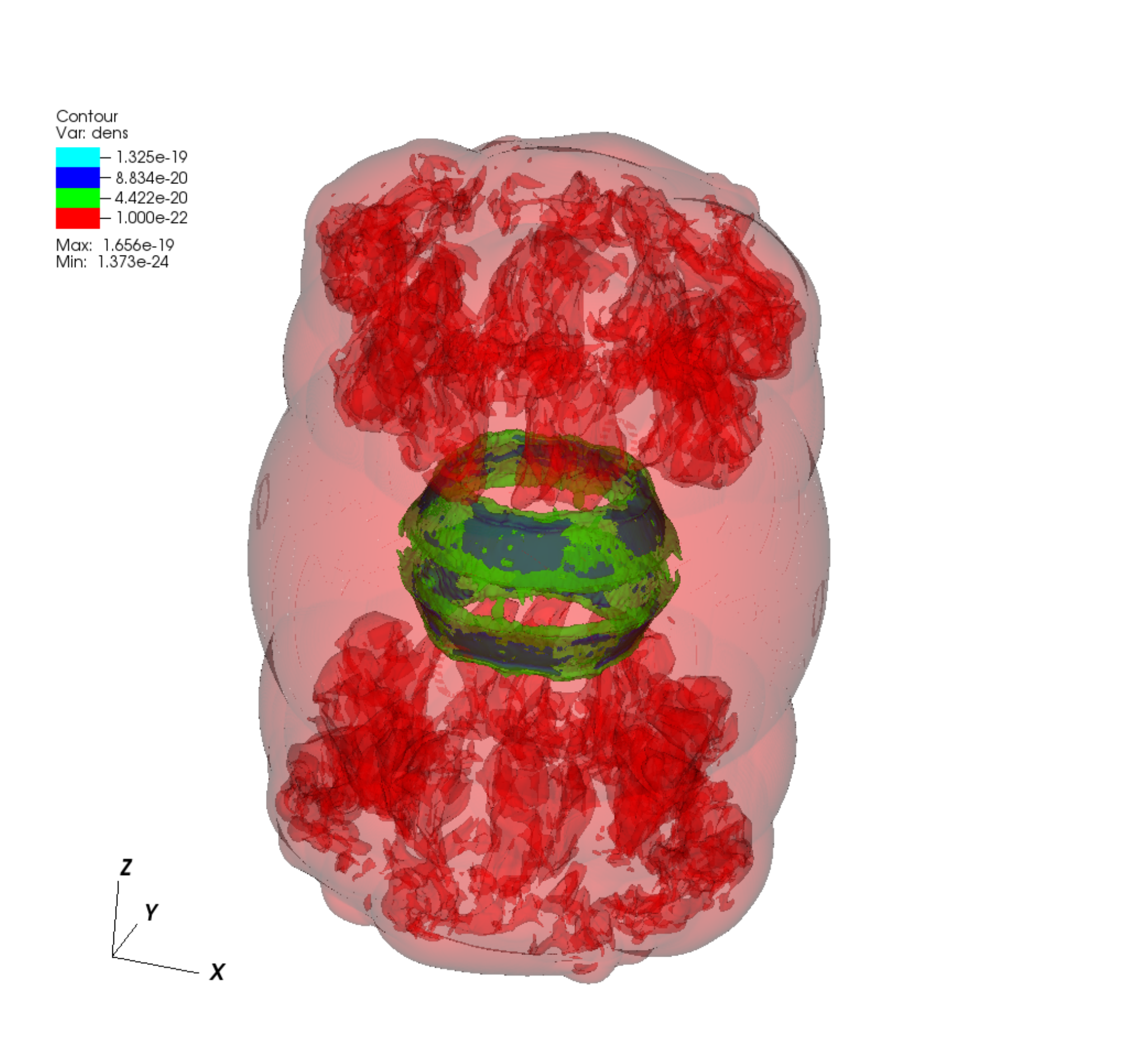}}
\vskip -0.3 cm
{\includegraphics[height=2.45in,width=2.9in,angle=0]{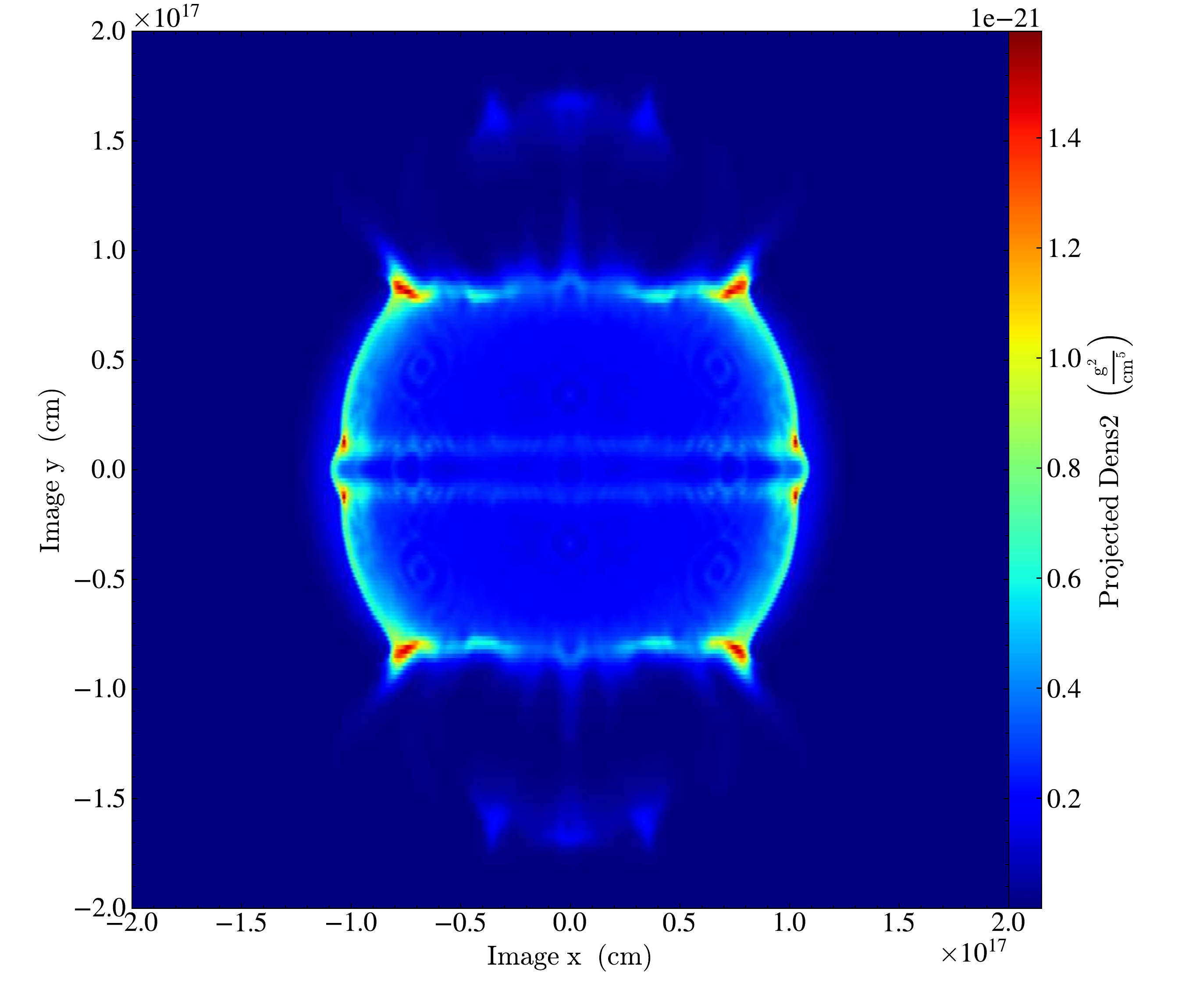}}
{\includegraphics[height=2.45in,width=2.9in,angle=0]{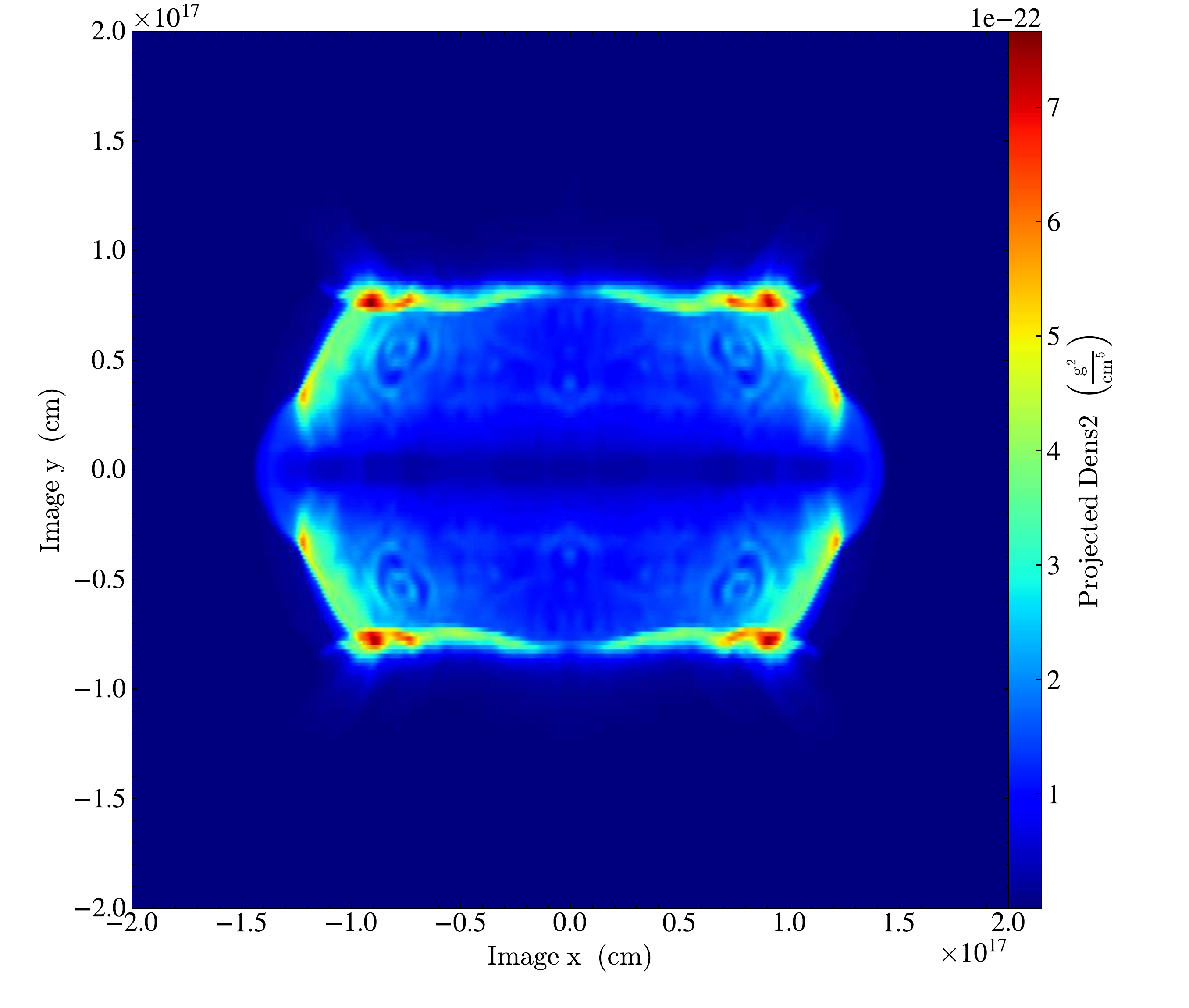}}
\vskip -0.3 cm
{\includegraphics[height=2.9in,width=2.9in,angle=0]{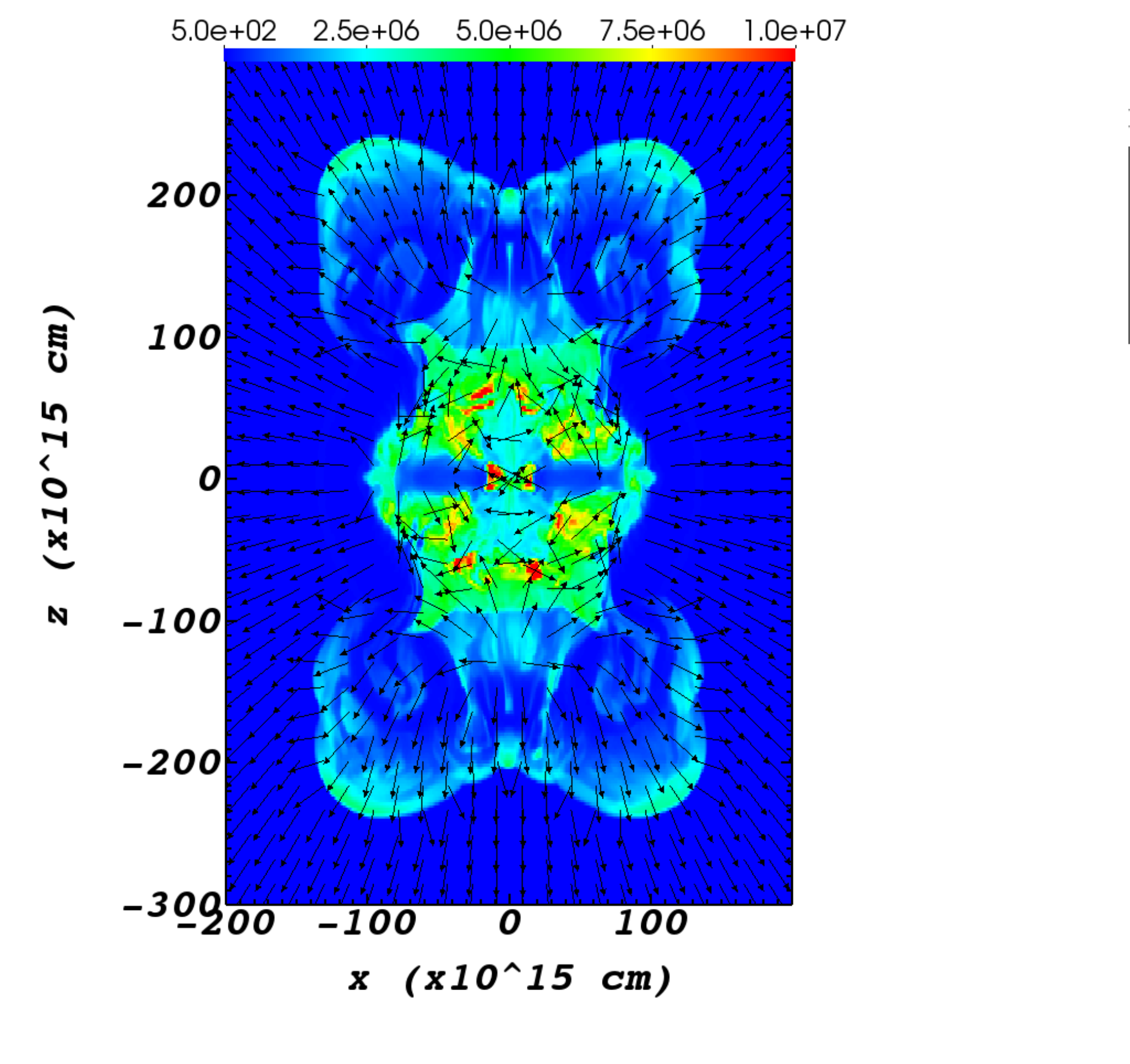}}
{\includegraphics[height=2.9in,width=2.9in,angle=0]{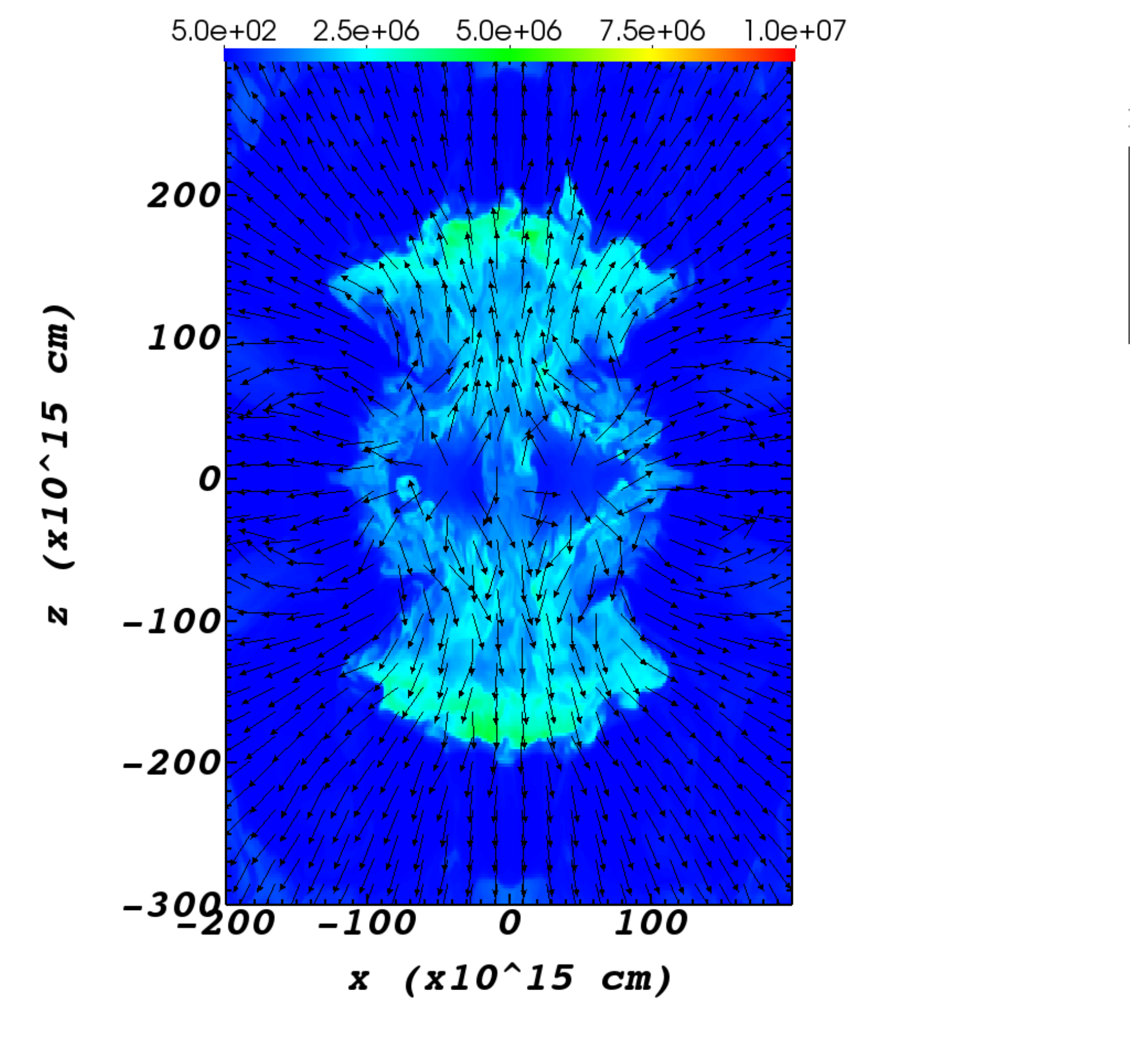}}
\vskip -0.8 cm
\caption{Like Fig. \ref{fig:R1} but for the run R5 and at later times, $t=209 \yr$ and $t=368 \yr$ in the left and right columns, respectively. The 4-colour coding for the upper left and upper right panels are 
$10^{-22}$ (red) $6.5 \times 10^{-20}$ (green), $1.3 \times 10^{-19}$ (blue), and $1.9 \times 10^{-19}$ (pale blue), and $10^{-22}$ (red) $4.4 \times 10^{-20}$ (green), $8.8 \times 10^{-20}$ (blue), and $1.3 \times 10^{-19}$ (pale blue), respectively. }
 \label{fig:R5}
\end{center}
\end{figure}

We show in Fig. \ref{fig:R5} the results of run R5 which has a large shell, but is otherwise like run R1. The green colours in the two upper panels reveal a clear barrel-like structure. This structure is clearly seen also in the middle-row panels. The lower right panel reveals that the interaction of the jets with the shell leads to vortices near the equatorial plane and a complicated flow in the polar directions.  
 
We end by presenting 3D structures for the same cases and times as we presented in the upper rows of Figs. \ref{fig:R1}-\ref{fig:R5}, but for a viewing angle that is tilted by $20^\circ$ to the equatorial plane. As expected, the general appearance changes with angle. While the H-shaped structure starts to disappear, the barrel-shaped structure does not change much.
\begin{figure}
\begin{center}
\vskip -0.3 cm
\subfigure{\includegraphics[height=1.3in,width=1.5in,angle=0]{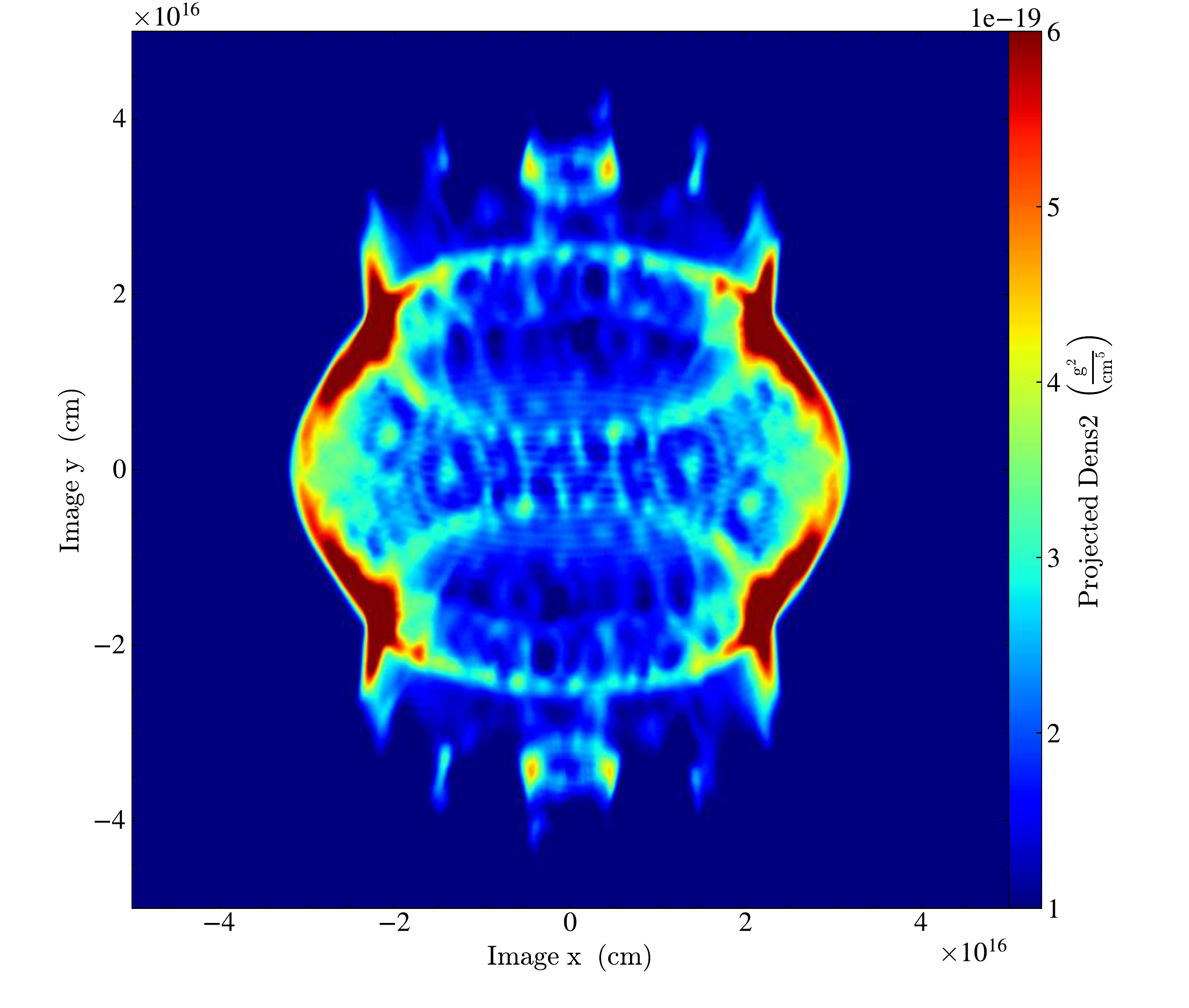}}
\subfigure{\includegraphics[height=1.3in,width=1.5in,angle=0]{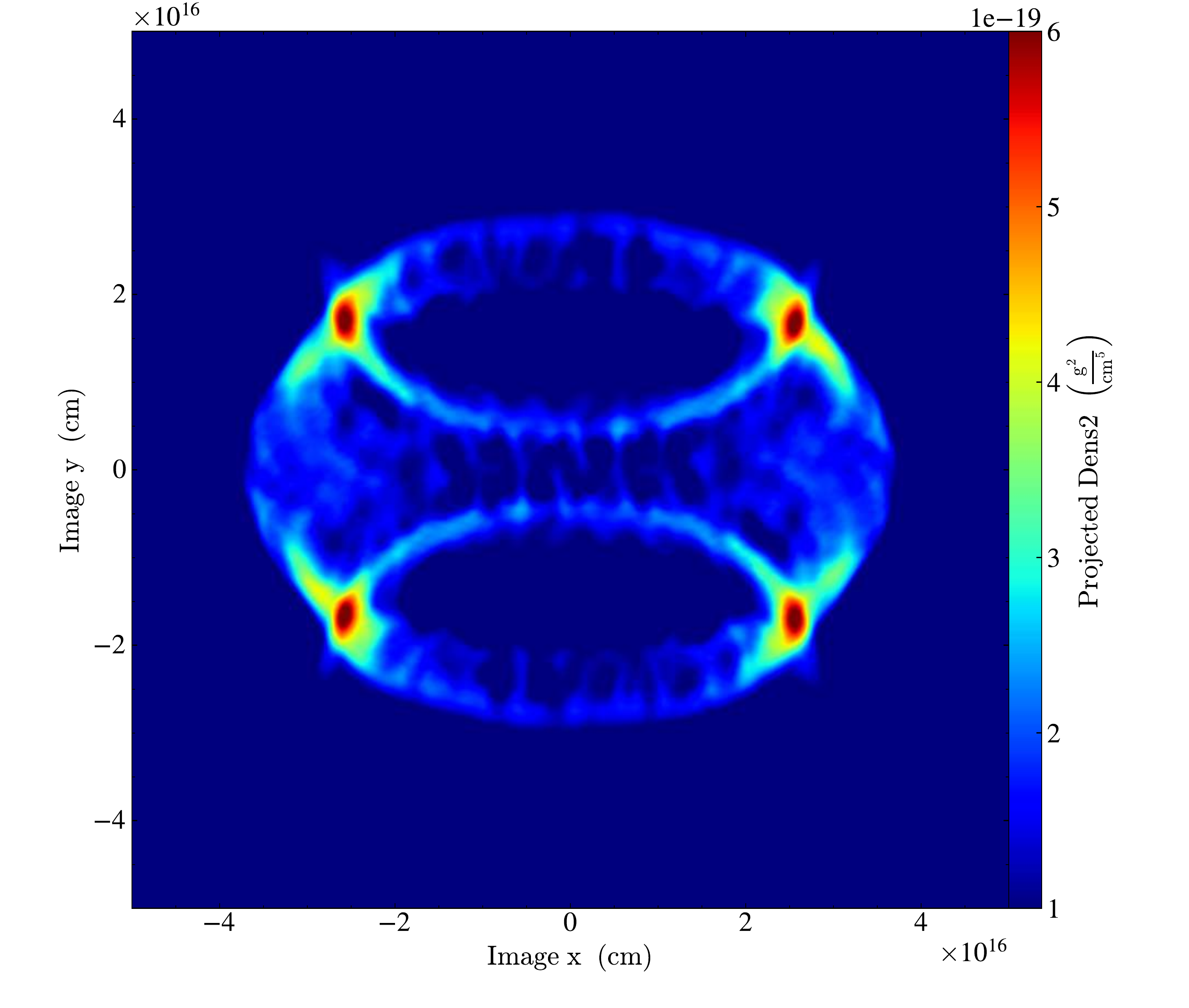}}

\subfigure{\includegraphics[height=1.3in,width=1.5in,angle=0]{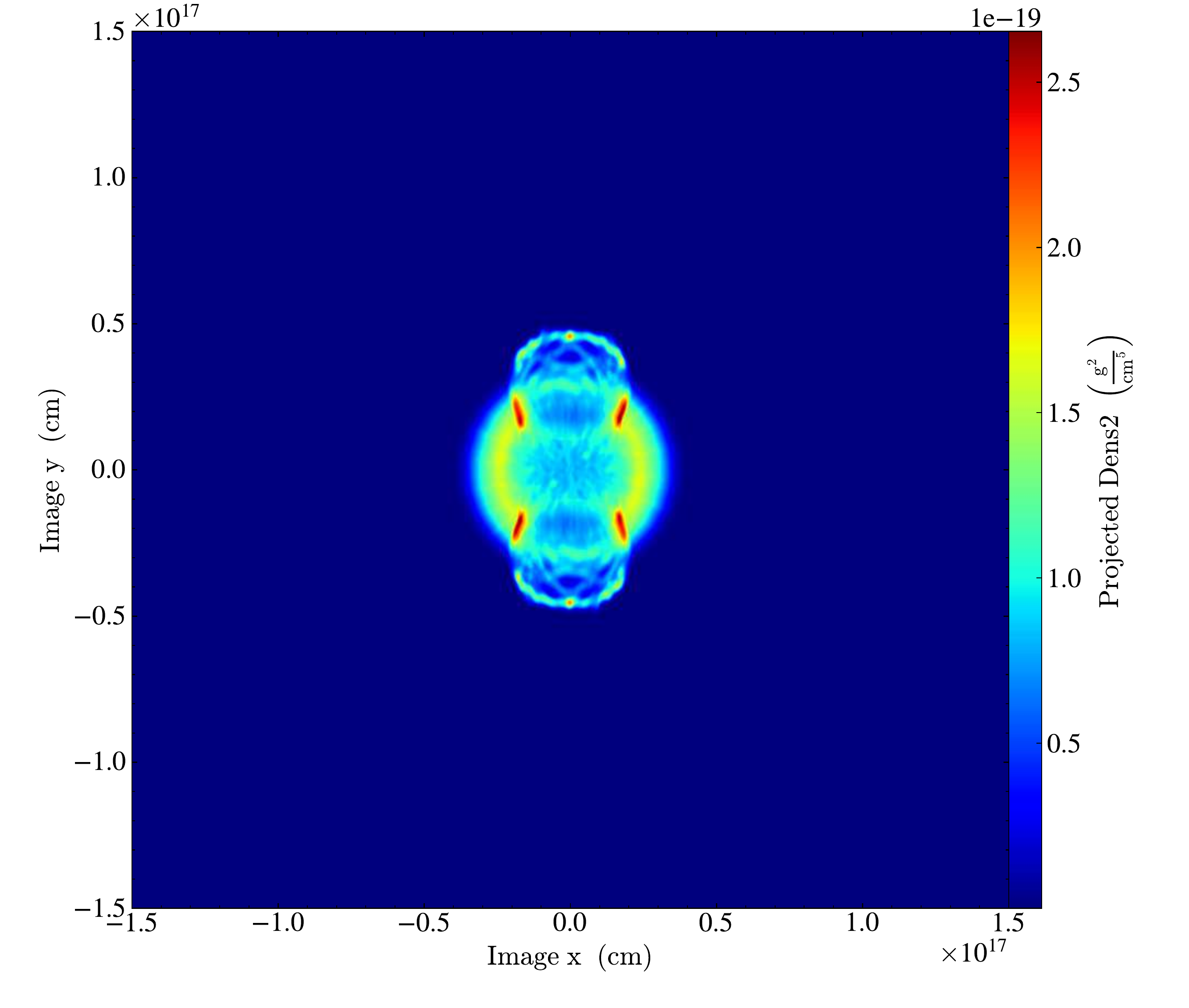}}
\subfigure{\includegraphics[height=1.3in,width=1.5in,angle=0]{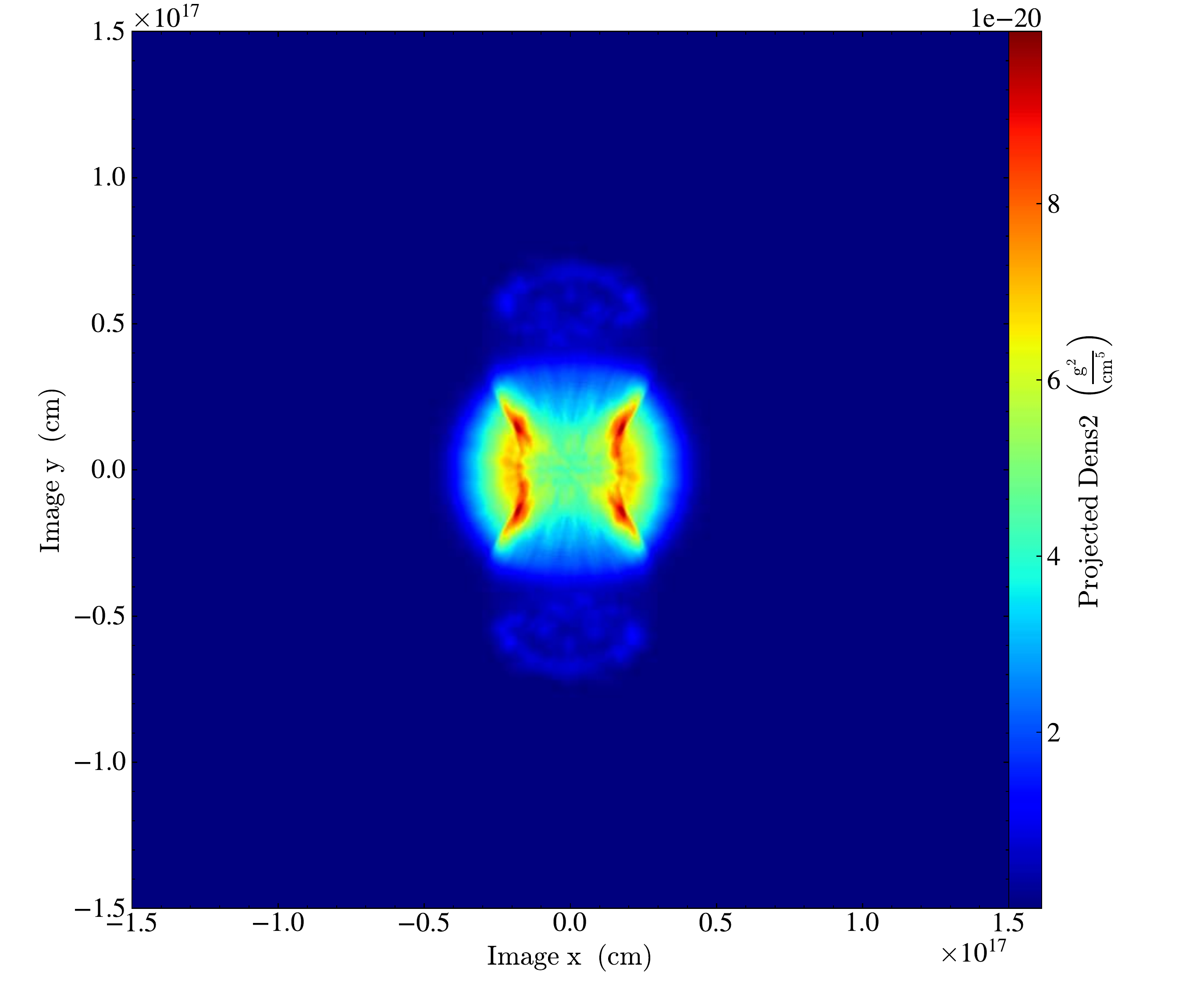}}

\subfigure{\includegraphics[height=1.3in,width=1.5in,angle=0]{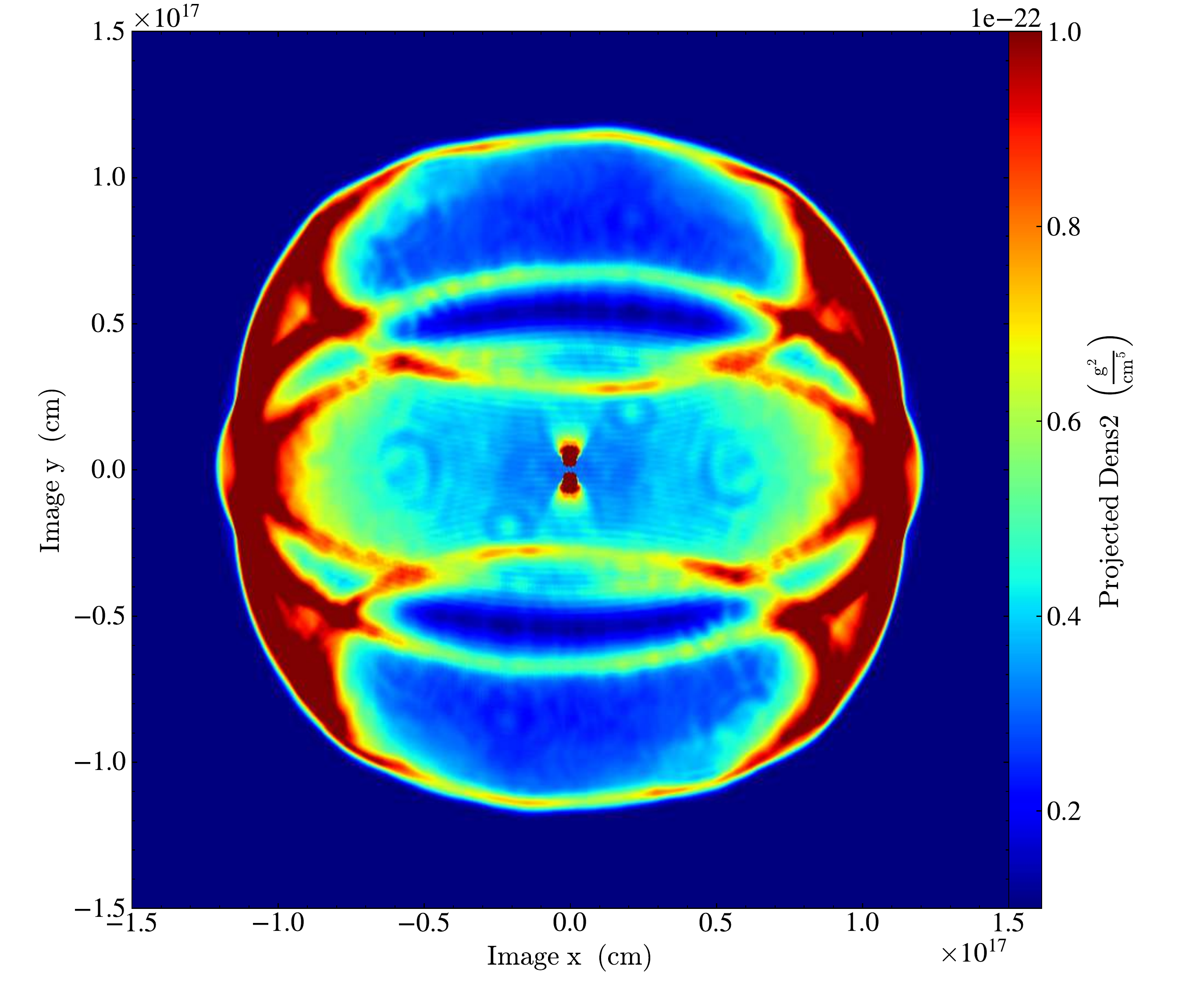}}
\subfigure{\includegraphics[height=1.3in,width=1.5in,angle=0]{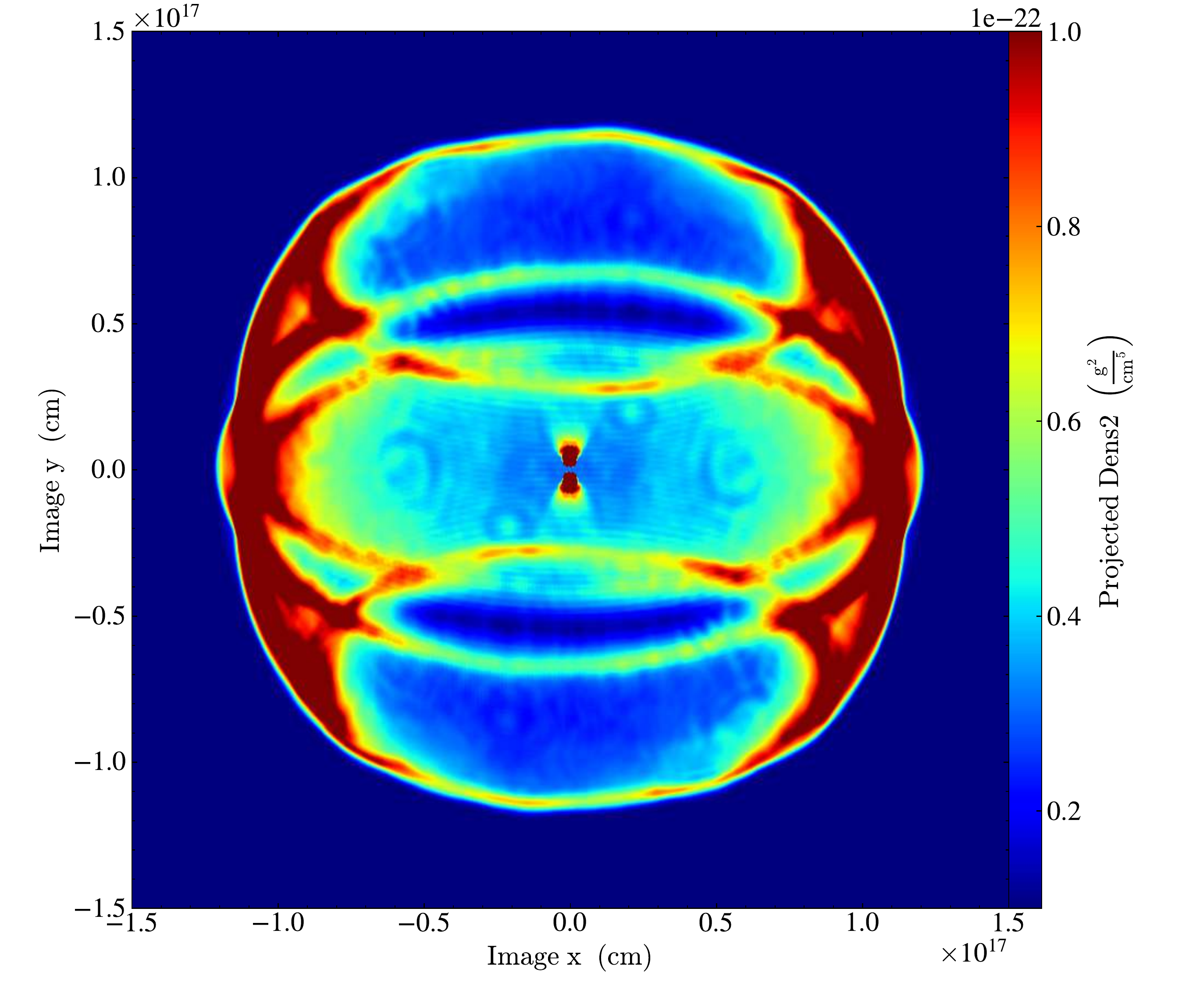}}

\subfigure{\includegraphics[height=1.3in,width=1.5in,angle=0]{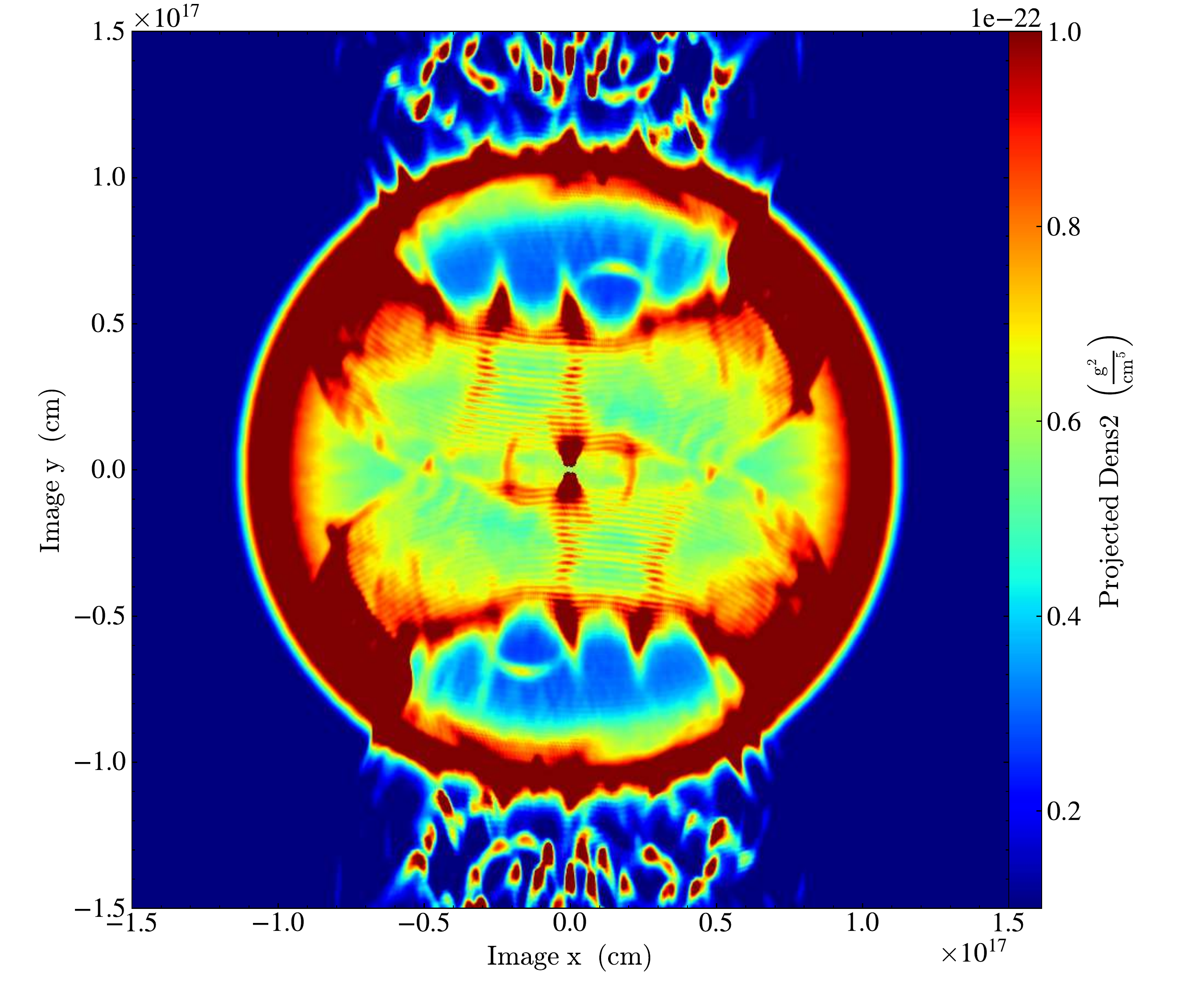}}
\subfigure{\includegraphics[height=1.3in,width=1.5in,angle=0]{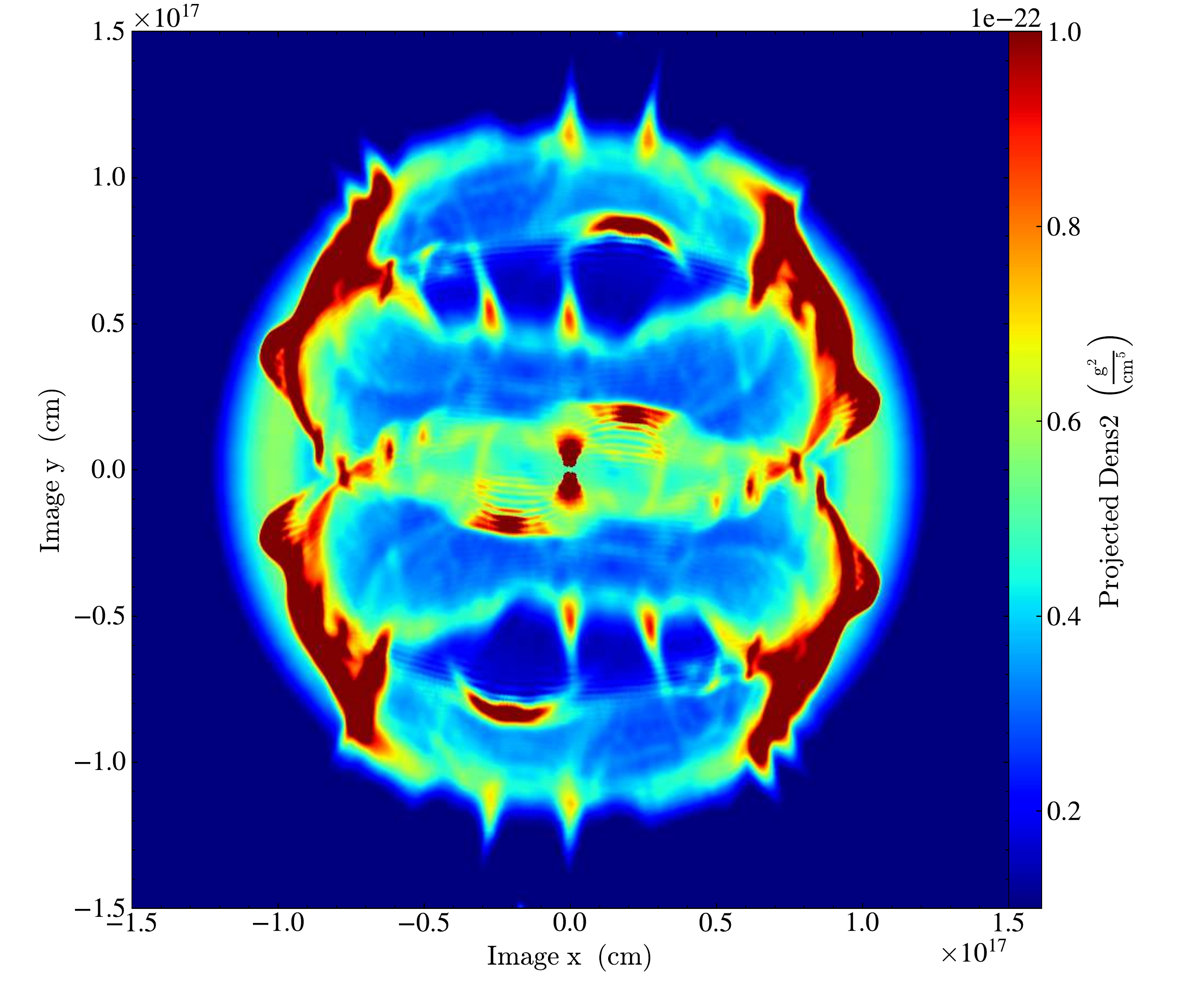}}

\subfigure{\includegraphics[height=1.3in,width=1.5in,angle=0]{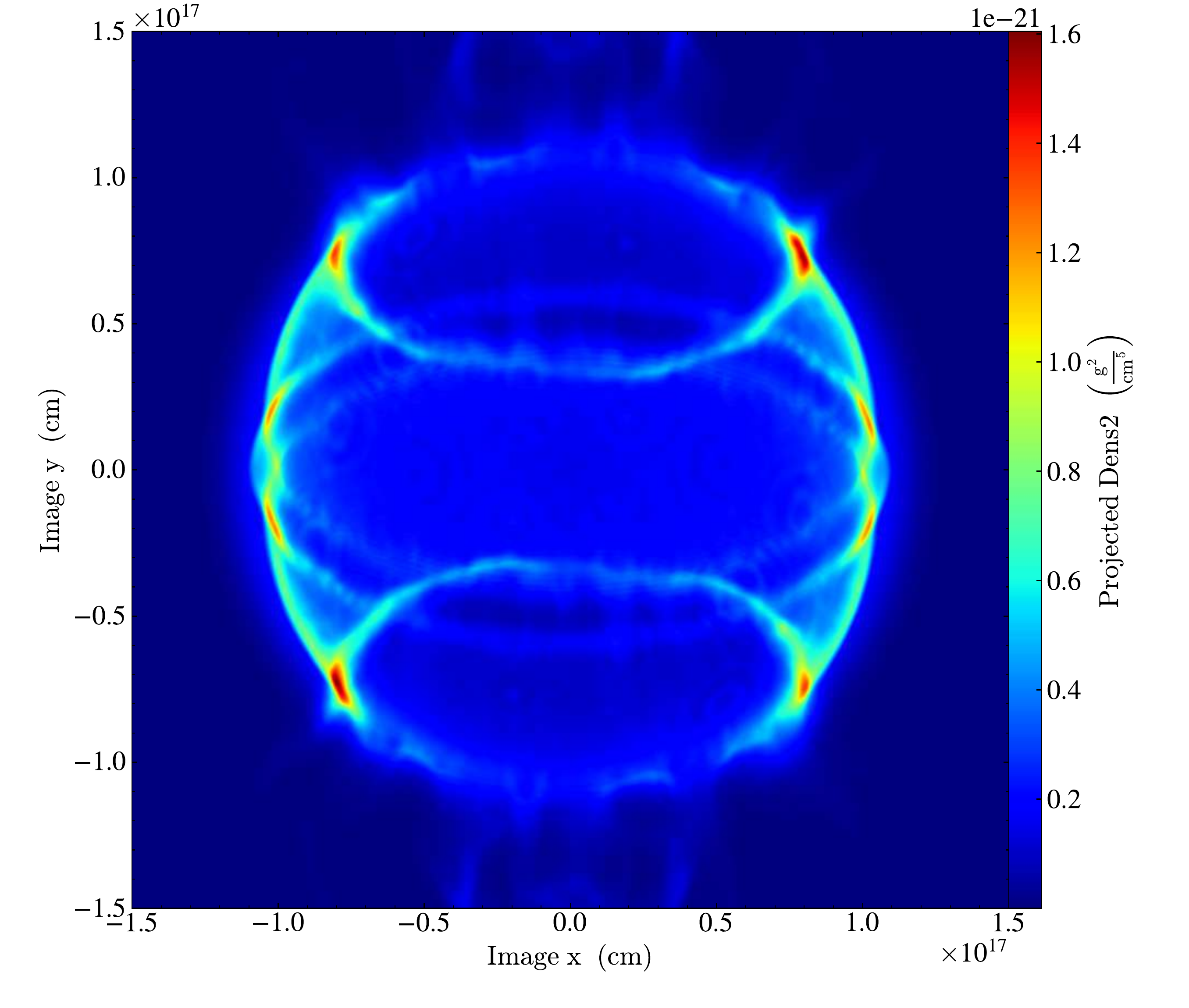}}
\subfigure{\includegraphics[height=1.3in,width=1.5in,angle=0]{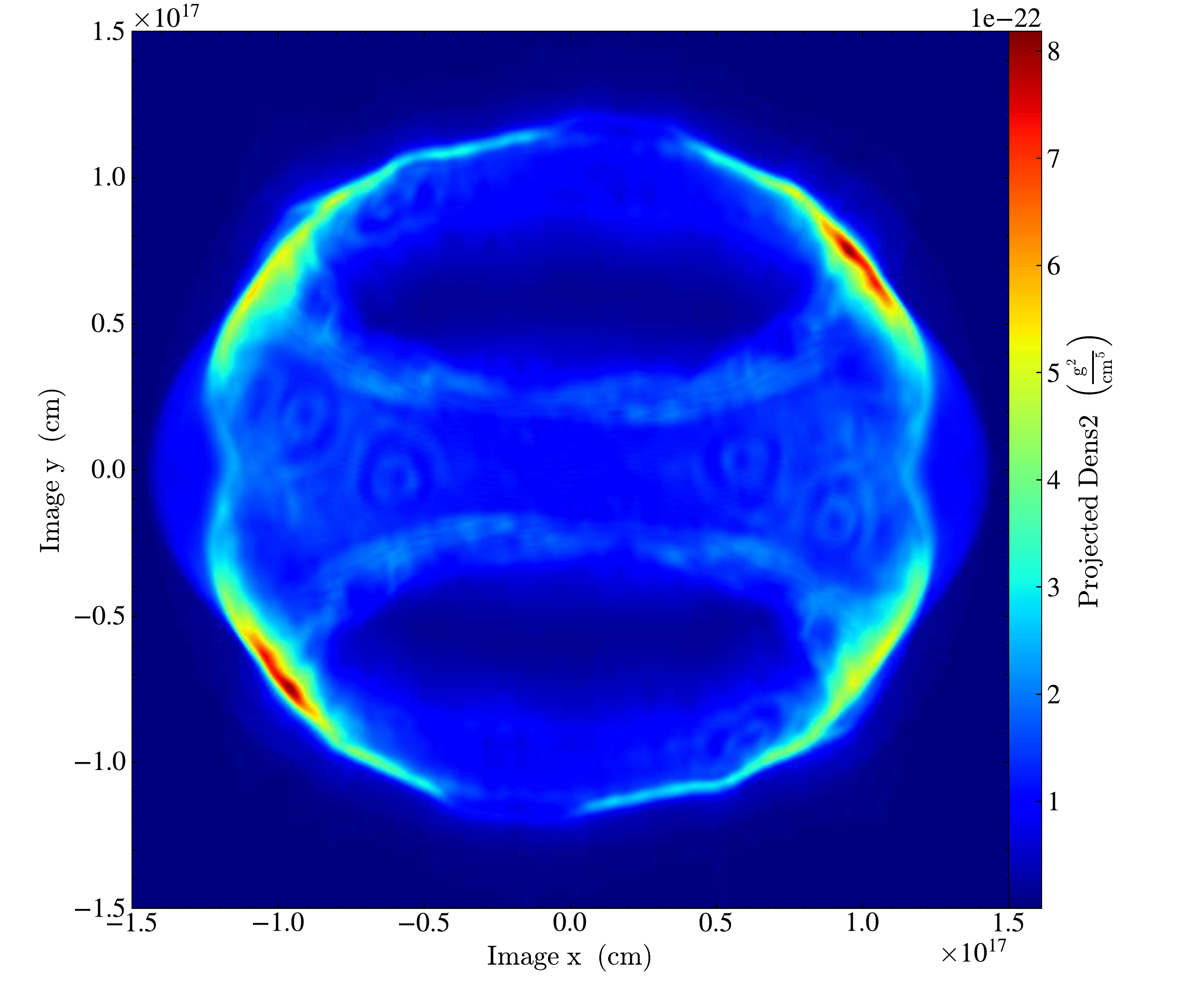}}

\caption{Projection maps from a viewing angle of $20^\circ$ relative to the equatorial plane. The rows from top to bottom corresponds to the upper row of Figs. \ref{fig:R1}-\ref{fig:R5}, respectively. }
 \label{fig:OFFAXIS}
\end{center}
\end{figure} 
 
The main new results from our simulations is the formations of barrel-like and H-like morphological structures as a result of the interaction of two opposite jets with a dense shell. These morphological features evolve during the short time of the simulations. After the jets cease and the gas cools the nebula will continue to freely expand and the morphological features will freeze. We can summarize that jet-shell interaction can form barrel-like and H-like nebulae. 

\section{COMPARISON TO OBSERVED PLANETARY NEBULAE}
 \label{sec:PNe}
  
In this section we only address some PNe that possess barrel-like or H-like shaped features. We present here only 12 PNe, but tens of other with similar morphological features exist. 
We consider only images presenting H-shaped and barrel-shaped morphological features, and we do not address other properties of these 12 PNe. We emphasize again that the flow setting we simulate here is not the only one that can lead to these morphological features (see section \ref{subsec:nebular}). 

\begin{figure}[!t]
\centering
\vskip -2.3 cm
\includegraphics[trim= 0.0cm 0.0cm 0.0cm 0.0cm,clip=true,width=1.0\textwidth]{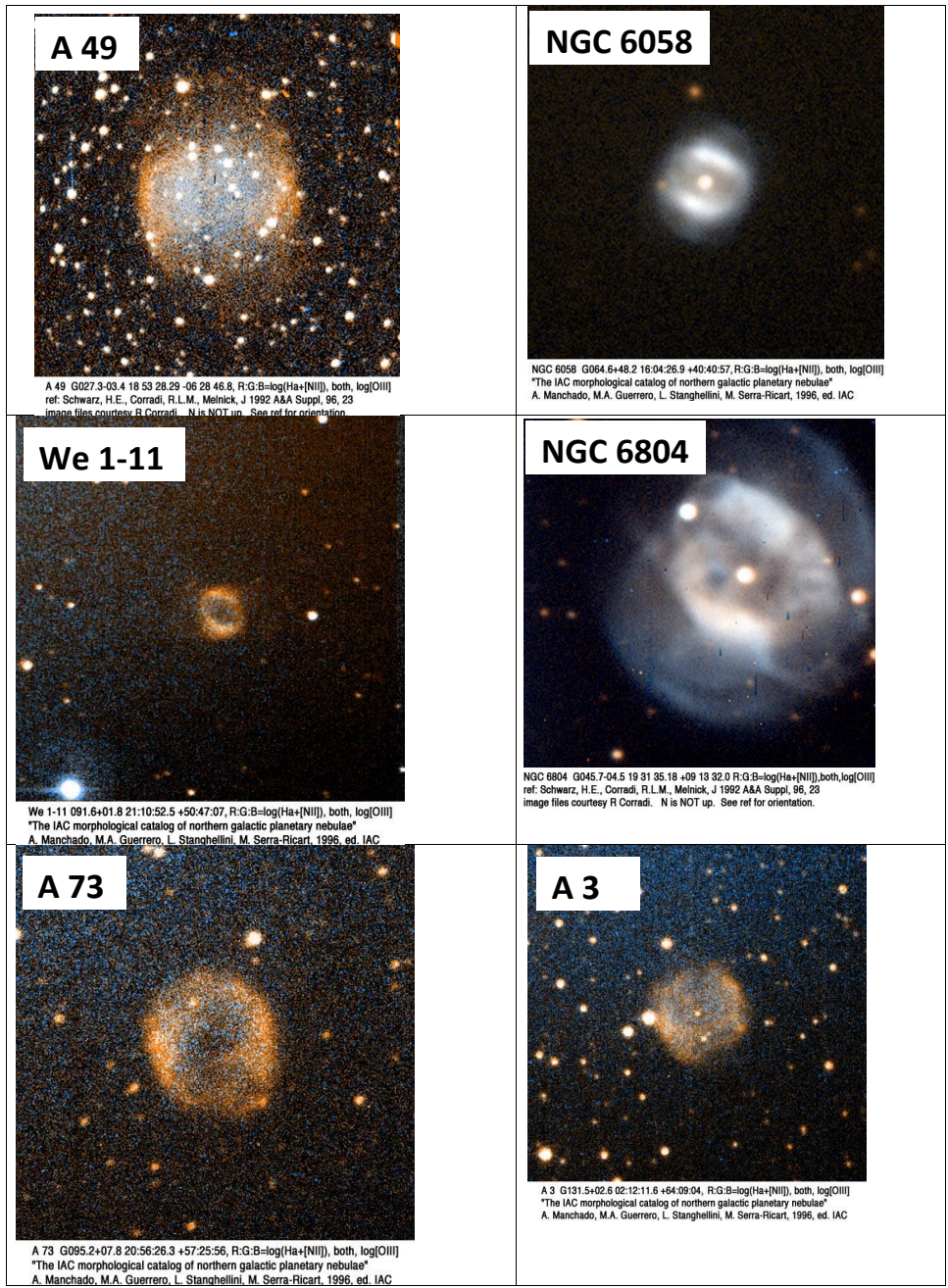}
\vskip -2.3 cm
\caption{The 6 panels in this figure show 6 PNe which share similar morphology,
the barrel-shaped structure. All pictures are taken from the catalog of Bruce Balick (based on \citealt{Balick1987}). Credit: A 49 (PN A66 49, G027.3-03.4)-\cite{Schwarzetal1992}; NGC 6058 (G064.6+48.2)-\cite{Manchadoetal1996}; We 1-11 (G091.6+01.8)-\cite{Manchadoetal1996}; NGC 6804 (G 045.7-04.5)-\cite{Schwarzetal1992}; A 73 (PN A66 73, G095.2+07.8)-\cite{Manchadoetal1996}; A 3 (PN A66 3, G131.5+02.6)-\cite{Manchadoetal1996}.}
  \label{fig:PNe_Lobes}
\end{figure}

Let us start with PNe that have a general barrel-shaped morphological feature, that we present in Fig. \ref{fig:PNe_Lobes}, and compare them to our simulations that we present in the two middle panels of Figs. \ref{fig:R1}, right middle panel of Fig. \ref{fig:R3}, and left middle panel of Fig. \ref{fig:R5}.
The general appearance of observed barrel-shaped nebulae and simulations is composed of two opposite bright arcs that are symmetrically trimmed on the both ends.  
We note that the core collapse supernovae remnant RCW~103 also has a general barrel-shape morphology, and \cite{Bearetal2017} assumed indeed that the structure of RCW~103 was formed by jets. Our results put on a solid ground the notion that barrel-like structures are formed by two opposite jets.  

Although our simulations can reproduce the general barrel-shaped morphologies, that appear as two opposite arcs trimmed on their ends, there are other features that require more ingredients in the simulations. These ingredients that can be added to the simulations strengthen the claim that jets can form most of the observed morphological features. (1) As evident already from the six PNe the we present in Fig. \ref{fig:PNe_Lobes} there is a rich variety of barrel-shaped morphologies. There are several parameters in our simulations that can lead to this variety of barrel-shapes. These include the properties of the shell (radius, width and mass), the mass-loss rate into the slow wind before and after the ejection of the shell, the velocity of the jets and their opening angle, and for most, the possibility that the slow wind and the shell depart from spherical symmetry.
In the present study we changed only the size of the shell, the duration of the jets, and the radiative cooling rate of the gas (see Table \ref{table:Runs}). More studies with variations of the other parameters are required.

(2) The second observed feature that requires ingredients that we did not include is the departure from axisymmetry. The two arcs in NGC 6804 and A~73 are not identical. In general a departure from pure-axisymmetrical structures might result from the orbital motion of the companion that launches the jets (e.g., \citealt{SokerHadar2002, GarciaArredondoFrank2004}), in particular if the orbit eccentricity is not zero. If the duration of the jet-launching episode is not much longer than the orbital period, and/or if the orbit is eccentric, then the axi-symmetrical symmetry is broken. This is the way the jet-ambient gas interaction can account for departure from axisymmetry. Actually, in the binary model that includes shaping by jets it is expected that some nebulae will depart from pure axisymmetry. 

(3) Our numerical results can explain the formation of clumps at the two ends of the barrel-shape (toward the lobes), e.g., the very faint regions of A 49, as coming from instabilities. What we cannot account for is the unequal shapes at the two opposite ends of the barrel, e.g., as in NGC 6804. The symmetry about the equatorial plane is broken. Interaction with the interstellar medium or the presence of a tertiary star might account for the broken mirror-symmetry.

\begin{figure}[!t]
\centering
\vskip -2.5 cm
\includegraphics[trim= 0.0cm 0.0cm 0.0cm 0.0cm,clip=true,width=1.0\textwidth]{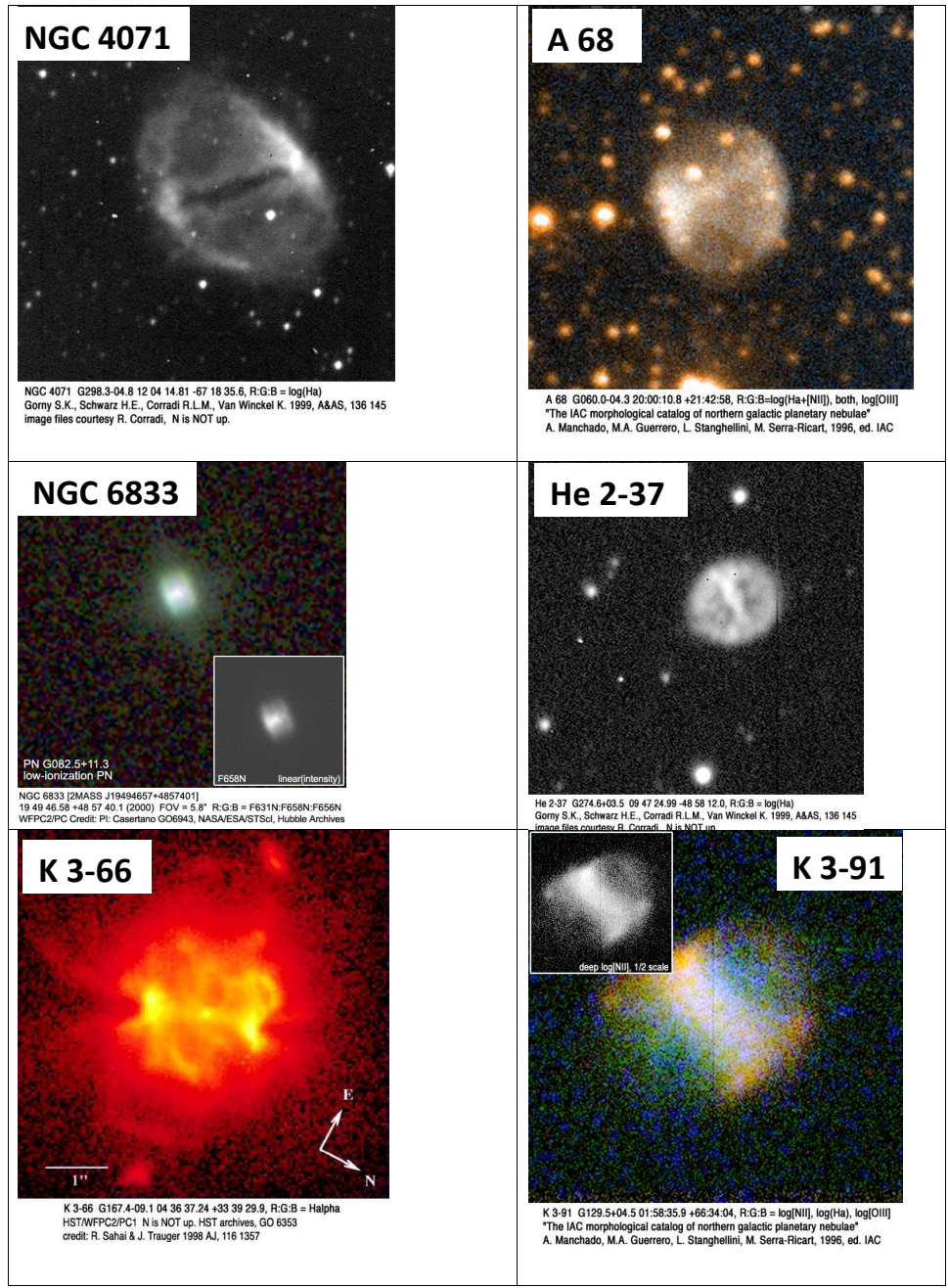}
\vskip -2.5 cm
\caption{The 6 panels in this figure show 6 PNe which share the 'H-shaped' morphology (from the catalog of Bruce Balick based on \citealt{Balick1987}). Credits: NGC 4071 (Hen 2-75, G298.3-04.8)-\cite{Gornyetal1999}; A 68 (PN A66 68, G060.0-04.3)-\cite{Manchadoetal1996}; NGC 6833 (2MASS J19494657$+$4857401)-\cite{Leeetal2010}; He 2-37 (WRAY 16-59, G274.6$+$03.5)-\cite{Gornyetal1999} ; 
K 3-66 (G167.4-09.1; PK 167-09 1)-\cite{SahaiTrauger1998}; 
K 3-91 (G129.5+04.5)-\cite{Manchadoetal1996}}
  \label{fig:PNe_HSHape}
\end{figure}

We presented the H-shaped morphological feature that we have obtained in our simulations in the two lower panels of Fig. \ref{fig:Flow1xz} and in the right panel in the fourth row of Fig. \ref{fig:OFFAXIS}. Traces of an H-shape are seen also in the middle panels of Fig. \ref{fig:R3}, and the right-middle panel of Fig. \ref{fig:R4}. 
We compare them to 6 PNe that we present in Fig. \ref{fig:PNe_HSHape}. In some cases the H-shaped feature is a dense structure embedded in a larger and fainter nebula. 
The signature of an H-shaped structure is a bright bar through the centre with two perpendicular bars attached to its edges.  
The H-shaped is not only present in PNe but can also be seen in some supernovae remnants such as W49B. \cite{BearSoker2017} assumed that the structure of W49B was shaped by jets. Our results show that two opposite jets can indeed form a bright H-shaped morphological feature within a larger nebula.
 
Let us elaborate on the observed H-shapes, and discuss what we can account for and what morphological features require more ingredients in the simulations. First point to make is that the general H-shaped structure is observed also in X-ray images of hot gas in some galaxies and clusters of galaxies, e.g., in the elliptical galaxy M84 \citep{FinoguenovJones2001}. In those cases radio emission teaches us that two opposite jets launched from an active galactic nucleus formed the H-shape. \cite{SokerBisker2006} compared the structure of the hot gas that is shaped by jets in galaxies and clusters of galaxies with PNe, and strengthened the case for jet-shaping in PNe. 
Our results show that jets can indeed form H-shaped nebulae. This similarity is one of our motivation to consider shaping by jets.
  
As in some barrel-shaped PNe, we can see in Fig. \ref{fig:PNe_HSHape} that NGC~4071, A~68, and K-3-91 possess departure from pure axi-symmetry or even from mirror symmetry in A~68. As discussed for barrel-shaped PNe, eccentric orbits and/or a tertiary star might account for the symmetry breaking.

Our simulations can account for instabilities, e.g., as appear in He~2-37. However, the large blobs and filaments seen in K~3-66 seem to require another ingredient. This might by instabilities in the mass transfer process from the AGB progenitor to the companion that launches the jets, and/or clumpy slow wind from the AGB progenitor. 
 
In this work we show from simulations that the morphological features of barrel-like and the H-like shapes can result from jets. We also compare them to observations of PNe for which two of them are already known to include jets, namely, K~3-66 \citep{SahaiTrauger1998} and possibly NGC~6833 \citep{Wrightetal2005, Leeetal2010}. These two PNe add to our motivation to consider shaping by jets.

\section{SUMMARY}
\label{sec:summary}
 
 The goal of our study is to extend the morphological features that jets can form. In the present paper we examined the formation of barrel-like and H-like shapes. For that we conducted 3D hydrodynamical simulations of fast jets interacting with a spherical dense shell. This interaction leads to complicated flow structures including instabilities and vortices, as we show in Figs. \ref{fig:Flow1xz}-\ref{fig:Flow1xy}.  
  
  In figures \ref{fig:R1}-\ref{fig:OFFAXIS} we present the morphologies at two times for each one of the five cases we have simulated. The main new results of our simulations are the formations of barrel-like and H-like morphological structures as a result of this interaction of two opposite jets with a dense shell.
The dense shell is formed by a short instability phase of the giant star. Since the system is assumed to be binary, a companion, through gravitational attraction and the orbital motion, might shape the dense shell into an elliptical shape. This will not change the qualitative results. The jets are then launched as the giant star transfers mass to the secondary star.
As the accretion takes place through an accretion disk, the secondary star launches two opposite jets. 
 We find that the H-shaped and barrel-shaped morphological features evolve with time, and that there are complicated flow patterns, such as vortices, instabilities, and caps moving ahead along the symmetry axis.
 
Barrel-like and/or H-like morphological features are observed in several types of nebulae (see section \ref{sec:intro}), mostly in PNe. Although we were not aiming at specific nebulae, but rather to obtain the general morphological features, we nonetheless present in section \ref{sec:PNe} twelve PNe with either H-like or Barrel-like shapes. We suggest that the jet-shell interaction that we have simulated in the present study can account for the barrel-like or H-like morphologies that are observed in these PNe. 

Our study adds to the rich variety of observed nebular structures and morphological features that the interaction of jets with an ambient medium can account for. On a wider scope, jets not only shape nebulae, but they can also energize and trigger the formation of nebulae, by, e.g., exploding massive star and ejecting common envelopes (see review by \citealt{Soker2016Rev} and references therein). As such, the present results further increase the general role that jets play in the evolution of stars. 
     
We thank an anonymous referee for very helpful comments. This research was supported by the Asher Fund for Space Research at the Technion and the Israel Science Foundation. 
N.S. is supported by the Charles Wolfson Academic Chair.


\label{lastpage}

\end{document}